\documentclass{aa}
\usepackage{graphicx,natbib,hyperref}
\usepackage[varg]{txfonts}
\usepackage{pdflscape}
\usepackage{afterpage}
\usepackage{color}
\usepackage{upgreek}
\usepackage{placeins}

\newcommand{\Prot}{\ensuremath{P_{\mathrm{rot}}}}
\newcommand{\vsi}{\ensuremath{v\,\sin i}}
\newcommand{\kms}{km\,s$^{-1}$}
\defcitealias{2020A&A...639A..31M}{Paper~I}

\bibpunct[ ]{(}{)}{;}{a}{}{,}

\begin{document}

\title{Long-period Ap stars discovered with TESS data: \\The northern ecliptic hemisphere}

\author{G.~Mathys\inst{1}
  \and D.~W.~Kurtz\inst{2,3}
  \and D.~L.~Holdsworth\inst{3}}

\institute{European Southern Observatory,
  Alonso de Cordova 3107, Vitacura, Santiago, Chile\\\email{gmathys@eso.org}
\and
Centre for Space Research, Physics Department, North West University, Mahikeng 2735, South Africa
\and
Jeremiah Horrocks Institute, University of Central Lancashire, Preston PR1 2HE, UK}

\date{Received $\ldots$ / Accepted $\ldots$}

\titlerunning{Long period Ap stars}

\abstract
{The rotation periods of the magnetic Ap stars span five to six orders
of magnitude. While it is well established that period differentiation
must have taken place at the pre-main sequence stage, the physical
processes that lead to it remain elusive. The existence of Ap stars
that have rotation periods of tens to hundreds of years is
particularly intriguing, and their study represents a promising avenue
to gain additional insight into the origin and evolution of rotation
in Ap stars.

Historically, almost all the longest period Ap stars known have been found to be
strongly magnetic; very few weakly magnetic Ap stars with very
long periods have been identified and studied. To remedy that,
we showed how a systematic search 
based on the analysis of TESS photometric data could be performed
to identify super-slowly rotating Ap (ssrAp) stars independently of
the strengths of their magnetic fields, with the intention to
characterise the distribution of the longest Ap star rotation periods
in an  unbiased manner. We successfully applied this method to the analysis of the
TESS 2-min cadence observations of Ap stars of the southern ecliptic
hemisphere. 

For our present study, we applied the same approach to the analysis of the TESS 2-min cadence 
observations of Ap stars of the northern ecliptic hemisphere. We
confirm that the technique leads to the reliable identification of ssrAp star
candidates in an unbiased manner. We find 67 Ap stars with
  no rotational variability in the northern
ecliptic hemisphere TESS data. Among them, 46 are newly
identified ssrAp star candidates, which is double the number found in the southern ecliptic hemisphere. 

We confirm that super-slow rotation tends to occur less frequently in
weakly magnetic Ap stars than in strongly magnetic stars. We present
new evidence of the existence of a gap between $\sim$2\,kG and
$\sim$3\,kG in the distribution of the magnetic field strengths of
long period Ap stars. We also confirm that the incidence of
roAp stars is higher than average in slowly rotating Ap stars. We
report the unexpected discovery of nine definite and five candidate
$\delta$~Sct stars, and of two eclipsing binaries. This work paves the
way for a systematic, unbiased study of the longest period Ap stars,
with a view to characterise the correlations between their rotational,
magnetic, and pulsational properties.
}

\keywords{stars: chemically peculiar --
  stars: magnetic field --
  stars: rotation --
  stars: oscillations}

\maketitle

\section{Introduction}
\label{sec:intro}
It is now well established that 5 to 10\% of all stars of spectral
types O to early F that have radiative atmospheres host strong,
predominantly dipolar large-scale organised magnetic fields
\citep{2017MNRAS.465.2432G,2017A&A...599A..66S,2019MNRAS.483.2300S}.
Among these
magnetic early-type stars, it is convenient to distinguish three
groups in order of increasing temperature: the chemically peculiar A
and B stars (Ap and Bp stars, which are often collectively referred to
simply as Ap stars), the magnetic early B stars (with spectral
types ranging from B5 to B0), and the magnetic O stars. The
distribution of the dipole field strengths is to first order similar
for all three types \citep{2019MNRAS.490..274S}.

Furthermore, the cumulative distributions of the rotation periods of
the Ap, magnetic early B, and magnetic O stars are also similar to
first order \citep{2018MNRAS.475.5144S}. All of these stars rotate, on
average, more slowly than their non-magnetic counterparts with similar
temperatures, and their rotation periods span five to six orders of
magnitudes, from $\sim$0.5\,d to at least $\sim$300\,yr, and possibly
as much as $\sim$1000\,yr
\citep{2017A&A...601A..14M}. Although recent 
observations suggest that rotation slows down over time in the
more massive stars that have higher mass loss, which
is qualitatively consistent with the occurrence of magnetospheric
braking \citep{2019MNRAS.490..274S}, such a loss of angular momentum
on the main sequence is at most marginal for Ap stars
\citep{2006A&A...450..763K,2007AN....328..475H}. Since evolutionary
changes to the rotation periods of Ap stars during their main-sequence lifetimes
are small, most of the period differentiation must have occurred at
the pre-main sequence stage. Accordingly, consideration of the present
distribution of the rotation periods provides valuable constraints towards
the theoretical understanding of the physical mechanisms responsible
for this differentiation.

\afterpage{\clearpage}
\begin{figure*}
  \centering
  \includegraphics[width=0.48\linewidth,angle=0]{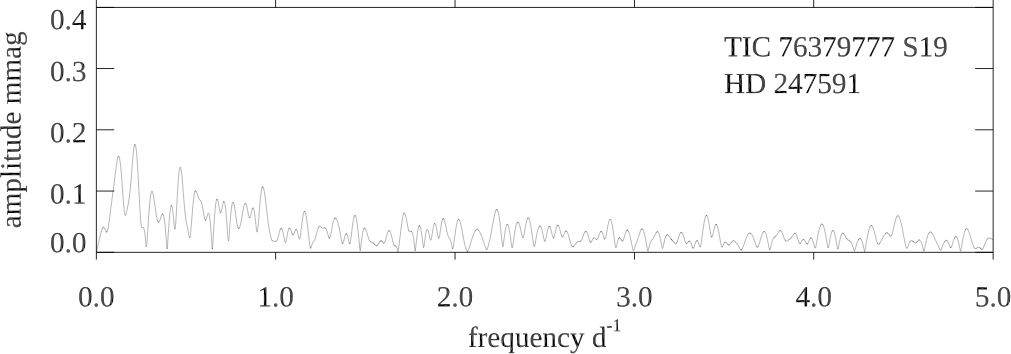}
  \includegraphics[width=0.48\linewidth,angle=0]{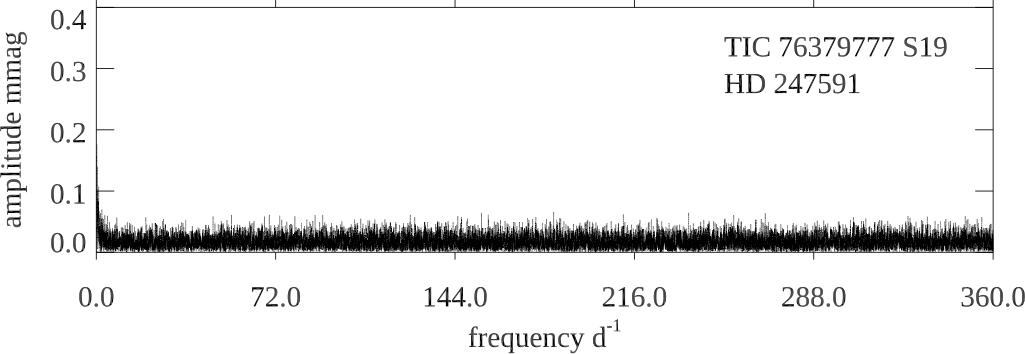}
  \includegraphics[width=0.48\linewidth,angle=0]{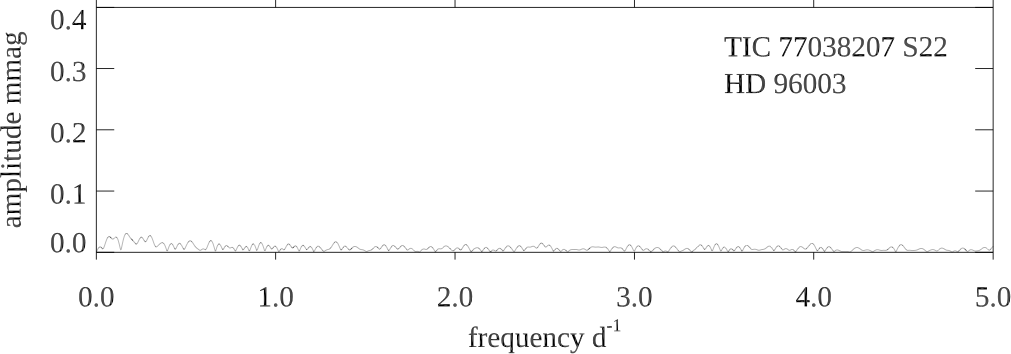}
  \includegraphics[width=0.48\linewidth,angle=0]{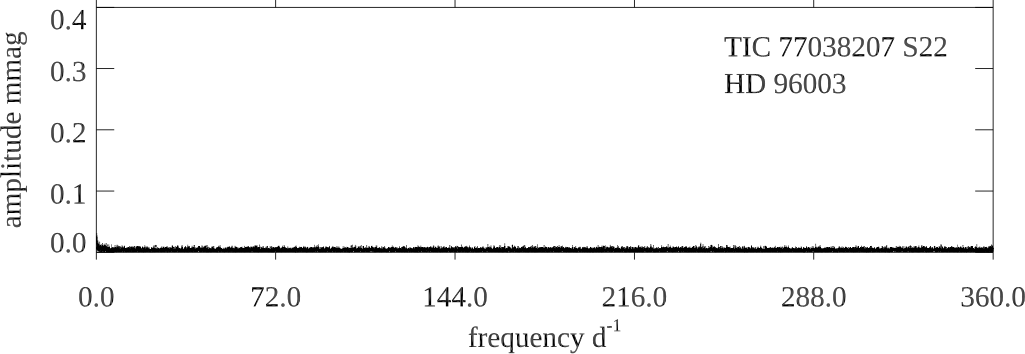}
  \includegraphics[width=0.48\linewidth,angle=0]{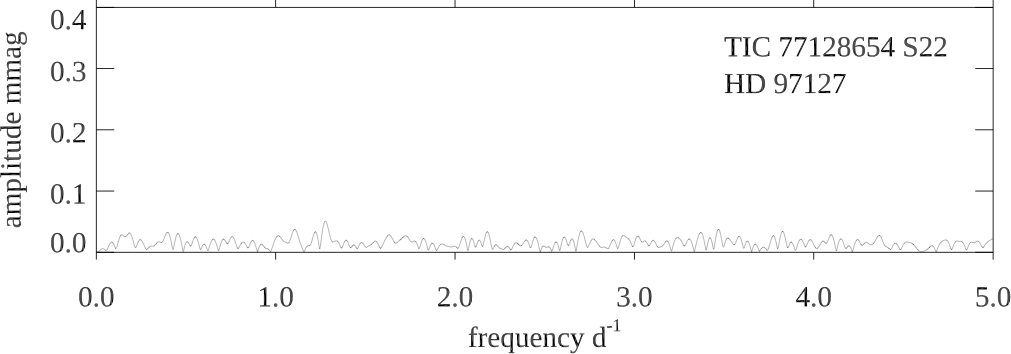}
  \includegraphics[width=0.48\linewidth,angle=0]{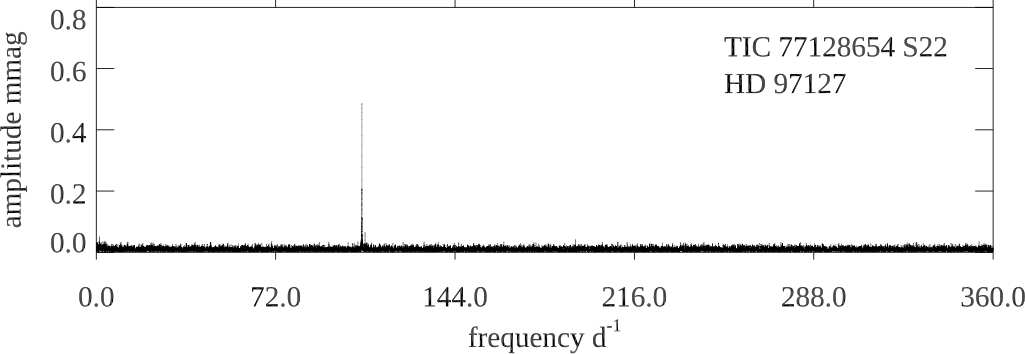}
  \includegraphics[width=0.48\linewidth,angle=0]{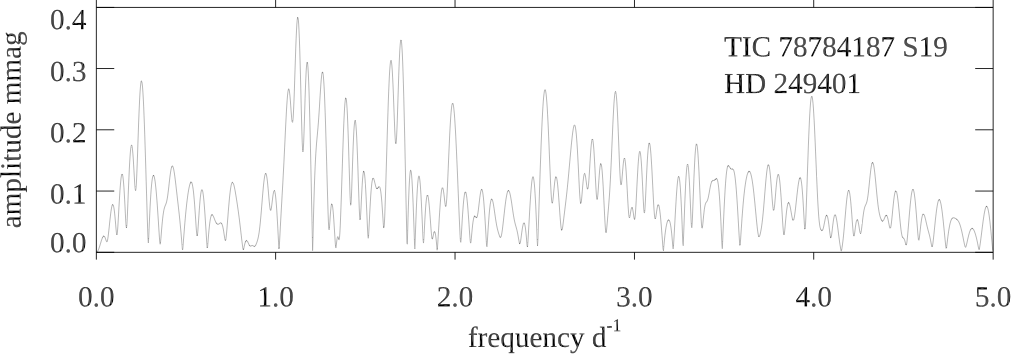}
  \includegraphics[width=0.48\linewidth,angle=0]{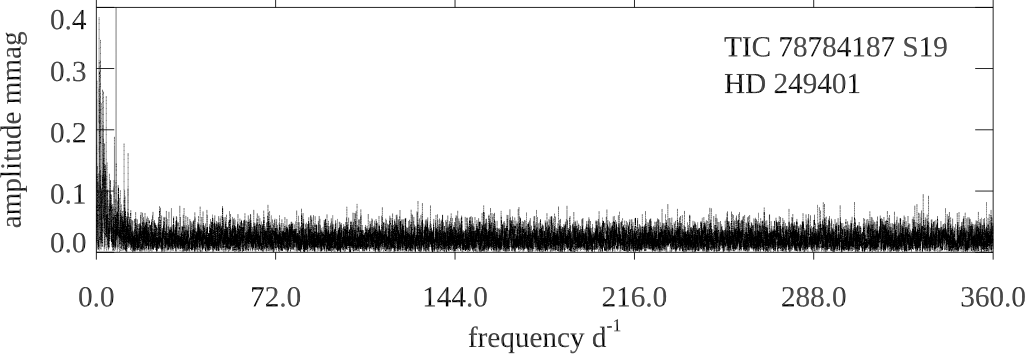}
  \includegraphics[width=0.48\linewidth,angle=0]{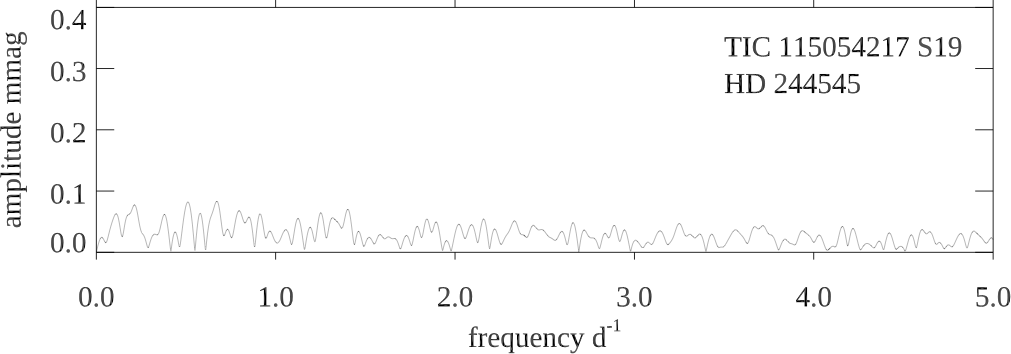}
  \includegraphics[width=0.48\linewidth,angle=0]{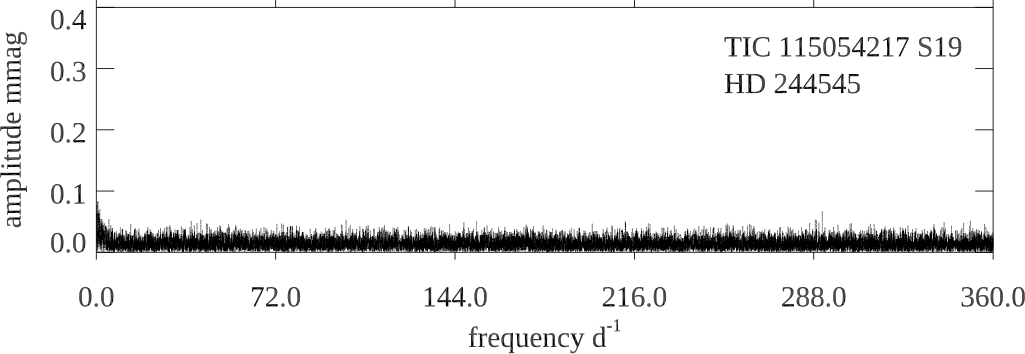}
  \includegraphics[width=0.48\linewidth,angle=0]{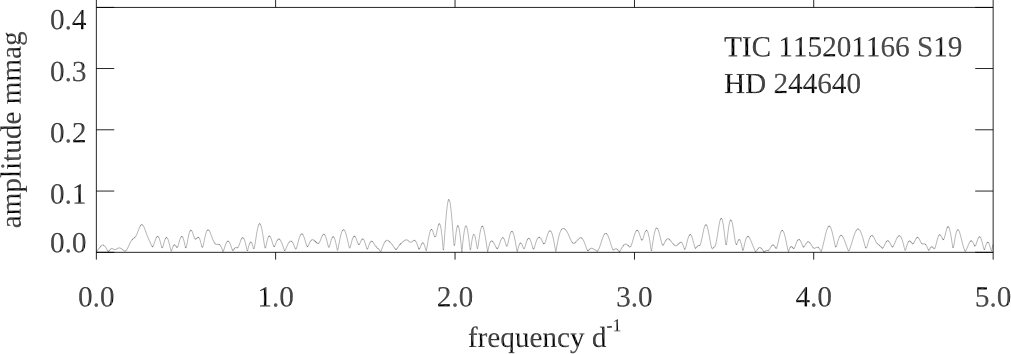}
  \includegraphics[width=0.48\linewidth,angle=0]{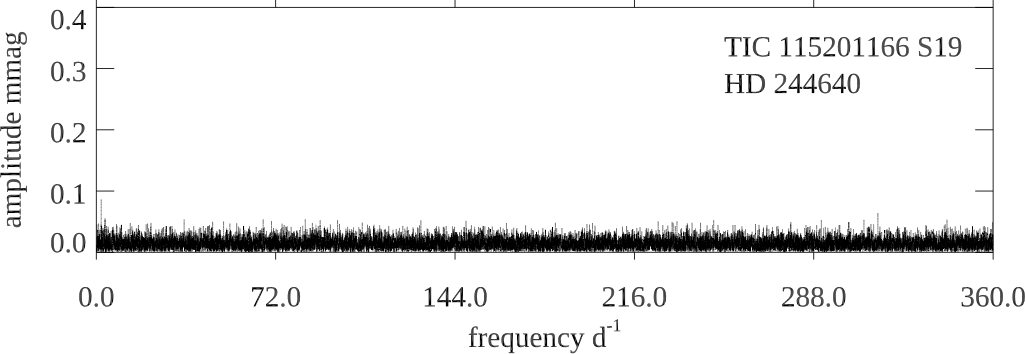}
  \includegraphics[width=0.48\linewidth,angle=0]{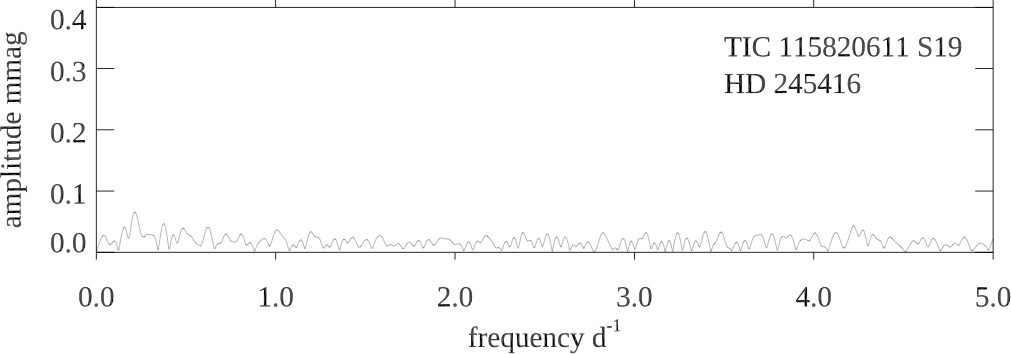}
  \includegraphics[width=0.48\linewidth,angle=0]{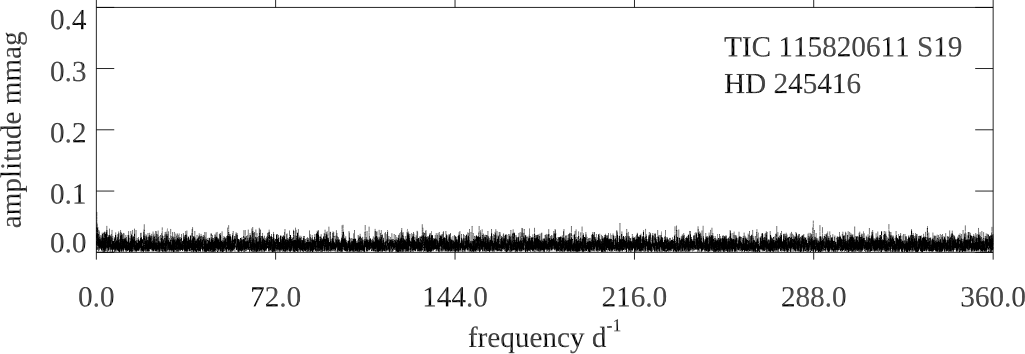}
  \caption{Amplitude spectra for the long-period Ap stars. Each row
    presents a low-frequency amplitude spectrum showing no rotational
    variation in the left panel, and a full amplitude spectrum to the
    Nyquist frequency of 360\,d$^{-1}$ in the right panel, which
    allows detection of $\delta$~Sct or roAp pulsation. We note the
    occasional changes of ordinate scale to accommodate pulsation
    peaks. TIC\,77128654 (HD\,97127) is an roAp star (section\,\ref{roAp}). TIC\,78784187 (HD\,249401) is a possible $\delta$~Sct star. }
  \label{fig:ssrAp2}
\end{figure*}

Even if the rotation rates of higher mass magnetic stars evolve more
on the main sequence than those of Ap stars, the arguments presented
by \citet[\!, hereafter Paper~I]{2020A&A...639A..31M} to support the
value of studying Ap stars for understanding the rotational
properties of the whole population of magnetic early-type stars remain
valid. Many more Ap stars are known than magnetic early B and
O stars; Ap stars as a group have been studied for much longer
than the other magnetic early-type stars, and they lend themselves
better than the latter to the study of rotational properties. However, the
present knowledge of the slow-rotation tail of their period
distribution remains incomplete. 

In particular, the occurrence of extremely slow rotation in weakly
magnetic Ap stars is poorly characterised (as discussed in Sect.~\ref{sec:bfield}). 
This results, at
least in part, from observational biases
\citepalias{2020A&A...639A..31M}, but may also reflect actual
differences in the distribution of the rotation rates of stars with
different magnetic field strengths. Establishing if the distribution
of the rotation periods is similar or not for the weakly and strongly
magnetic stars represents an essential element for the theoretical
understanding of the rotational behaviour of Ap stars. 

In \citetalias{2020A&A...639A..31M}, we showed how TESS data can be
exploited to carry out a systematic search for long-period Ap star
candidates, in a way that is fully independent of their magnetic field
strengths. Following \citet{2020pase.conf...35M}, we define long
  period to be greater than 50\,d, and we 
  refer to these stars as `super slowly rotating Ap (ssrAp) stars'. 
The approach that is used for this search takes advantage of
the fact that Ap stars are oblique rotators, with a non-uniform
surface brightness whose distribution, which is stable over timescales
much longer than the rotation periods, 
  presents a certain degree of
symmetry about the predominantly dipolar component of the magnetic
field. As the latter is, in general, inclined with respect to the
rotation axis, rotationally induced periodic photometric variations
are observed, with typical amplitudes of a few hundredths of a
magnitude. 

Most of the Ap stars observed with TESS clearly show such
low-frequency  ($\nu_{\rm rot} \la 2$\,d$^{-1}$; $P_{\rm rot} \ga 0.5$\,d) photometric variations due to
rotational modulation, 
sometimes together with superimposed high-frequency photometric
variations caused by pulsation. 
See Paper~I for some examples.
However, some do not show any
low-frequency variations over one, or several, TESS 27-d
sector(s). Almost all of those are slow rotators,
with periods $\Prot\ga27$\,d. The only exceptions are those stars for
which the 
inclination of the rotation axis to the line of sight is too small
($i\la5\degr$) or the obliquity of the magnetic axis with respect to
the rotation axis is too small ($\beta\la5\degr$). As discussed in
\citetalias{2020A&A...639A..31M}, under the assumption of random
  distribution of the inclinations and obliquities, stars for which
one or both of these 
conditions are fulfilled should be rare, representing less than 1\% of
all Ap stars. Admittedly, the distribution of the magnetic
  obliquities has not yet been uniquely constrained. The
  results of some works suggest that it may not be
  random, although their conclusions are discrepant, while other recent
  studies yield results consistent with random obliquities
  \citepalias[see][for more details]{2020A&A...639A..31M}.

\citetalias{2020A&A...639A..31M} was devoted to the analysis of the
TESS 2-min cadence observations of the Ap stars from the southern
ecliptic hemisphere (TESS Sectors 1--13). Of the 1014 stars for which
TESS SPOC (Science Processing Operations Center;
\citealt{2016SPIE.9913E..3EJ}) data were available, there were 60 for
which no low-frequency photometric variations were observed. For 31 of
them, existing constraints on the rotation period and/or $\vsi$ are
consistent with (very) slow rotation. For only six of the non-variable
stars, the available spectroscopic observations showed broad spectral
lines indicative of short rotation periods. These are probably
(relatively) fast rotators with nearly aligned magnetic and rotation
axes. Their rate of occurrence is compatible with the expected one for
low obliquity ($\beta\la5\degr$). The 23 remaining non-variable stars,
whose rotation periods are unknown and for which no high-resolution
spectra have been obtained, are newly identified long-period
candidates. Spectroscopic confirmation of their low $\vsi$ will
further establish the validity of the search technique introduced in
\citetalias{2020A&A...639A..31M}.

Hereafter, we apply this technique to the analysis of the TESS 2-min
cadence SPOC data of the Ap stars from the northern ecliptic
hemisphere (TESS Sectors 14--26). 

\section{TESS data}
\label{sec:TESS}

The sample of stars that was analysed consists of 492 stars that were classified as, or suspected to be, Ap
stars that had 2-min cadence data available. By running an automated detection algorithm
\citepalias[see][for details]{2020A&A...639A..31M} on this sample, a
preliminary list of 188 stars that did not show definitive photometric
variations over a time scale of 27\,d (or longer when more than one sector is available) was established. In a
second step, this list was critically reviewed to
restrict it to bona fide Ap stars. To this effect, we used as
a primary reference the Catalogue of
Ap, HgMn and Am stars \citep{2009A&A...498..961R}. We excluded from
our list those stars that are not present in the Catalogue
with an Ap spectral classification, as well as those stars whose
peculiar nature is flagged as doubtful (`?' in the first column in
the Catalogue) or that  have been improperly considered to have an
Ap nature (`/' in the first column). However, in a second step, we
reinstated in the list six (mostly faint) stars that are not in the
Catalogue but were subsequently identified as definite rapidly
oscillating Ap (roAp) stars. Then, we carried out a literature
survey using the SIMBAD Astronomical Database to check for possible
cases of stars from the Catalogue whose Ap nature had been disproved
in studies published more recently. We also removed these stars from
the list, which as a result was reduced to 95 entries.

As a final selection step, the TESS data obtained for the 95 remaining
stars were inspected manually to confirm the probable lack of
low-frequency variability diagnosed by application of the detection
procedure. This led to the elimination of another 28 stars, whose
light curves are similar to the typical rotational light curves of
short-period ($\Prot<27$\,d) Ap stars. The final
list, which contains 67 long-period Ap star candidates, is presented in
Table~\ref{tab:lpn} (Appendix~\ref{sec:ssrAplist}).

Amplitude spectra for some of the stars
that we have identified as probable long rotation period Ap stars are
shown in Fig.~\ref{fig:ssrAp2}. The amplitude spectra of the other
long-period Ap star candidates are shown in Appendix~\ref{sec:amp_sp}. 
The ordinate axis was chosen to have a maximum amplitude of 0.4\,mmag
for most stars to show easily the lack of a low frequency peak
associated with rotation in Ap stars at $\nu_{\rm rot} \le
2$\,d$^{-1}$. In the left column the maximum of the abscissa was
chosen to be 5\,d$^{-1}$ to encompass the low frequency range of Ap
rotation, to show some additional range so it can be seen that the
noise settles to be white by 5\,d$^{-1}$, and to show the typical
harmonic series for Ap stars that do show rotation. Of course, those
examples are not in the figures of this paper -- since they do not show
rotational variation -- but they were part of the selection process. The
right column has a maximum for the abscissa of 360\,d$^{-1}$, which is
the Nyquist frequency for the TESS 2-min data. These plots show that
the higher frequency noise is white, that the highest noise peaks are
usually (except for fainter stars) under 0.05\,mmag, and they show
$\delta$~Sct,  $\gamma$~Dor, and roAp pulsation, where it is
present. 

The list of Ap stars from the northern ecliptic hemisphere (TESS
Sectors 14--26) that show no rotational variation in the TESS data
contains 67 entries. These 67 stars are likely to have rotation
periods longer than 27\,d.

Eight of these stars show resolved magnetically split lines: seven
were listed in Table~1 of \citet{2017A&A...601A..14M}; the eighth one
(BD+44~3063) was reported by \citet{2011MNRAS.410..517B}. Of the
stars with resolved magnetically split lines that appear in
Table~\ref{tab:lpn}, the latter is the only one whose rotation period is not 
accurately known. Furthermore,
the list includes seven stars whose unresolved spectral lines appear
sharp, or rather sharp, in our high-resolution AURELIE spectra. Additionally,
there are three more stars, HD~17330, HD~92728, and HD~96003, with line
shapes that do not 
significantly differ from the instrumental profile in medium-high
resolution spectra according to published information, which indicates
an upper limit of $\sim$20\,km\,s$^{-1}$ for $v\,\sin i$. 

Accurate periods of variation are available in the literature for 14
of the stars of our list. For eleven of them, the periods are
definitely longer than $\sim$27\,d. The three exceptions
are HD~22860, for which we cannot exclude a short period, albeit
inconsistent with the published value \citep{2012MNRAS.420..757W},
HD~89069, for which there is no evidence in the TESS data of the value
of the period reported by \citet{2017MNRAS.468.2745N}, and HD~148330,
for which the value derived by \citet{1990BAICz..41..118Z} may be
spurious, as the amplitude of the photometric variations from which it
is determined, $\sim$0.02\,mag peak-to-peak, is inconsistent with the
lack of variability of the TESS data. For three more of the stars
whose periods are accurately known, the values are less than
50\,d: HD~14437 ($P_{\rm rot}=26\fd87$), HD\,12288 ($P_{\rm
  rot}=34\fd9$), and HD~335238 ($P_{\rm rot}=48\fd7$). These three
stars are not strictly speaking ssrAp stars as they do not fulfil the
criterion of $P_{\rm rot}>50$\,d defined by \citet{2020pase.conf...35M}. However,
this threshold is arbitrary, and unrelated to the detection threshold
of the present search, which is set by the 27-d length of a TESS
sector. The published periods are  consistent with the lack of
variability over a 27-d TESS sector. The three stars considered are
definitely slow rotators, 
albeit not so extreme as many others.

At least some magnetic field measurements exist for 17 of the 67 stars
of Table~\ref{tab:lpn}. Eight of the 17
stars have resolved magnetically split lines; seven have sharp,
unresolved lines; and for two, HD~17330 and HD~96003, $\vsi$ is at most of
the order of the instrumental profile width  (20\,\kms).

Out of the 67 stars of the list, five are roAp, and eight more may also
be. Assuming that they can be confirmed, 19\%\ of the {ssrAp} candidates 
of the present survey would be roAp stars -- less than the 22\% found
among the ssrAp candidates of  the southern ecliptic hemisphere
survey, but still considerably 
more than the overall occurrence rate among all the southern ecliptic
hemisphere Ap stars observed by TESS \citepalias{2020A&A...639A..31M}.

Unexpectedly, nine of the 67 stars of Table~\ref{tab:lpn} are definite
$\delta$~Scuti stars, and five more may also be. This is unusual for Ap
stars. Furthermore, two stars may be part of eclipsing binaries
-- also a rare occurrence among Ap stars. We look at these special
cases in more detail in the coming sections.  

\subsection{The roAp stars}
\label{roAp}

Our goal is to understand the origin and impact of super slow rotation in Ap stars. Our discovery of an apparently high incidence of roAp stars amongst the ssrAp stars suggests that super slow rotation may be conducive to, but not necessary for, pulsation driving in roAp stars. This is an unanswered question that our studies of ssrAp stars will address. We therefore assess the roAp stars individually here. The roAp candidates typically have a single peak in the roAp pulsation frequency range with at least a 5$\sigma$ confidence in amplitude. 

\subsubsection{TIC\,26749633, KIC\,11031749} 

TIC\,26749633 is a roAp star discovered by \citet{2019MNRAS.488...18H} in {\it Kepler} long cadence data using superNyquist asteroseismology \citep{2013MNRAS.430.2986M}. They found a single, unmodulated frequency at 118.6\,d$^{-1}$ with an amplitude of only 26\,$\upmu$mag. It is therefore unsurprising that we do not detect this pulsation in the TESS S14-15 data (Fig.\,A1), as the highest noise peaks are about  180\,$\upmu$mag as a consequence of the relative faintness of this star ($V=12.50$). \citet{2019MNRAS.488...18H} did find evidence of binarity using the phase modulation (PM) technique \citep{2018MNRAS.474.4322M}. They obtained an orbital period of 1036\,d and a mild eccentricity of $e = 0.2$. The mass function gives $M_2 \sin i = 0.1$\,M$_\odot$, indicating a lower main-sequence companion. Visual binaries are known among the roAp stars \citep{2012A&A...545A..38S}, including some of the brightest and well-known examples, such as $\alpha$\,Cir, $\gamma$\,Equ, and HR\,3831. For TIC\,26749633 an orbital period of near 3\,yr has easily escaped attention, showing the power of the PM technique with {\it Kepler} data. A wide binary such as this with a lower main-sequence secondary is not expected to have any impact on the formation of the Ap star. 
                                            
\subsubsection{TIC\,77128654, HD\,97127} 

TIC\,77128654 is a known roAp star discovered in SuperWASP data by \citet{2014MNRAS.439.2078H}, who referred to it in a brief form as J1110. They found a pulsation frequency at 106.61\,d$^{-1}$ with an amplitude of 0.66\,mmag through the WASP filter. In Fig.\,1 we also find a peak at 106.61\,d$^{-1}$ with an amplitude of 0.49\,mmag through the redder TESS filter, in good agreement with the WASP result. We also detect a second pulsation frequency at 107.88\,d$^{-1}$ with an amplitude of $60 \pm 9$\,$\upmu$mag. The frequency separation is 15\,$\upmu$Hz, which is possibly half the large separation. \citet{2014MNRAS.439.2078H} classified the star as F3p with $T_{\rm eff} = 6500$\,K, similar to the TIC $T_{\rm eff} = 6700$\,K given in Table\,1. This is one of the coolest roAp stars.    

\subsubsection{TIC\,158216369, KIC\,7018170}
 
TIC\,158216369 is a known roAp star discovered by \citet{2019MNRAS.488...18H} in {\it Kepler} long cadence data using superNyquist asteroseismology \citep{2013MNRAS.430.2986M}. They found one frequency quintuplet and two triplets equally split by 0.159\,$\upmu$Hz, which they inferred to be the rotation frequency, giving a rotation period of $P_{\rm rot} = 72.7$\,d. The amplitudes of all the frequencies they detected are too low for detection in the TESS data, where the noise peaks are around 400\,$\upmu$mag as a consequence of the faintness of the star ($V=13.26$). 

\subsubsection{TIC\,158275114, BD+44 3063}

TIC\,158275114 (KIC\,8677585) was observed for the full 4\,yr of the {\it Kepler} main mission in long cadence and for 830\,d in short cadence (1\,min). \citet{2013MNRAS.432.2808B} studied this star in detail, finding frequency variability and an interesting, unexplained low frequency at 3.142\,d$^{-1}$. The TESS data do show several peaks in the frequency range of those found by \citet{2013MNRAS.432.2808B}, but do not add to the much higher frequency resolution, higher signal-to-noise {\it Kepler} data.

\begin{figure*}
  \centering
 \includegraphics[width=0.48\linewidth,angle=0]{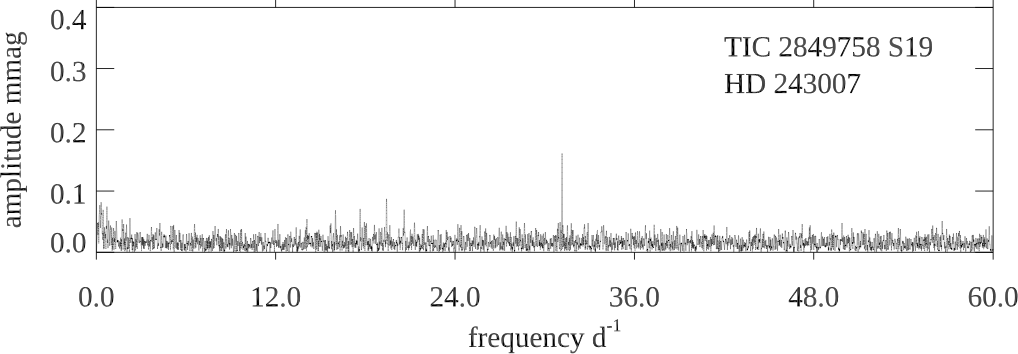}
 \includegraphics[width=0.48\linewidth,angle=0]{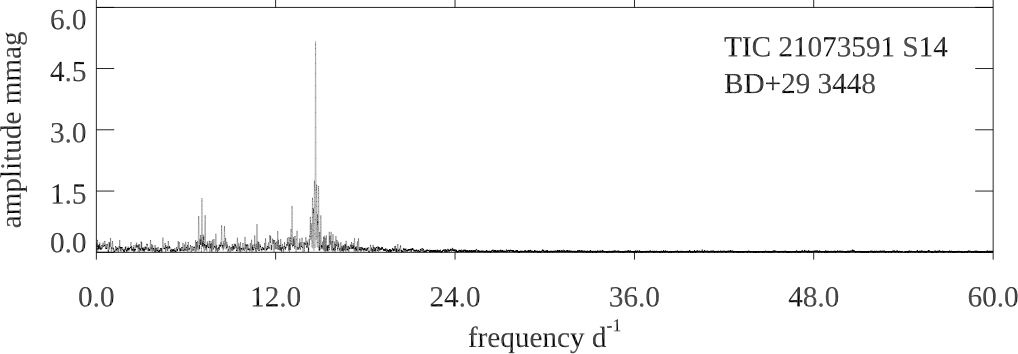}
 \includegraphics[width=0.48\linewidth,angle=0]{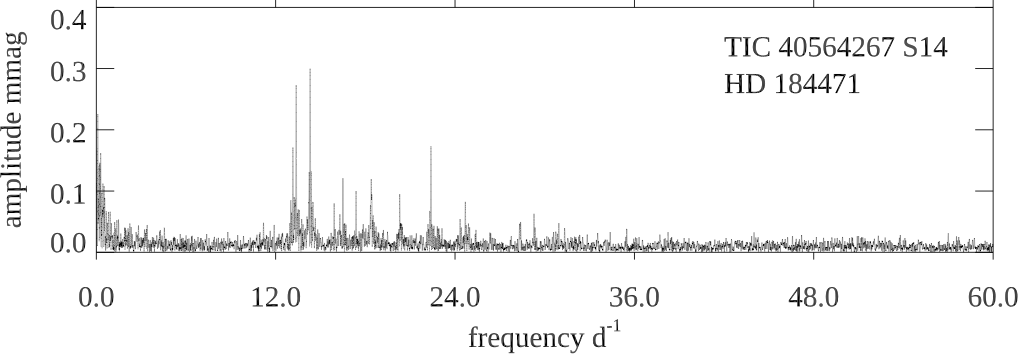}
 \includegraphics[width=0.48\linewidth,angle=0]{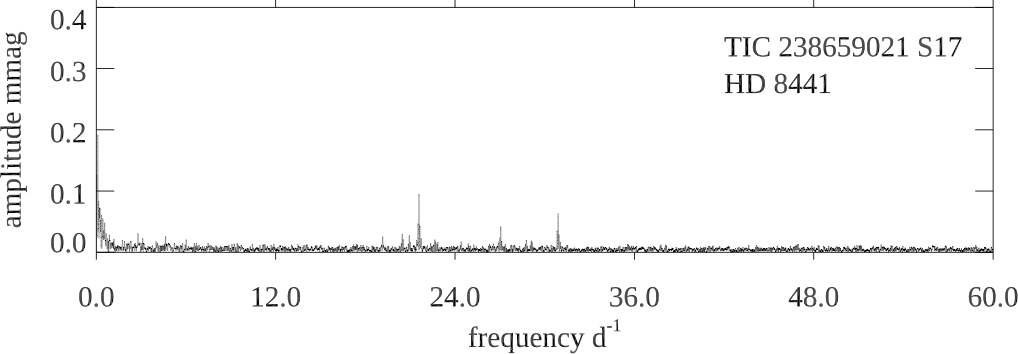}
 \includegraphics[width=0.48\linewidth,angle=0]{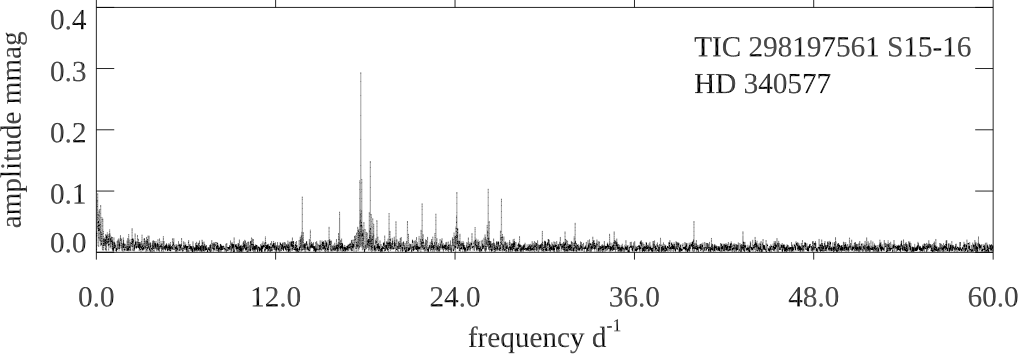}
 \includegraphics[width=0.48\linewidth,angle=0]{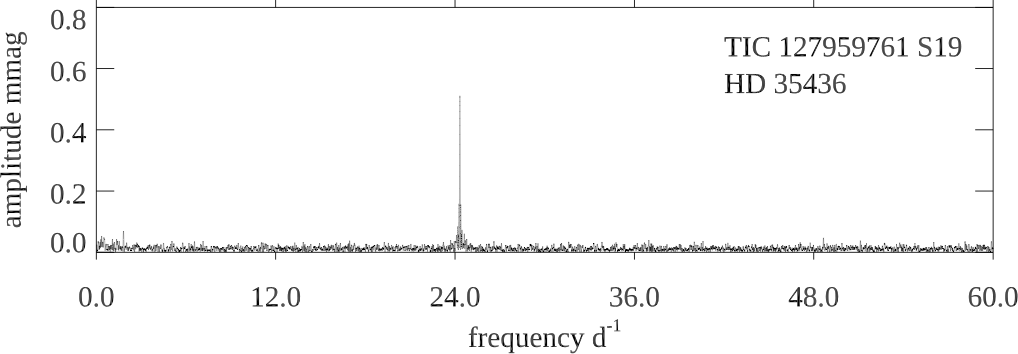}
 \includegraphics[width=0.48\linewidth,angle=0]{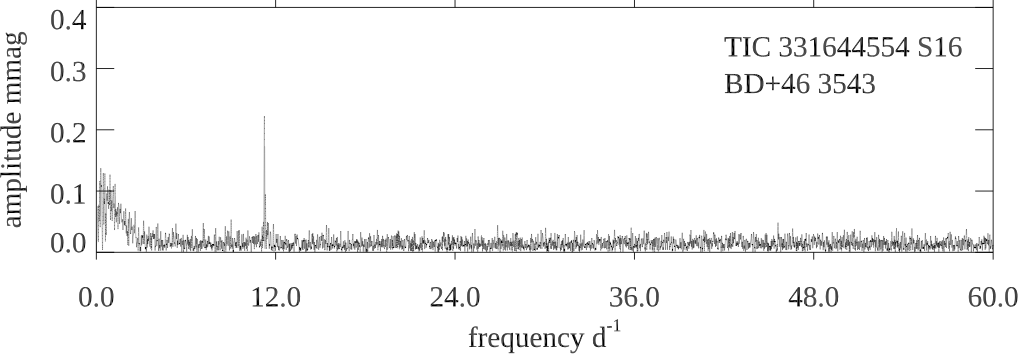}
 \includegraphics[width=0.48\linewidth,angle=0]{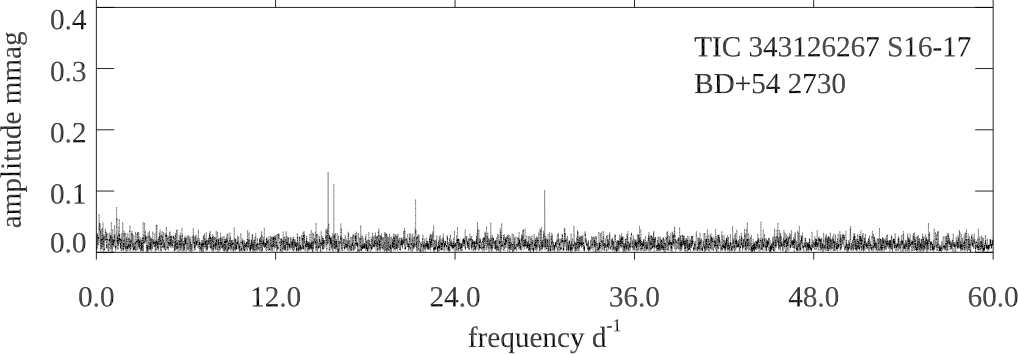}
 \includegraphics[width=0.48\linewidth,angle=0]{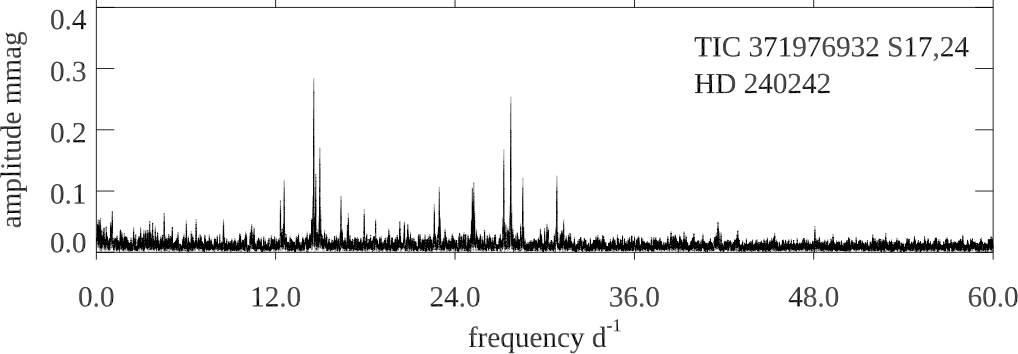}
  \caption{Amplitude spectra for the nine $\delta$~Sct stars. We note the different ordinate scales for TIC\,21073591 and TIC\,127959761. With the exception of TIC\,21073591, they are all very low amplitude pulsators, as is predicted for low-overtone $\delta$~Sct pulsation in Ap stars.  }
  \label{fig:deltasct}
\end{figure*}

\begin{figure*}
  \centering
 \includegraphics[width=0.48\linewidth,angle=0]{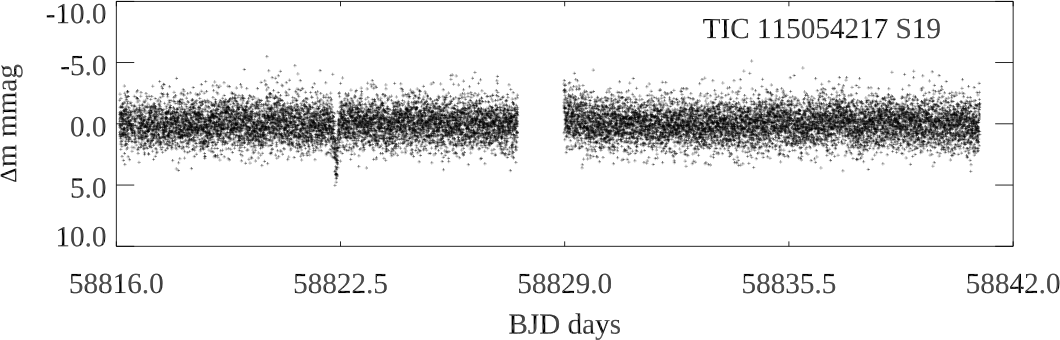}
 \includegraphics[width=0.48\linewidth,angle=0]{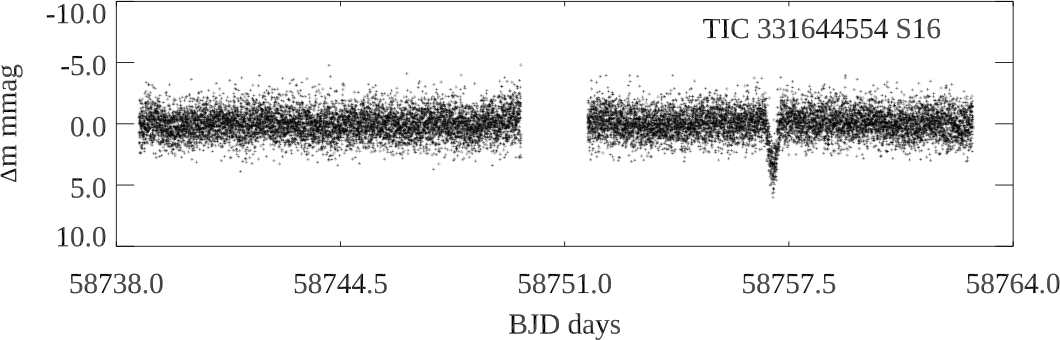}
  \caption{Light curves of two eclipsing binaries among the ssrAp
    stars. There is only one eclipse for each star, hence the orbital
    period is longer than the TESS sector; i.e. longer than 27\,d, in
    these cases.} 
  \label{fig:lc}
\end{figure*}

\subsubsection{TIC\,286992225, TYC\,2553-480-1}

TIC\,286992225 is a known roAp star discovered in SuperWASP data by
\citet{2014MNRAS.439.2078H}, who referred to it in a brief form as
J1430. They found a pulsation at 235.5\,d$^{-1}$ ($P = 6.1$\,min)
which is at the high frequency end of the roAp star range. The TESS
data clearly show two pulsation frequencies (Fig.\,A5) at
$\nu_1 = 235.541$\,d$^{-1}$, $A_1 = 545$\,$\upmu$mag
 and $\nu_2 = 232.525$\,d$^{-1}$, $A_2 = 323$\,$\upmu$mag. The separation between them
is 35\,$\upmu$Hz. Using the Gaia EDR3 value of
  parallax and a bolometric correction of 0.02
gives $L=10.37$\,L$_\odot$ and a scaled $\Delta\nu$ of $66$\,$\muup$Hz. 
The separation between the two frequencies of 35\,$\upmu$Hz is therefore half the large separation. 
The relative amplitudes suggest that $\nu_2$ may be from a quadrupole mode and $\nu_1$ from a dipole mode. However, roAp stars do not show equipartition of mode energy, hence this is speculative.

\subsection{The $\delta$~Sct stars}
\label{deltasct}

\citet{2020MNRAS.498.4272M} showed from models that $\delta$~Sct
low-overtone p~mode pulsations, and $\gamma$~Dor high-overtone g~mode
pulsations, are damped in Ap stars with field strengths above about
1\,kG. They showed a case study of KIC\,11296437, the first known roAp
star that shows both high-overtone p~modes typical of roAp stars, and
low-overtone p~modes as in $\delta$~Sct stars. Magnetic Ap stars with
$\delta$~Sct pulsations are rare, hence our finding of at least 9
among our 67 ssrAp stars is notable. TIC\,127959761 and
TIC\,331644544 are unusual for $\delta$~Sct stars, and especially for
such low-amplitude $\delta$~Sct stars, in being singly
periodic. Perhaps that is related to the magnetic nature of these
stars. Remarkably, TIC\,331644544 is also an unusual eclipsing binary
(Section\,\ref{sec:EB}). The $\delta$~Sct frequency range for these
stars is illustrated in Fig.\,\ref{fig:deltasct}. 

 We note in Fig.\,\ref{fig:deltasct} that there are no obvious g~modes
 excited. That is consistent with Hideyuki Saio's magnetic models
 \citep{2020MNRAS.498.4272M} that show high order g~modes, as in
 $\gamma$~Dor stars, suppressed in Ap stars. Those same models do
 allow low overtone p~modes to be excited for field strengths less
 than about 1\,kG, and the fundamental radial mode to be excited for a
 field strength up to 4\,kG. It is further notable that all but one of
 the nine stars presented have exceedingly low amplitudes -- under
 0.6\,mmag. These stars would not have been found in ground-based
 observations, hence it is unsurprising that we only now are
 discovering them with TESS data, with their combination of
   higher precision and 
 longer duration, and more $\delta$~Sct stars being observed.
 \citet{2020MNRAS.498.4272M} pointed out, `Other roAp stars with low
 field strengths should also be found with $\delta$~Sct pulsations,
 similar to KIC 11296437.' As a by-product of our search for ssrAp
 stars, that appears to be what we have found with these nine (and
 possibly more) $\delta$~Sct stars.  

\subsubsection{Caveats} 

It is important to determine the magnetic field strengths of all of
the ssrAp stars, including these $\delta$~Sct stars. Only
TIC\,40564267 has a literature measurement (Table\,1), and that is
$\langle B_z \rangle_{\mathrm{rms}} = 0.5$\,kG, consistent with
the models of \citet{2020MNRAS.498.4272M}. We do not know
whether any of these Ap -- $\delta$~Sct stars 
are binaries with one Ap and one $\delta$~Sct
star. \citet{2020MNRAS.498.4272M} showed that was not the case for
KIC\,11296437. Spectra for the stars shown in Fig.\,\ref{fig:deltasct}
will resolve this, as in such a pair both components should be
visible. As those spectra will be obtained for magnetic field
measurement, this question will be answered.  

\subsection{Eclipsing binaries}
\label{sec:EB}

Magnetic Ap (or late Bp) stars in eclipsing binaries are rare. To the
best of our knowledge, only five have been reported in the literature:
HD~123335 \citep[Bp\,He\,wk\,Sr;][]{Hensberge:2007oz}, HD~66051
\citep[A0p\,Si;][]{2018MNRAS.478.1749K}, HD~99458
  \citep{2019MNRAS.487.4230S}, HD~62658
  \citep[B9p\,Si;][]{2019MNRAS.490.4154S}, and HD~102797
  \citepalias[B9p\,Si;][]{2020A&A...639A..31M}. The spectral types are
  from \citet{2009A&A...498..961R}, except for HD~99458, whose
  peculiarity was noted by \citet{2019MNRAS.487.4230S}. For HD~102797
  we cannot definitely rule out the possibility that the eclipses may
  originate from a contaminating star in the TESS field.
  
The low rate of occurrence of eclipsing binaries among Ap stars is not very
surprising, since the incidence of close binaries with Ap stars is
very low, and eclipses are low probability in the longer period Ap
binaries. Yet we have found two, long period, eclipsing
systems amongst the ssrAp stars. Their light curves are shown in
Fig.\,\ref{fig:lc} where only one eclipse is seen in each case, hence
the orbital periods are longer than 27\,d. In both cases the eclipse
depths are $3 - 4$\,mmag, giving a ratio of radii of $R_2/R_1 \approx
0.06$, assuming that these are transits, and not grazing eclipses. For
a roughly estimated radius for the primaries of $R_1 \approx
2$\,R$_\odot$, both of the companions have radii around $R_2 \approx
0.12$\,R$_\odot$; that is they are lower main-sequence companions, which
is plausible. We examined the target pixels files for these two
  stars and find that the eclipse signals for both stars come from the
  brightest pixels centred on the targets. All Gaia sources within
  these pixels are at least 5\,mag fainter. It is unlikely that the
  eclipses arise from a background contaminator.

With a rough estimate of the primary mass and radius for both of these binaries to be $2$\,M$_\odot$ and $2$\,R$_\odot$, respectively, the duration of the transits, which is the time it takes the secondary to cross a distance of $2$\,R$_\odot$  assuming a circular orbit, allows an estimate of the orbital velocity of the secondary, which then can be translated into a rough estimate of the semi-major axis of the orbit and the orbital period. While this is crude, it gives approximate orbital periods of 50\,d for TIC\,115054217 and 450\,d for TIC\,331644544. It is unlikely that the companions of these stars have any impact on
the Ap star.

With orbital periods of that order, compliance of these two systems with
the conjecture of \citet{2017A&A...601A..14M} that, barring
synchronisation, short rotational and orbital periods are mutually
exclusive, would require the rotation periods of the Ap components to
be longer than $\sim$50\,d. This is not implausible given the apparent
lack of rotation-like variability over a TESS sector. The two systems
would then lie in the upper left quadrant of Fig.~16 of
\citet{2017A&A...601A..14M}.

\section{Discussion}
\label{sec:disc}

For 22 of the 67 stars of Table~\ref{tab:lpn}, there exists some
independent information about the variability and/or rotation, such as
published values of the rotation period and of $\vsi$, as well as line
profile characterisation from either the literature or our own
collection of high-resolution spectra.

There are two stars that show sharp lines in high-resolution spectra,
hence that have low projected equatorial velocities $\vsi$, for which
variation periods shorter than 27\,d have been reported in the
literature: HD~89069 and HD~148330. The 18\,d value of the period of
the former that was published by \citet{2017MNRAS.468.2745N} can be
reconciled with the sharpness of its spectral lines even if neither
the inclination $i$ of the rotation axis to the line of sight nor the
obliquity $\beta$ of the magnetic axis are exceptionally small. Other similar
occurrences are known, such as the prototypical Ap star $\beta$~CrB (=
HD~137909), which has clearly resolved magnetically split lines while
showing rather large amplitude variations (including a reversing mean
longitudinal magnetic field) along its $18\fd5$\,d rotation period
\citep{1970ApJ...160.1049W}. While there is a low frequency bump in
the amplitude spectrum of HD~89069 (see Fig.~\ref{fig:ssrAp7}), it is
not typical of Ap rotation. 
For HD~148330, the TESS photometry sets a very stringent
upper limit of the variability, 3 to 4 orders of magnitude lower than
the peak-to-peak amplitude of the data on which the $4\fd28$ value of
the period proposed by \citet{1990BAICz..41..118Z} is based. This
leads to the suspicion that this value is spurious. The lack of
photometric variability of the star over a TESS sector and the
sharpness of its spectral lines consistently suggest that its rotation
period is (very) long.

In contrast with the stars discussed above, we could not find any
information about the profiles of the spectral lines of HD~22860. The
highest amplitude low-frequency peak in the TESS data is at $\nu_{\rm
  rot}=0.192$\,d$^{-1}$. The corresponding value of the rotation
period, $\Prot=5\fd21$, is typical of Ap rotation, but it differs by
1\,d from the published value
\citep[$\Prot=6\fd2$][]{2012MNRAS.420..757W}, of which it is not an
alias, and the variation amplitude is very low. This star represents a
good illustration of the need for high resolution spectroscopy to
complement the present, photometry-based search. Because the periodic
variability that is suggested by the low-frequency peaks in the
amplitude spectra of the TESS observations (see
Fig.~\ref{fig:ssrAp8}) is just above the threshold of
significance, we cannot confidently 
decide if HD~22860 has a rotation period of a few days, typical of the
majority of the Ap stars, or if it may be rotating very slowly. In the
former case, the low photometric amplitude could indicate that either
the inclination $i$ or the obliquity $\beta$ is small. With a small
$\beta$, the spectral lines should show significant broadening (unless
$i$ is also small, but the probability that both $i$ and $\beta$ are
small is low).

For the remaining 18 stars of Table~\ref{tab:lpn} for which
complementary information about the periods and/or the line profiles
is available, this information is fully compatible with the
interpretation that the lack of variability over a TESS sector (or
more) reflects the occurrence of (very) slow rotation. As there are
plausible explanations for the apparent inconsistencies affecting the
three cases discussed above, the present analysis of the northern
ecliptic hemisphere TESS observations supports the conclusion drawn
from the consideration of the southern ecliptic hemisphere part of the
survey \citepalias{2020A&A...639A..31M}, that most of the constant
stars that are identified with our search technique belong, with a
high probability, to the group of the ssrAp stars.

In \citetalias{2020A&A...639A..31M}, we showed that some stars that
were expected to be constant over 27\,d were missed by our
search. The degree of incompleteness of the achieved detections could
not be fully assessed due to the small number statistics. Eleven of
the 33 Ap stars known to have an accurately determined rotation period
longer than 50\,d according to the critical compilation of ssrAp
stars by \citet{2020pase.conf...35M} are located in the northern
ecliptic hemisphere. Of these 11 stars, seven were found to be constant
over 27\,d. The four that were not are HD~5797, HD~158919, HD~188041,
and HD~200311. For three of these stars, no 2-min cadence data are available:
HD~158919 and 
HD~188041 fell outside the detector limits, while HD~200311 was only
observed in 30-min cadence. Observations of HD~5797 (TIC\,256106349) were obtained in three
sectors. Three clear frequency peaks are visible. These are possibly
caused by g~mode pulsations, but their periods and period separations 
do not obviously fit the $P - \Delta P$ theoretical relations of \citet{2017MNRAS.465.2294O}, and g~modes are previously not known, and are not expected in Ap stars, making this star unusual.
While these low frequencies led to the rejection of the
star in our analysis, there is no indication of variability with the
published period, $\Prot=68\fd05$ \citep{2018PASP..130d4202D} in the
pipeline reduced data. However, in the data that have not been
pipeline reduced (SAP data), there are significant variations that
could plausibly be the rotation signal. This case probably represents
an example of removal of an astrophysical signal by the
pipeline. See also Fig.~4 of \citet{2021MNRAS.506.1073H} for
  another example of the occasional interference of the pipeline with the
  astrophysical signal in TESS observations.

In summary, of the eight known ssrAp stars of the northern ecliptic
hemisphere that have been observed in 2-min cadence by TESS, seven were
duly detected in our search. The eighth one was missed only because of
a reduction  artefact. This is consistent with the conclusion reached
in \citetalias{2020A&A...639A..31M}, that while our survey does not
return a complete list of the most slowly rotating Ap stars, its
incompleteness is not related to the physical properties of the missed
stars. Accordingly, its outcome should lend itself to unbiased studies
of possible relations between slow rotation and other physical
properties of Ap stars.

On the other hand, while for all the southern ecliptic hemisphere
candidates identified in \citetalias{2020A&A...639A..31M} whose
rotation periods were accurately known, these periods were all (much)
longer than 50\,d (the shortest one was $\sim155$\,d), three of the
candidates identified in the present study have periods between 27\,d
and 49\,d, and and four more have periods shorter than 120\,d. This
illustrates the suitability of our technique also to identify
moderately slow rotators.

\section{Magnetic fields}
\label{sec:bfield}

\begin{figure}
\resizebox{\hsize}{!}{\includegraphics{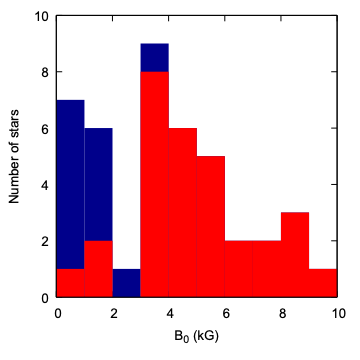}}
\caption{Distribution of the phase-averaged magnetic field strength
  $B_0$ for the long-period stars of Table~\ref{tab:lpn} of this paper
  and of Table~1 of \citetalias{2020A&A...639A..31M}. The red
  part of the histogram corresponds to the stars for which
  measurements of the mean magnetic field modulus or of the mean
  quadratic magnetic field are available; for the remaining stars
  (blue part of the histogram), a
  lower limit of $B_0$ was inferred from the existing mean
  longitudinal magnetic field measurements.} 
\label{fig:bavhist}
\end{figure}

Figure~\ref{fig:bavhist} shows the distribution of the phase-averaged
mean magnetic field modulus $B_0$ for all of the stars that were identified in
\citetalias{2020A&A...639A..31M} and this paper as not showing photometric
variability over a TESS sector, and for which magnetic field measurements
have been obtained. As in \citetalias{2020A&A...639A..31M}, for those
stars for which no mean magnetic field modulus determinations are
available, an estimate of $B_0$ was computed by dividing the
phase-averaged value of the mean quadratic field, $Q_0$, by 1.28
\citep[the slope of the correlation between these two field
  moments, according to][]{2017A&A...601A..14M}; 
when only mean longitudinal magnetic field measurements had been
obtained, we used $B_0=3\,\langle B_z\rangle_{\mathrm{rms}}$, which is
actually a lower limit. To reflect this difference, the corresponding
parts of the histogram are filled in blue, while those parts based on
measured values of $\langle B\rangle$ or $\langle 
B_{\mathrm{q}}\rangle$ are filled in red.

Besides the obvious implications of the above-mentioned usage of lower
limits of $B_0$ 
for those stars for which $\langle B_z\rangle$ is the only magnetic
moment that has been measured, some additional caution is required
when drawing 
conclusions from consideration of Fig.~\ref{fig:bavhist}. The
correlation between $B_0$ and $Q_0$ is sufficiently tight \citep[see
Fig.~12 of][]{2017A&A...601A..14M} that the usage of $Q_0/1.28$
instead of $B_0$ in the absence of $\langle B\rangle$ measurements
should at most have a moderate impact on the overall
distribution. Of more concern is the fact that magnetic measurements
have been obtained over at least a full rotation cycle for only 12 of
the 42 stars on which the histogram is based. These 12 stars are the
only ones for which $B_0$ can indeed be regarded as an average over
all the phases; for the remaining 30 stars, it is only an average over
the existing measurements, with an incomplete and uneven phase
coverage. The resulting uncertainty on the actual value of $B_0$ is
limited for those stars for which measurements of the mean magnetic
field modulus or of the mean quadratic magnetic field are available,
since the ratio between the extrema of $\langle B\rangle$ has never
been observed to exceed 2, and in most cases is lower than 1.3
\citep[see Fig.~9 of][]{2017A&A...601A..14M}. But
the mean longitudinal magnetic field can have very large variation
amplitudes, so that the lower limits of the value of $B_0$ that are
inferred from its consideration should not be over-interpreted. This is
why a blue filling of the histogram was used in Fig.~\ref{fig:bavhist}
to distinguish the 12 stars for which only $\langle B_z\rangle$
measurements have been obtained. However, the fact that, as a group,
these 12 stars are concentrated in the lower bins of the histogram
(especially the two lowest ones) is consistent with the conjecture that
the magnetic field of most of them is too weak to lead to the
resolution of magnetically split spectral lines. 

 Figure~\ref{fig:bavhist} represents an augmented version of Fig.~3 of
\citetalias{2020A&A...639A..31M}. Similar to the latter, it shows a
lack of stars with $B_0>10$\,kG, which is consistent with the
conclusion that the strongest magnetic fields ($B_0\ga7.5$\,kG) occur
only  in those Ap
stars with periods shorter than 150\,d. It is actually
noticeable, although inconclusive, that the three
stars of Table~\ref{tab:lpn} in which the strongest magnetic fields
have been measured are also the only three stars of this table that
are known to have periods shorter than 50\,d, hence that are not ssrAp
stars in the strictest sense. 

On the other hand, while the lowest bins of the histogram
($B_0\leq2$\,kG) are relatively more populated than for the southern
ecliptic survey alone, the rate of occurrence of long-period
candidates still remains less in the 0--3\,kG $B_0$ range
than in the 3--6\,kG range. The result that for most of the stars in the lower
range, only mean longitudinal magnetic field measurements have been
obtained, hence the values of $B_0$ used in this histogram are lower
limits, only strengthens this result. As the distribution of the
root-mean-square longitudinal magnetic fields among all Ap stars
sharply peaks at the lowest values \citep{2009MNRAS.394.1338B}, the
conclusion that the rate of occurrence of super-slow rotation is
considerably lower in weakly magnetic stars than in strongly magnetic
ones seems inescapable. This conclusion remains valid even if we take
into account six stars of \citetalias{2020A&A...639A..31M} for which
no magnetic measurements have been obtained but whose line
profiles (five sharp unresolved ones and one showing only marginal
magnetic resolution) indicate that the fields must be considerably
lower than 3~kG. (There are no such stars in Table~\ref{tab:lpn}.)

\begin{figure}
\resizebox{\hsize}{!}{\includegraphics{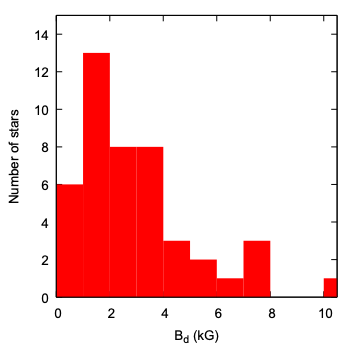}}
\caption{Distribution of the magnetic dipole field strength
  $B_{\rm d}$ for the Ap stars of the volume-limited sample of
  \citet{2019MNRAS.483.3127S}.} 
\label{fig:bavhist_sik}
\end{figure}

 Admittedly, the mean longitudinal magnetic field measurements compiled
in the catalogue of \citep{2009MNRAS.394.1338B} were extracted from a
very inhomogeneous collection of sources and they include without
distinction $\langle B_z\rangle$ values that are below the threshold of formal
significance, so that one may suspect the low field strength
population to be overestimated. To assess the
extent to which such a bias could affect the above conclusion about the
comparatively low rate of occurrence of super-slow rotation among
weakly magnetic stars, we further compare the present sample of ssrAp
candidates with the volume-limited sample of Ap stars studied by
\citet{2019MNRAS.483.3127S}. We plotted in Fig.~\ref{fig:bavhist_sik} the
distribution of the dipole field strengths $B_{\rm d}$ determined by
those authors in the same format as the $B_0$ distribution that we
derived for the ssrAp candidates (Fig.~\ref{fig:bavhist}). Although
the values of the phase-averaged magnetic field strength $B_0$ and of
the dipole field strength $B_{\rm d}$ (which cannot be determined for
most of the ssrAp candidates) are different from each other, the
histograms of Figs.~\ref{fig:bavhist} and \ref{fig:bavhist_sik} still
lend themselves to meaningful comparison, in the following way.

\begin{table*}
  \caption{Summary.}
  \begin{tabular*}{12.5cm}[]{@{}@{\extracolsep{0pt}}l@{\extracolsep{\fill}}rrr@{\extracolsep{0pt}}}
    \hline\hline\\[-4pt]
    Star type&South&North&Total\\[4pt]
    \hline\hline\\[-4pt]
    Original sample                       &1014&492& 1506\\
    ssrAp candidates                      &60&67&127\\[4pt]
    Resolved magnetically split lines     &19& 8&27\\
    Sharp unresolved lines                 &10& 7&17\\
    $\vsi\la20$\,\kms                     & 0& 3& 3\\
    Broad lines / $\vsi>20$\,\kms         & 6& 0& 6\\[4pt]
    Accurate rotation period values       & 6&14&20\\
    $\Prot>50$\,d                         & 6&11&17\\
    26\,d$<\Prot<50$\,d                   & 0& 3& 3\\
    $\Prot<26$\,d?                        & 0& 3& 3\\[4pt]
    Stars with magnetic field measurements&24&17&41\\[4pt]
    Definite roAp stars                   &12& 5&17\\
    Possible roAp stars                   & 1& 8& 9\\
    Fraction of roAp stars                &22\%&$\la19$\%&$\la20$\%\\[4pt]
    Definite $\delta$~Sct stars           & 2& 9&11\\
    Possible $\delta$~Sct stars           & 1& 5& 6\\
    Possible eclipsing binaries           & 1& 2& 3\\[4pt]
    \hline\\[-4pt]
    \label{tab:summary}
\end{tabular*}
\end{table*}

The sample of \citet{2019MNRAS.483.3127S} comprises 27 stars in the
range $0\leq B_{\rm d} <3$\,kG and 13 stars in the range $3\leq
B_{\rm d}<6$\,kG. The ssrAp candidates for which magnetic field data
exist include 14 stars with $0\leq B_0<3$\,kG and 20 stars with $3\leq
B_0<6$\,kG. One should keep in mind that for most of the stars in the
range $[0,3]$\,kG, the value of $B_0$ is an
underestimate of the actual phase-averaged magnetic field strength, so
that based on the latter,  several of the stars that appear in the
blue parts of the $[0,3]$\,kG interval in the histogram of
  Fig.~\ref{fig:bavhist} may have $B_0\geq3$\,kG.

The ranges of magnetic field values do not correspond to the same
physical quantity in the two samples shown in Figs.~\ref{fig:bavhist}
and \ref{fig:bavhist_sik}, so that the absolute numbers of
stars in them are not directly comparable. But the difference
between the relative populations of the lower and higher  ranges of
magnetic field strengths in
the two samples can be meaningfully compared. Namely, in the 
sample of \citet{2019MNRAS.483.3127S}, the lower range ($[0,3]$\,kG)
is considerably 
more populated than the higher  
one ($[3,6]$\,kG). This is the opposite for the ssrAp candidates
sample, for which 
the lower range is definitely less populated than the higher one,
probably even more so than the plain counts suggest, if one makes
allowance for the fact that a number of the stars seemingly in the
lower sample likely belong to the higher one.
Thus, comparison of our sample with the one of
\citet{2019MNRAS.483.3127S} unambiguously confirms the lower rate of
occurrence of weakly magnetic stars than of strongly magnetic stars
among the ssrAp candidates. 
  
A particularly intriguing feature of the histogram of
Fig.~\ref{fig:bavhist} is the apparent gap in the $B_0$ distribution
between 2 and 3\,kG. Admittedly, the reality of this gap will need to
be confirmed by obtaining more precise constraints on the actual value
of $B_0$ than the uncertain lower limits based on mostly incomplete
sets of measurements of the mean longitudinal magnetic field that
prevail in the three lowest bins of the histogram. But its coincidence
with the apparent bimodality of the mean magnetic field modulus
distribution advocated by
\citet{2017A&A...601A..14M} is noteworthy. 
As already noted by \citet{1997A&AS..123..353M}, \citet{2017A&A...601A..14M} confirmed that
for none of the known Ap stars with resolved magnetically split lines
is the average $B_0$ of the mean magnetic field modulus over a rotation
cycle lower than $\sim$2.8\,kG, although it should be possible to
resolve magnetically split lines down to $\langle B\rangle\sim1.7$\,kG
in most of the spectra analysed in the considered
studies. He then argued that there exists a
distinct population of Ap stars, separate from that of the Ap stars
with resolved magnetically split lines, whose mean magnetic field
moduli, averaged over a rotation period, are weaker than
$\sim$2\,kG. This conclusion is based on the consideration of
published estimates of the magnetic fields that are essentially
comparable to the mean magnetic field modulus or to the mean quadratic
magnetic field. The double-peaked histogram shown in
Fig.~\ref{fig:bavhist} is strikingly consistent with the bimodal
distribution suggested by \citet{2017A&A...601A..14M}. Confirming the
existence of a gap in the distribution of the magnetic field strengths
of the ssrAp stars between $\sim$2\,kG and $\sim$3\,kG is one of the
important objectives of the systematic study of the magnetic fields of
the candidate ssrAp stars that we intend to carry out. 

\section{Conclusion}

Table~\ref{tab:summary} summarises the results of the search for Ap
stars that show no low-frequency variability over a TESS sector in the
southern ecliptic hemisphere \citepalias{2020A&A...639A..31M} and in
the northern ecliptic hemisphere (this work). A total of 127 such
stars were identified, split almost evenly between the two
hemispheres. Spectroscopic information is available for 53 of them
($\sim$2/3 of which are in the southern ecliptic hemisphere),
either in the literature or in our own collection of spectra. Of these
53, 27 have spectral lines resolved into their magnetically split
components, and 17 show sharp spectral lines that are unresolved at
high resolving powers ($5\times10^4\la R\la10^5$). Almost all of these
44 ($=27+17$) 
stars likely are genuine slow rotators. Three more stars 
plausibly belong to this group, as their line profiles do not show
significant rotational broadening in medium resolution spectra,
indicating values of $\vsi\la20$\,km\,s$^{-1}$. The remaining six stars
for which we have spectroscopic information (all from the southern
ecliptic hemisphere) have broad or very broad spectral lines. If the magnetic
obliquities of Ap stars are randomly distributed, the
interpretation that we proposed in \citetalias{2020A&A...639A..31M} --
that these six stars are fast-rotating stars in which the angle between the
magnetic and rotation axes is small -- is consistent with
their rate of occurrence, six out of 1506 stars in the
entire sample that was analysed, or 0.40\%.

Values of the rotation period have been derived for four of the stars for which
we have no line profile information \citepalias[one
from][]{2020A&A...639A..31M}. Three of them exceed 
70\,d; the published period of the fourth star, $\Prot=6\fd27$, is not
supported by the TESS observations. 

In summary, the available line profile and period information is
consistent with the view that Ap stars that show no variability over a
TESS sector are (very) slow rotators for 50 of the 57 stars for which
such information was found. The few exceptions can be reasonably
justified, mostly in terms of low magnetic obliquity.
This result, based on an increased sample, strengthens the conclusion from
\citetalias{2020A&A...639A..31M}  that the search technique that we
applied is effective and reliable. Accordingly, we expect almost all
the stars of Table~\ref{tab:lpn} for which neither line profile nor
period constraints have been published yet to be genuine long-period
stars. In the present study, we identify in this way 46 new ssrAp star
candidates. Together with the 23 candidates from
\citetalias{2020A&A...639A..31M}, they open the prospect of increasing
by a factor of almost three the number of known ssrAp stars, in a mostly
bias-free way. In particular, with the technique discussed here, the
identification of ssrAp stars is fully independent of their magnetic
fields, which makes it possible to study the existence of possible
dependences between the magnetic field strengths and the rotation
periods.

\begin{figure}
\resizebox{\hsize}{!}{\includegraphics{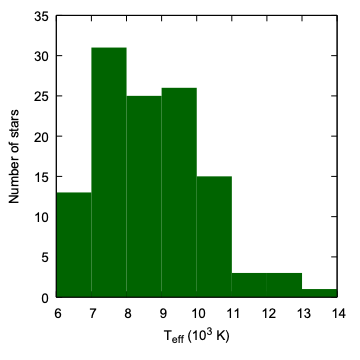}}
\caption{Distribution of the effective temperatures of the ssrAp
  candidates from Table~\ref{tab:lpn} and from Table~1 of
  \citetalias{2020A&A...639A..31M}.} 
\label{fig:teffhist}
\end{figure}

The next, critical step in this study is to obtain high-resolution
spectra of the 69 ssrAp star candidates, to confirm that they have low
projected equatorial velocities $\vsi$ and to obtain preliminary
estimates of their magnetic fields. Once this is done, multi-epoch
spectroscopic or spectropolarimetric observations of the confirmed
candidates will have to be obtained so as to constrain their rotation
periods and characterise their magnetic variations, either from
consideration of their mean magnetic field moduli or of their mean
longitudinal magnetic fields (or both).

Despite the incompleteness of the rotation and magnetic information
that has been acquired until now, the present study strengthens the
tentative conclusion reached in \citetalias{2020A&A...639A..31M} that
the rate of occurrence of super-slow rotation appears significantly
lower in weakly magnetic Ap stars than in those that are strongly
magnetic. One of the main objectives of follow-up studies will be to
characterise the low end of the magnetic field strength distribution
among ssrAp stars. This is an area of the parameter space that has
remained almost unexplored until recently. It will be of particular
interest to test the possible existence of a gap between $\sim$2\,kG
and $\sim$3\,kG in which no ssrAp star is found. The existence of such
a gap, which was first proposed by \citet{2017A&A...601A..14M} as part
of a general study of the Ap stars with resolved magnetically split
lines, and which appears further supported by the results of our
present search for ssrAp stars, would potentially have profound
implications for the understanding of the origin of super-slow
rotation in Ap stars. Indeed, it would suggest the existence of two
distinct populations, consisting of weakly and strongly
magnetic Ap stars, respectively, whose rotational evolution would follow different
paths and involve different physical mechanisms.

Another result from \citetalias{2020A&A...639A..31M} that is further
confirmed by the present study is the high rate of occurrence of roAp
stars among ssrAp candidates. A definitive quantitative
characterisation of this rate cannot be given yet:
Table~\ref{tab:lpn} includes five definite roAp stars and eight candidates
that have not been fully confirmed yet. We expect that most of them will
eventually be, but even being conservative, the 17 definite roAp stars
summed over the two hemispheres represent 14\% of the 120 long-period
Ap stars from the combined lists of \citetalias{2020A&A...639A..31M}
and of the present study. Even this lower limit is considerably higher
than the fraction of roAp stars reported for the $>1000$ Ap stars
observed by TESS in Sectors~1--13
\citep{2019MNRAS.487.3523C,2021MNRAS.506.1073H}. The
  incidence of the roAp phenomenon is low among Ap stars in general,
  but definitely higher for ssrAp stars.

One might possibly suspect
  that this difference is not directly related to slow rotation
  itself, but to some other distinctive property of the ssrAp stars as
  a group. One obvious candidate for such a connection is the
  distribution of the effective temperatures. From consideration of
  the roAp star lists of \citet{2015MNRAS.452.3334S} and of
  \citet{2021MNRAS.506.1073H} (which partly overlap with each other), 
one can see that these stars are found
  in a narrow $T_{\rm eff}$ range, 6,000--9,000\,K, with a
  distribution sharply peaked towards the centre of this interval. This
  contrasts with the $T_{\rm eff}$ distribution of the ssrAp
  candidates identified in \citetalias{2020A&A...639A..31M} and in the
  present study, as shown in Fig.~\ref{fig:teffhist}. This
  distribution is much broader and shallower, with 94\% of the stars
  in the $T_{\rm eff}$ range 6,000--11,000\,K. It is almost flat from
  7,000 to 10,000\,K, an interval that accounts for 70\% of the ssrAp
  candidates. In summary, the ssrAp candidates span a much wider range
  of temperatures than the roAp stars, in a more uniform way, so that
  the comparatively high 
  rate of occurrence of roAp stars among ssrAp stars does not appear
  to be primarily temperature related. What causes it remains to be
  determined.

Also surprising on the asteroseismology front is the
discovery that nine of the 66 long-period Ap stars from
Table~\ref{tab:lpn} are definite $\delta$~Sct stars, and five more are
candidate $\delta$~Sct stars. This is not only unexpected, but also in
marked contrast with the southern ecliptic hemisphere
statistics. Indeed, there were only two definite $\delta$~Sct stars and
one candidate in Table~1 of \citetalias{2020A&A...639A..31M}.

These findings raise intriguing questions about the connections
between rotation, pulsation and magnetic field in Ap stars. They
represent an additional incentive to pursue systematic spectroscopic
studies of the candidate ssrAp stars identified here and in
\citetalias{2020A&A...639A..31M}, with a view to constraining their
rotational and magnetic properties, as well as to exploit the
next TESS data releases to further increase the number of 
ssrAp star candidates.

 \begin{acknowledgements} 
This research has made use of the SIMBAD database, operated at CDS,
Strasbourg, France.
This paper includes data collected by the TESS mission,
which are publicly available from the Mikulski Archive for Space
Telescopes (MAST). Funding for the TESS mission is provided
by NASA's Science Mission directorate. Funding for the TESS
Asteroseismic Science Operations Centre is provided by the Danish
National Research Foundation (Grant agreement no.: DNRF106),
ESA PRODEX (PEA 4000119301) and Stellar Astrophysics Centre
(SAC) at Aarhus University. We thank the TESS team and staff and
TASC/TASOC for their support of this work. We also thank the referee,
Gregg Wade,  
for helpful comments that led to improvements in this paper. 
\end{acknowledgements}

\bibliographystyle{aa}
\bibliography{lpn2}

\appendix

\section{List of long rotation period Ap star candidates}
\label{sec:ssrAplist}
Table~\ref{tab:lpn} lists the 67 long rotation Ap star candidates that
were identified as described in Sect.~\ref{sec:TESS}. It is ordered by
the TIC numbers, which are given in 
Column~1. Column~2 gives an alternate identifier: HD number as first
choice, then DM number, for stars having these; other identifiers
otherwise. The spectral types listed in Column~3 are from the
Catalogue of Ap, HgMn and Am stars \citep{2009A&A...498..961R}. For
those stars that are not in this catalogue, the spectral type entry
was left blank. As mentioned in Sect.~\ref{sec:TESS}, these stars were kept in the
list only on the basis of more recently published compelling evidence
of their Ap nature. The $V$ magnitudes in Column~4 were extracted from
the SIMBAD database, while the $T_{\rm eff}$ and $\log g$ values in
Columns~5 and 6 were taken from the TIC.
Column~7 identifies the known or suspected roAp stars; unconfirmed
cases are flagged with a question mark. Column~8 gives information
about the measured magnetic field strengths. All three values
correspond to the mean over a rotation period (if known) or over the
existing observations (otherwise) of the range spanned by each of the
following magnetic field moments: $\langle B_z\rangle_{\rm rms}$ is
the root-mean-square longitudinal field, as defined by
\citet{1993A&A...269..355B}. It is essentially the quadratic mean of
the mean longitudinal magnetic field (the line-intensity weighted
average over the stellar disc of the component of the magnetic vector
along the line of sight). We denote by $B_0$ the average
value over a rotation cycle of 
the mean magnetic field modulus $\langle B\rangle$ (the line-intensity
weighted average over the stellar disc of the modulus of the magnetic
vector); this is the same 
quantity as listed in Column~3 of Table~13 of 
\citet{2017A&A...601A..14M}. The notation $Q_0$ refers to the average
value over a rotation cycle of the mean quadratic magnetic field
$\langle B_{\rm q}\rangle$; this is the same quantity as listed in
Column~10 of Table~13 of \citet{2017A&A...601A..14M}.  The values
indicated in Column~8 are based on the references specified in
Column~9. Column~10 gives the published values of the rotation periods,
from the references appearing in the following column. Column~12 gives
an indication of the width or the magnetic resolution of the spectral
lines (as a letter) and the $\vsi$ value (as a number). The resolution
information and $\vsi$ values are extracted from the references
indicated in Column~13. The width information for unresolved lines is
based on their appearance in high-resolution spectra from our
collection (Reference~3). These spectra were obtained at the Observatoire de
Haute-Provence with the AURELIE spectrograph fed by the 1.52-m
telescope \citep[see][for details]{1997A&AS..123..353M}. Column~14,
the penultimate column, contains various notes, in particular about
variability due to oscillations and binarity. The final column
indicates in which TESS 27-d sector(s) the star was observed.

\begin{landscape}
  \setlength\textwidth{705pt}
  \setlength\textheight{184mm}
\begin{table*}
\small
\caption{List of long-period Ap stars found by our technique in the TESS Sectors $14-26$ data, obtained in the second year of mission operations and covering the northern ecliptic hemisphere. }
\begin{tabular*}{\textwidth}[]{@{}@{\extracolsep{\fill}}rllrrrcclrlclll}
\hline\hline\\[-4pt]
  &  &Spectral&   &  &    &   &   &   &  &
  &\multicolumn{1}{c}{Lines\tablefootmark{a}}   &  & &TESS  \\
  \multicolumn{1}{c}{TIC}  &  \multicolumn{1}{c}{HD/Other id} &type&  \multicolumn{1}{c}{$V$}  & \multicolumn{1}{c}{$T_{\rm  eff}$}   & \multicolumn{1}{c}{$\log g$}   &  \multicolumn{1}{c}{roAp}  &  \multicolumn{1}{c}{$\langle B_z\rangle_{\mathrm{rms}}$/$B_0$/$Q_0$} &  Refs  &  \multicolumn{1}{c}{$P_{\rm rot}$}  &  Refs  & \multicolumn{1}{c}{$\vsi$}  & Refs & Notes &sectors  \\
  &  &  &  \multicolumn{1}{c}{(mag)}  & \multicolumn{1}{c}{(K)}   & \multicolumn{1}{c}{(cm\,s$^{-2}$)}   &   &  \multicolumn{1}{c}{(kG)}  &   &  \multicolumn{1}{c}{(d)}  &    & \multicolumn{1}{c}{(\kms)} &  &  \\[4pt]
  \hline\\[-4pt]
2689133&HD~242800&A0p Si&10.55&8500&4.0&&&&&&&&&S19\\
2849758&HD~243007&A1p Si&10.2\phantom{0}&9000&3.6&&&&&&&&$\delta$\,Sct star&S19\\
21073591&BD+29 3448&A9p Sr&9.54&7600&3.6&&&&&&&&$\delta$\,Sct star&S14\\
26749633&KIC~11031749 &~&12.50&7000&3.9&roAp&&&&&&&&S14,15\\
29715050&TYC~2653-2787-1&A0p HgSi&9.08&10450&4.2&&&&&&&&&S14\\
40564267&HD~184471&A9p SrCrEu&9.00&7750&3.6&&0.5/--/--&4&50.8&4&s&3&$\delta$\,Sct star&S14\\
44488067&HD~2453&A1p SrEuCr&6.91&8900&3.8&&0.8/3.7/4.3&1&521&12&r&1&&S17\\
76379777&HD~247591&A1p SiCrSr&10.47&9900&3.8&&&&&&&&&S19\\
77038207&HD~96003&A3p SrCr&6.87&9500&3.9&&0.2/--/--&5, 6&&&$<20$&5&&S22\\
77128654&HD~97127&~&9.43&6700&4.0&roAp&&&&&&&&S22\\
78784187&HD~249401&A2p Si&10.83&8150&3.6&&&&&&&&$\delta$\,Sct/$\gamma$\,Dor star?&S19\\
115054217&HD~244545&A1p Si&10.37&8350&&roAp?&&&&&&&EB&S19\\
115201166&HD~244640&A0p SiSr&9.94&9650&4.0&roAp?&&&&&&&&S19\\
115820611&HD~245416&A2p CrSi&9.71&8150&3.7&&&&&&&&&S19\\
116144208&HD~245726&A1p SiSr&10.8\phantom{0}&8750&3.6&&&&&&&&&S19\\
116995376&TYC~2413-476-1&A0p Sr&10.82&9050&3.8&&&&&&&&$\delta$\,Sct star?&S19\\
127755719&HD~243321&A0p Si&9.61&10000&3.9&&&&&&&&&S19\\
127755856&BD+32 976&A2p Sr&10.28&7600&4.2&roAp?&&&&&&&&S19\\
127959761&HD~35436&A1p SiSr&9.55&8000&3.5&&&&&&&&$\delta$\,Sct star&S19\\
154786038&HD~96571&A2p SrEu&7.31&8100&3.6&&&&&&&&$\delta$\,Sct star?&S14,19,20,26\\
155667362&HD~276625&A7p CrEu&9.94&7100&3.9&&&&&&&&&S19\\
158216369&KIC~7018170&~&13.28\rlap{\tablefootmark{b}}&6950&4.0&roAp&&&72.7&13&&&&S14\\
158275114&BD+44 3063&A5p EuCr&10.19&7450&4.0&roAp&--/3.2/--&7&&&r; 4.2&2&&S14,26\\
163801263&HD~203922&A2p SrCrEu&8.50&7600&3.6&&&&&&&&&S15\\
165327084&HD~110066&A1p SrCrEu&6.38\rlap{5}&9000&3.9&&0.1/4.1/--&1&4900?&14&r&1&&S22\\
165446000&BD+39 4435&F0p SiSrCr&9.3\phantom{0}&7700&&&&&&&&&&S15\\
176902576&HD~222416&A0p SiSr&7.73&9850&3.9&&&&&&&&&S17\\
185716313&HD~191742&A5p SrCr&8.16&8400&3.6&&0.6/--/1.8&8, 9&&&s; 6&3, 9&$\delta$\,Sct star?&S14,15\\
198781718&BD+29 3427&A6p CrSr&9.66&8000&4.2&&&&&&&&&S14\\
202899762&BD+46 570&Ap SrEuCr&9.63&7350&3.6&&&&&&&&&S18\\
207468665&HD~148330&A2p SiSr&5.73\rlap{4}&9700&3.8&&0.3/--/--&10&4.288\rlap{\tablefootmark{c}}&10&(s); 11.6&3, 10&&S16,19,23,25\\
233539061&HD~174016&\tablefootmark{d}&7.45&6225&&&&&&&&&&All but S18\\
238659021&HD~8441&A2p Sr&6.67\rlap{6}&9200&&&0.1/--/--&11&69.51&15&s&3&$\delta$\,Sct star&S17\\
239801694&HD~247628&A5p SiSr&10.59&9050&4.1&&&&&&&&&S19\\[4pt]
\hline\\[-4pt]
\end{tabular*}
\tablefoottext{a}{r = resolved; s = sharp; (s) = rather sharp.}
\tablefoottext{b}{$V$ magnitude from \citet{2004AAS...205.4815Z}.}
\tablefoottext{c}{TESS photometry definitely does not show any
  low-frequency variability; the published period may be spurious.}
\tablefoottext{d}{SB2 system; HD~174017 (secondary) is the Ap
  component (A0p SrCrEu); HD~174016 is a G6III star
  \citep{2009A&A...498..961R}, unresolved by 
  speckle interferometry 
  \citep{1984PASP...96..105H,2015AJ....150..151H}.}
\tablefoottext{e}{TESS photometry shows variations  
  with a period of $5\fd21$, which is typical of rotation, but the
  amplitude is very low and the period is not an alias of the
  published value.}
 \addtocounter{table}{-1}
 \label{tab:lpn}
\end{table*}
\end{landscape}	

\begin{landscape}
  \setlength\textwidth{705pt}
  \setlength\textheight{184mm}
\begin{table*}
\small
\caption{continued.}
\begin{tabular*}{\textwidth}[]{@{}@{\extracolsep{\fill}}rllrrrcclrlclll}
\hline\hline\\[-4pt]
  &  &Spectral&   &  &    &   &   &   &  &
  &\multicolumn{1}{c}{Lines\tablefootmark{a}}   &  & &TESS  \\
  \multicolumn{1}{c}{TIC}  &  \multicolumn{1}{c}{HD/Other id} &type&  \multicolumn{1}{c}{$V$}  & \multicolumn{1}{c}{$T_{\rm  eff}$}   & \multicolumn{1}{c}{$\log g$}   &  \multicolumn{1}{c}{roAp}  &  \multicolumn{1}{c}{$\langle B_z\rangle_{\mathrm{rms}}$/$B_0$/$Q_0$} &  Refs  &  \multicolumn{1}{c}{$P_{\rm rot}$}  &  Refs  & \multicolumn{1}{c}{$\vsi$}  & Refs & Notes &sectors  \\
  &  &  &  \multicolumn{1}{c}{(mag)}  & \multicolumn{1}{c}{(K)}   & \multicolumn{1}{c}{(cm\,s$^{-2}$)}   &   &  \multicolumn{1}{c}{(kG)}  &   &  \multicolumn{1}{c}{(d)}  &    & \multicolumn{1}{c}{(\kms)} &  &  \\[4pt]
 \hline\\[-4pt]
251282995&HD~18078&A0p SrCr&8.27&8500&3.4&roAp?&0.7/3.5/--&&1358&16&r&1&&S18\\
251976407&HD~221568&A1p SrCrEu&7.55&9500&3.9&&0.7/--/--&6&159.1&15&s&3&&S17,24\\
274242788&HD~92728&A0p Si&5.78\rlap{2}&10050&3.9&&&&&&22&21&&S21\\
286462756&HD~335238&A1p CrEu&9.26&9350&4.2&&1.4/9.3/11.9&1&48.7&1&r&1&&S15\\
286965228&HD~127304&B9p Si?&6.05&9950&4.1&&&&&&&&&S23\\
286992225&TYC~2553-480-1&A2p SrEu&11.56&7500&4.3&roAp&&&&&&&&S23\\
289548946&BD+47 3253&A0p Si&9.55&8800&4.0&&&&&&&&&S15,16\\
298197561&HD~340577&A3p SrCrEu&9.09&8450&3.5&&&&116.7&17&&&$\delta$\,Sct star&S14,15\\
301918605&HD~17330&B7p Si&7.11&10250&&&0.6/--/--&5, 6&&&23&6&&S18\\
301946105&HD~7410&A7p CrEu&9.07&7800&3.5&&&&&&&&&S17\\
312221714&HD~248727&A0p MnSiCr&10.34&9800&3.9&&&&&&&&$\delta$~Sct star?&S19\\
327293700&HD~9996&B9p CrEuSi&6.38&10500&4.0&&0.7/4.7/--&1&7850&15&r&1&&S18\\
331644554&BD+46 3543&A2p Si&9.75&8450&4.3&&&&&&&&$\delta$\,Sct star; EB&S16\\
341616734&HD~89069&A0p SrCrEu&8.42&9550&4.0&&0.5/--/--&5&19&18&s; $<20$&3, 5&&S14,20,21,26\\
343126267&BD+54 2730&F0p SiSr&10.46&8000&4.2&&&&&&&&$\delta$\,Sct star&S16,17\\
346595127&HD~22860&B9p Si&6.86&10700&4.1&&&&6.27\rlap{\tablefootmark{e}}&19&&&&S18\\
347202840&HD~236298&A0p Si&9.45&10700&4.2&&&&&&&&&S17,24\\
352787151&BD+35 5094&F0p SrEu&9.08&6900&3.5&roAp?&&&&&&&&S17\\
353371885&HD~249931&A0p SiSr&10.62&9100&4.3&&&&&&&&&S19\\
368073692&HD~278204&B5p Si&10.45&10850&4.1&&&&&&&&&S19\\
371976932&HD~240242&A4p Si&10.15&&&&&&&&&&$\delta$\,Sct star&S17,24\\
373378410&BD+35 4488&A2p SiSr&9.83&8750&4.0&&&&&&&&&S15\\
379529618&HD~12288&A2p CrSi&7.74&8800&3.8&&1.6/8.0/--&1&34.9&20&r&1&&S18,19,25\\
389105232&BD+57 2636&Ap SrCrEu&9.55&7900&3.8&&&&&&&&&S16,17\\
391462523&HD~187400&A0p Si&8.63&9100&4.1&&&&&&&&&S14\\
403625657&HD~11187&A0p SiCrSr&7.94&10750&3.5&&0.6/--/--&&&&(s)&3&&S18\\
408211646&HD~332312&A4p Sr&9.72&8650&3.5&&&&&&&&&S14,15\\
409522234&HD~14437&B9p CrEuSr&7.27&10450&3.9&&1.9/7.6/--&1&26.87&20&r&1&&S18\\
430355895&BD+52 3124&A0p Si&10.12&9050&4.0&roAp?&&&&&&&&S16,17\\
450217536&TYC~3729-816-1&F0p SrSi&11.35&6650&3.6&&&&&&&&&S19\\
468507699&HD~206977&A5p SrEu&8.98&8650&3.8&&&&&&&&&S15,16\\
470837956&BD+61 2565&A8p SrEuCr&9.9\phantom{0}&8200&3.4&roAp?&&&&&&&&S17,18\\
470878580&BD+65 1981&Ap Si&10.09&10550&4.3&roAp?&&&&&&&&S17,18,24\\[4pt]
\hline\\[-4pt]
\end{tabular*}
\tablebib{(1)~\citet{2017A&A...601A..14M};
  (2)~\citet{2011MNRAS.410..517B}; (3)~this paper;
  (4)~\citet{2012AN....333...41K}; (5)~\citet{2017AstBu..72..391R};
  (6)~\citet{2018AstBu..73..178R}; (7)~\citet{2011MNRAS.410..517B};
  (8)~\citet{1958ApJS....3..141B}; (9)~\citet{1971ApJ...164..309P};
  (10)~\citet{1990BAICz..41..118Z}; (11)~\citet{2007A&A...475.1053A};
  (12)~\citet{1997A&AS..123..353M};
  (13)~\citet{2019MNRAS.488...18H}; (14)~\citet{1981A&AS...44..265A};
  (15)~\citet{2017PASP..129j4203P}; (16)~\citet{2016A&A...586A..85M};
  (17)~\citet{2016AJ....152..104H}; (18)~\citet{2017MNRAS.468.2745N};
  (19)~\citet{2012MNRAS.420..757W}; (20)~\citet{2000A&A...355.1080W};
  (21)~\citet{2012A&A...537A.120Z}.}
\end{table*}
\end{landscape}

\onecolumn

\section{Amplitude spectra of long rotation period Ap star candidates}
\label{sec:amp_sp}
\begin{figure*}[h!]
  \centering
  \includegraphics[width=0.45\linewidth,angle=0]{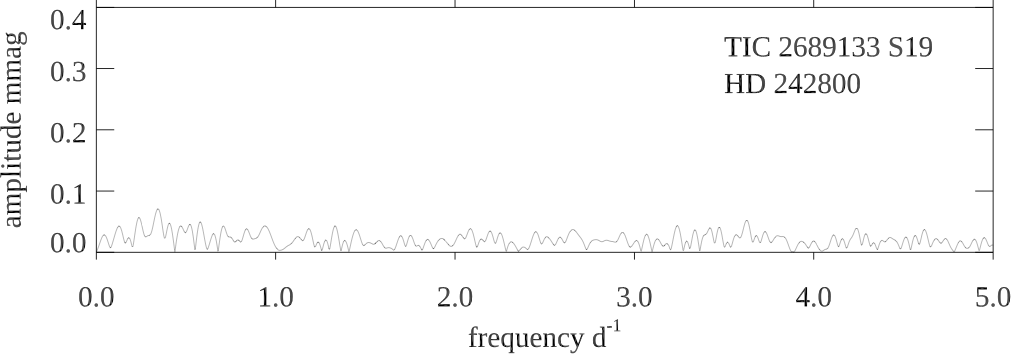}
  \includegraphics[width=0.45\linewidth,angle=0]{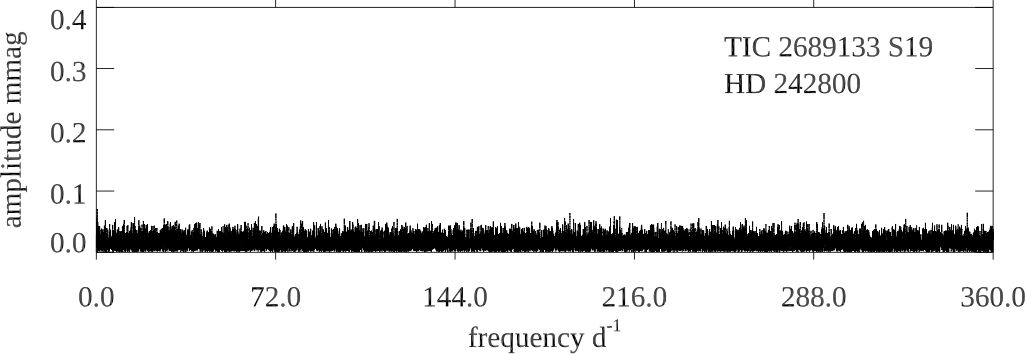}
  \includegraphics[width=0.45\linewidth,angle=0]{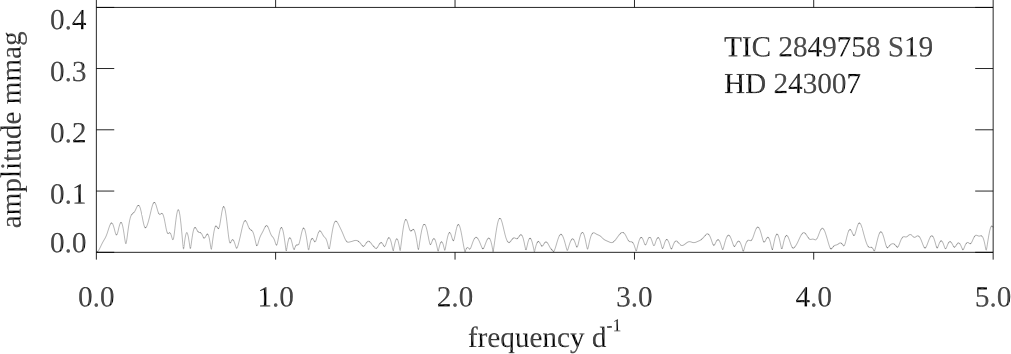}
  \includegraphics[width=0.45\linewidth,angle=0]{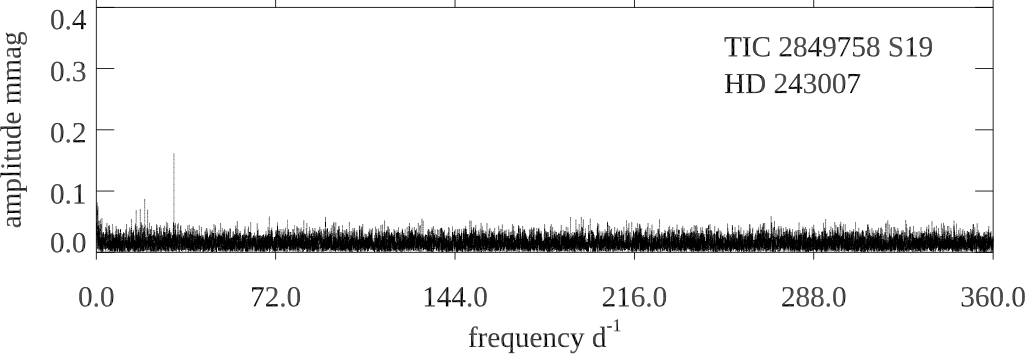}
  \includegraphics[width=0.45\linewidth,angle=0]{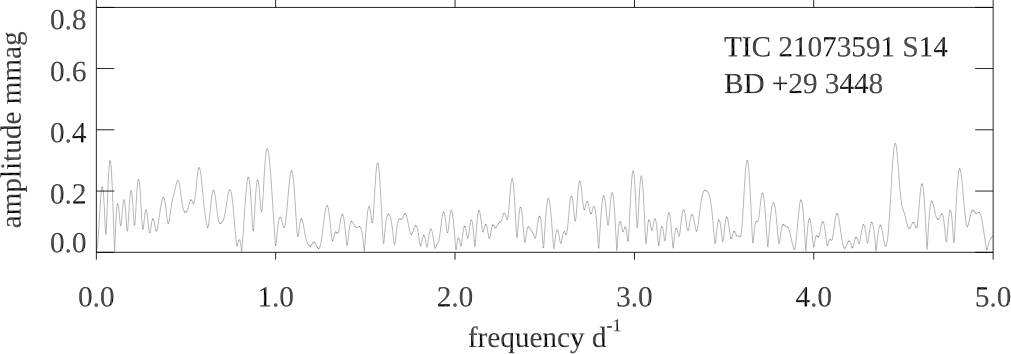}
  \includegraphics[width=0.45\linewidth,angle=0]{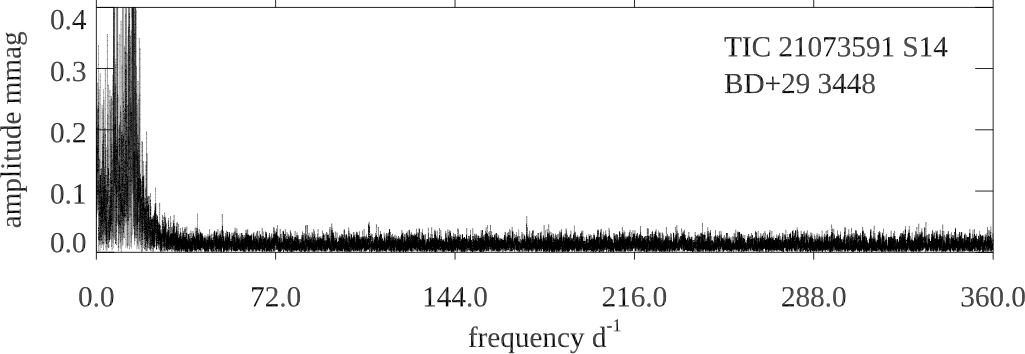}
  \includegraphics[width=0.45\linewidth,angle=0]{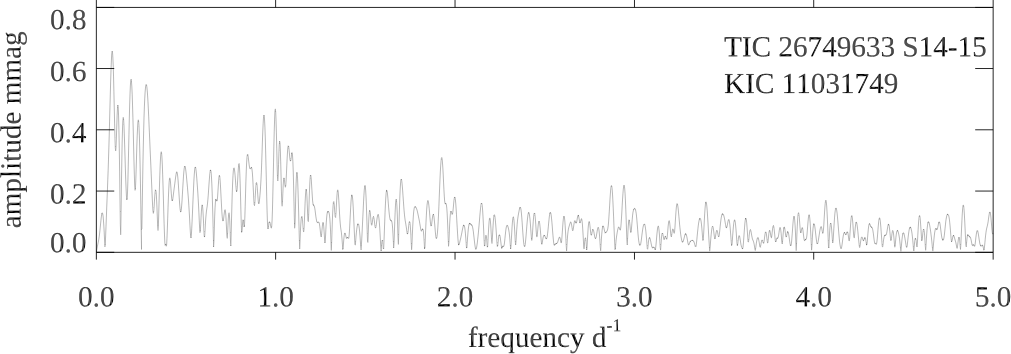}
  \includegraphics[width=0.45\linewidth,angle=0]{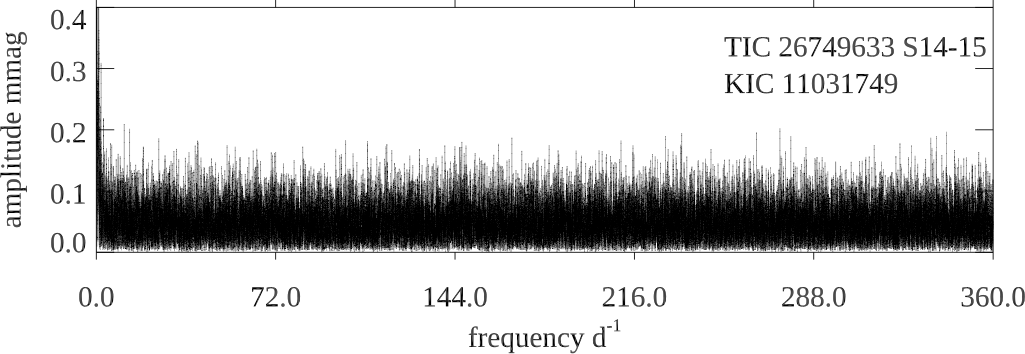}
  \includegraphics[width=0.45\linewidth,angle=0]{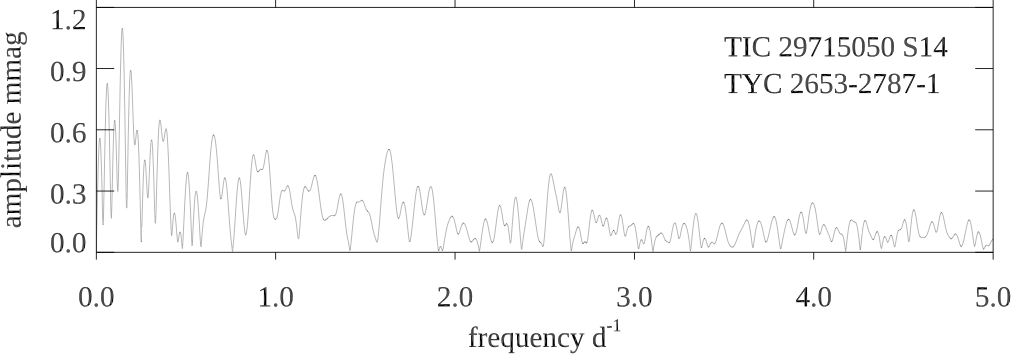}
  \includegraphics[width=0.45\linewidth,angle=0]{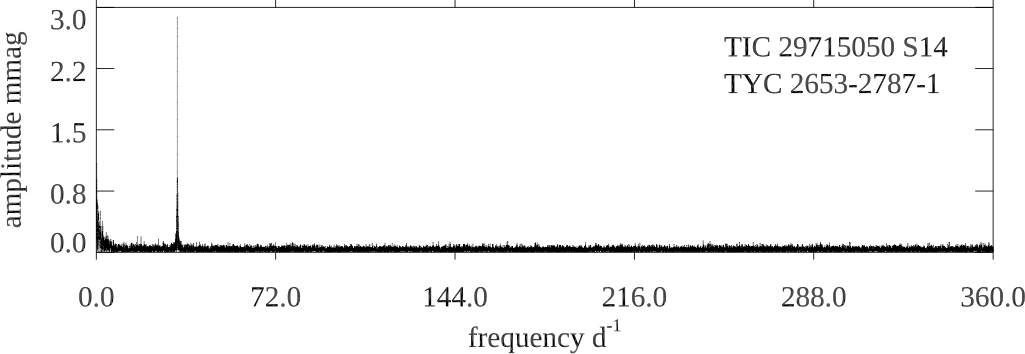}
  \includegraphics[width=0.45\linewidth,angle=0]{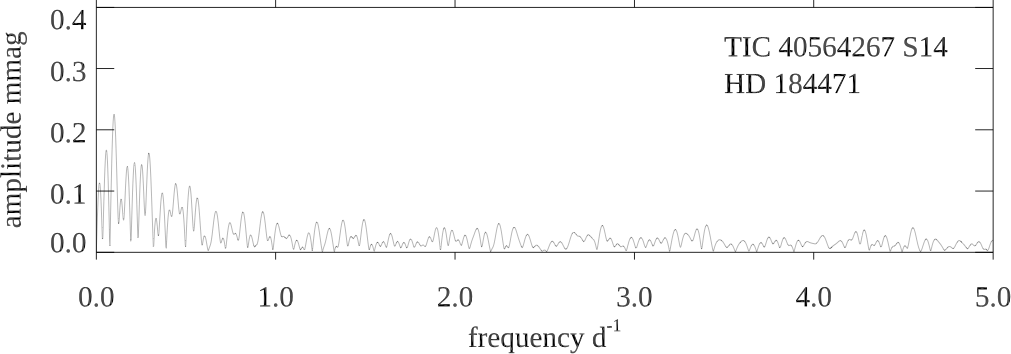}
  \includegraphics[width=0.45\linewidth,angle=0]{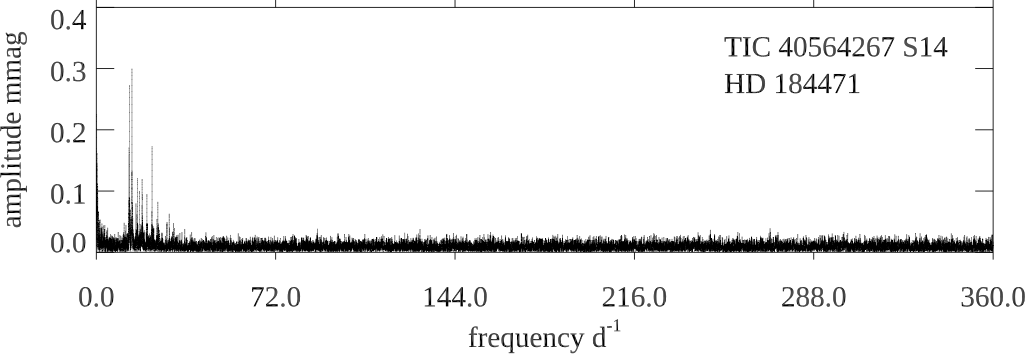}
  \includegraphics[width=0.45\linewidth,angle=0]{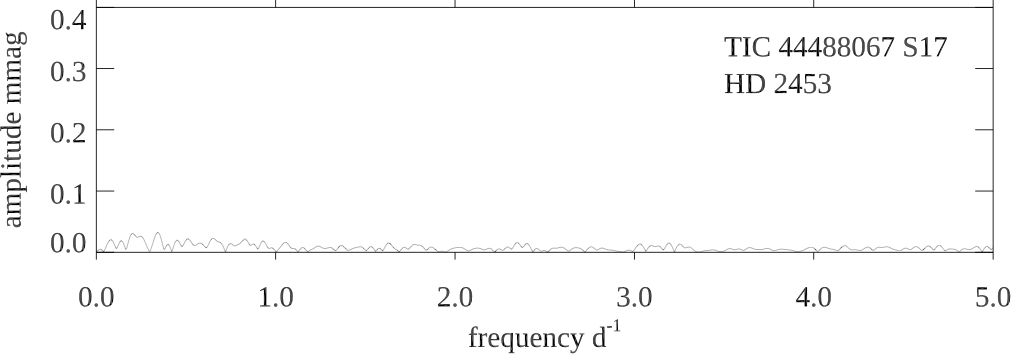}
  \includegraphics[width=0.45\linewidth,angle=0]{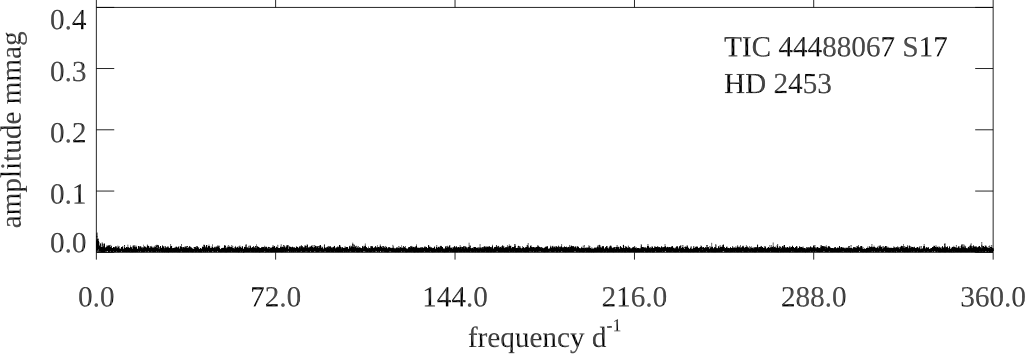}
  \caption{Amplitude spectra for the long-period Ap stars, continued. TIC\,21073591 and TIC\,40564267 are $\delta$~Sct stars (Section\,\ref{deltasct} and Fig.\,\ref{fig:deltasct}). TIC\,29715050  is an roAp star (section\,\ref{roAp}).}
  \label{fig:ssrAp1}
  \end{figure*}

\begin{figure*}[p]
  \centering
  \includegraphics[width=0.45\linewidth,angle=0]{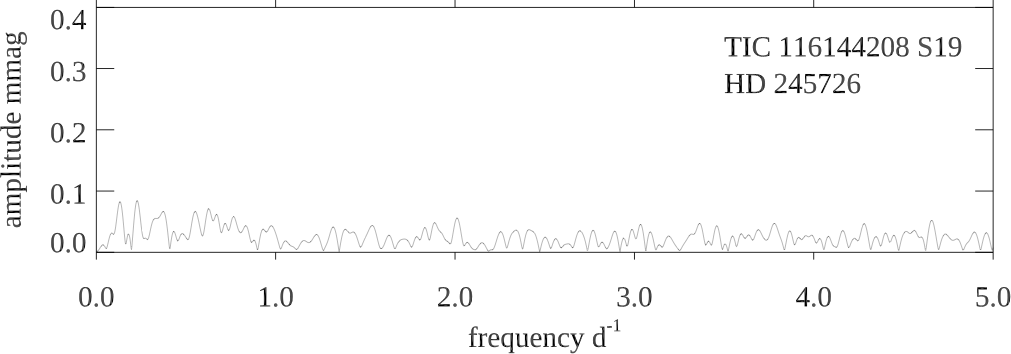}
  \includegraphics[width=0.45\linewidth,angle=0]{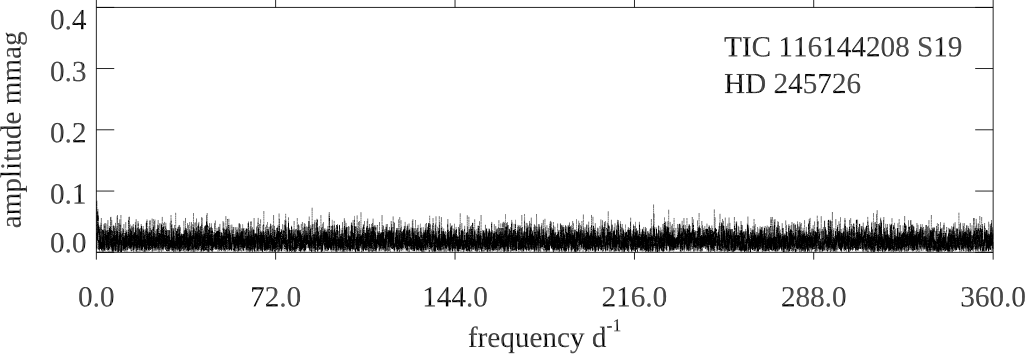}
  \includegraphics[width=0.45\linewidth,angle=0]{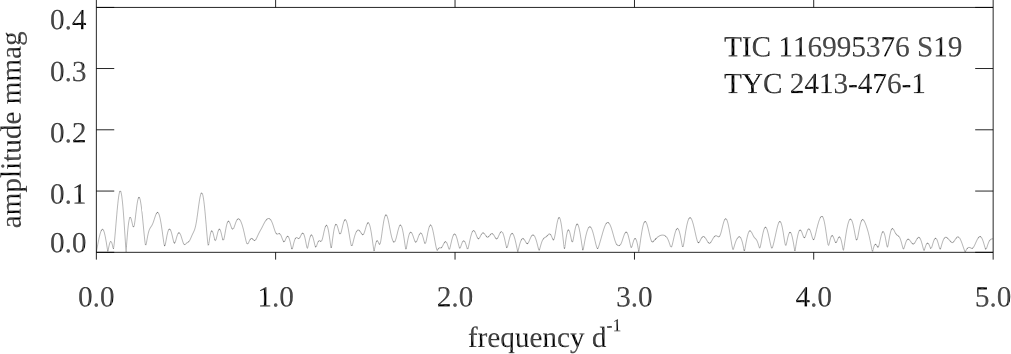}
  \includegraphics[width=0.45\linewidth,angle=0]{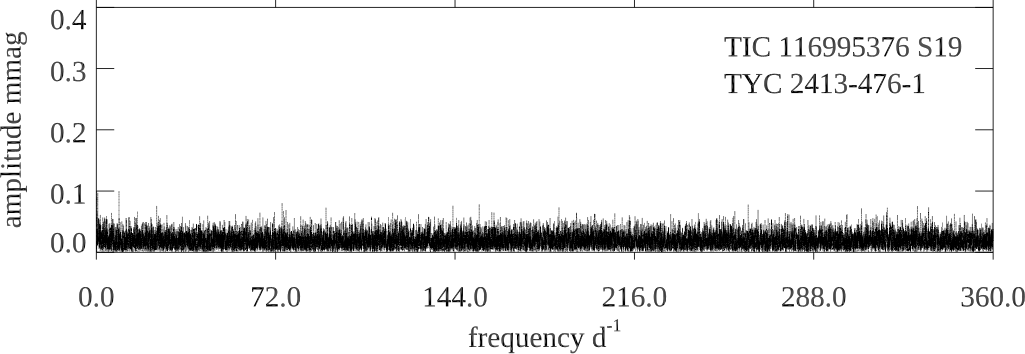}
  \includegraphics[width=0.45\linewidth,angle=0]{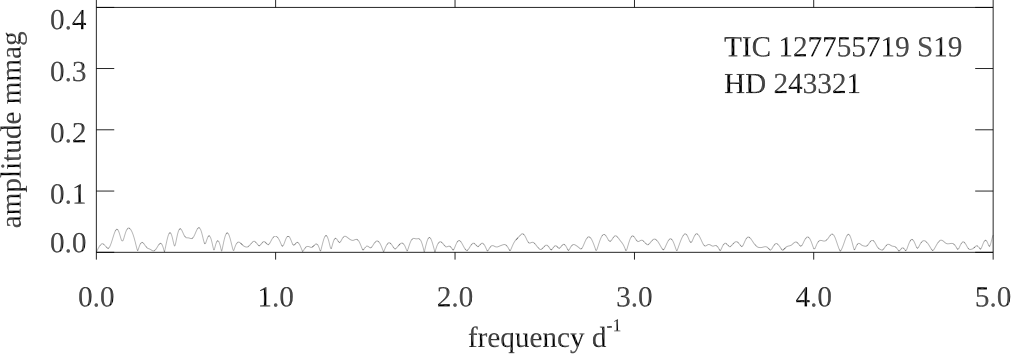}
  \includegraphics[width=0.45\linewidth,angle=0]{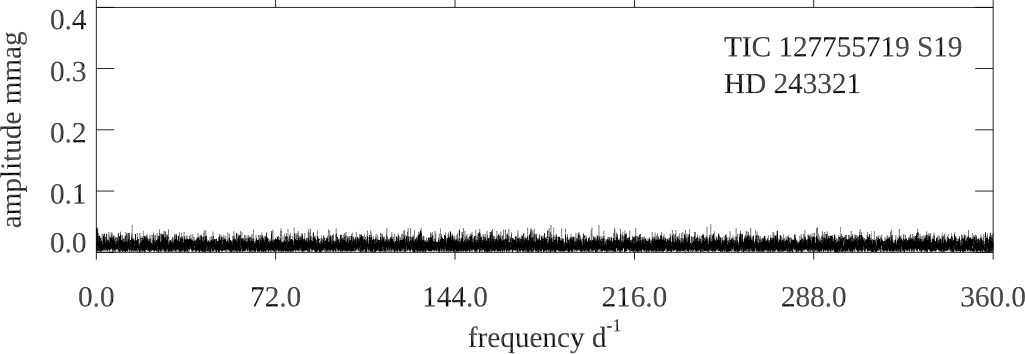}
  \includegraphics[width=0.45\linewidth,angle=0]{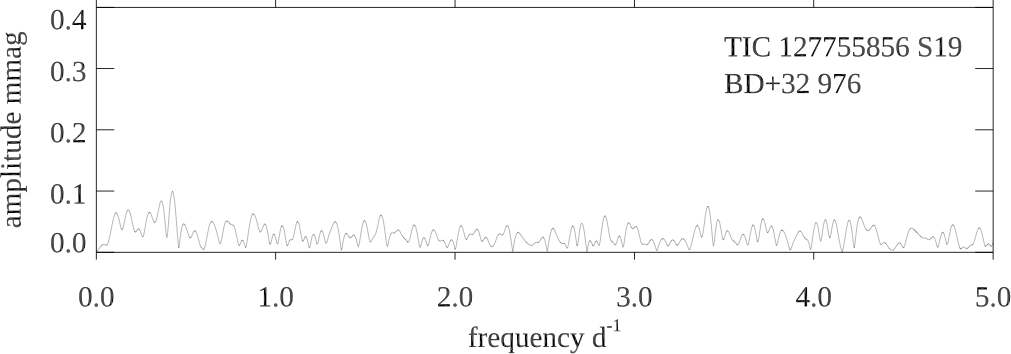}
  \includegraphics[width=0.45\linewidth,angle=0]{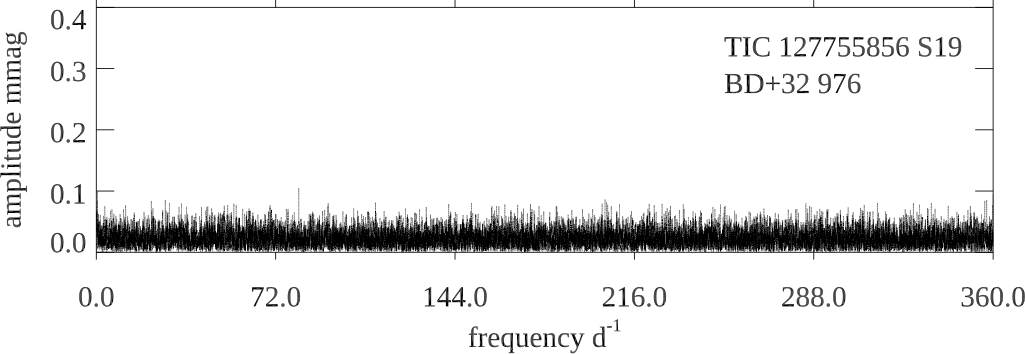}
  \includegraphics[width=0.45\linewidth,angle=0]{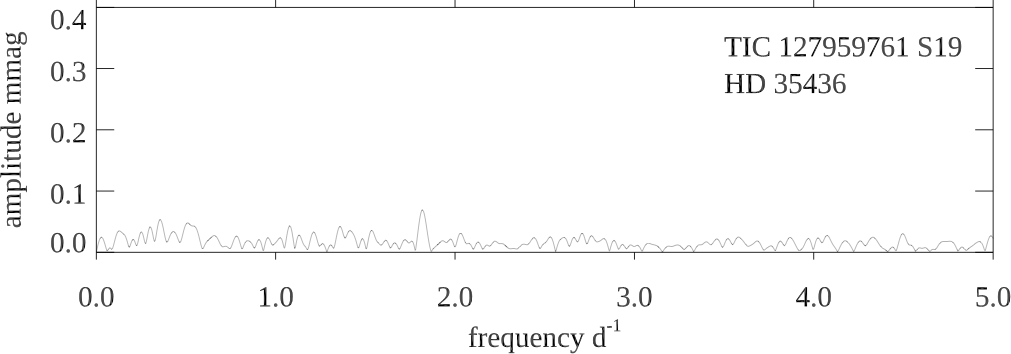}
  \includegraphics[width=0.45\linewidth,angle=0]{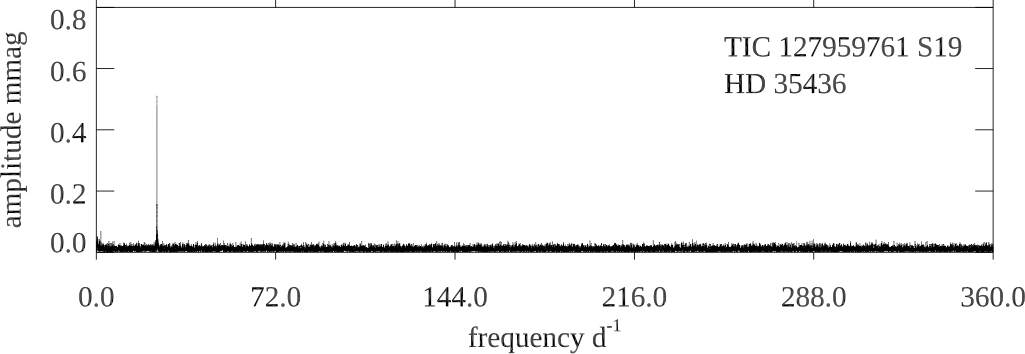}
  \includegraphics[width=0.45\linewidth,angle=0]{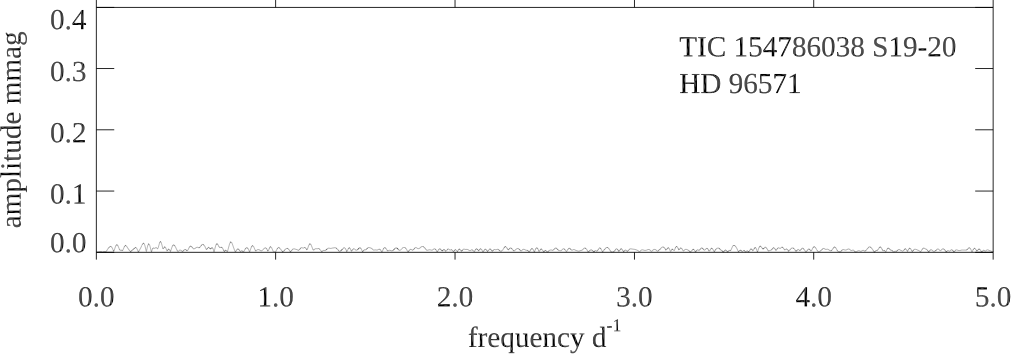}
  \includegraphics[width=0.45\linewidth,angle=0]{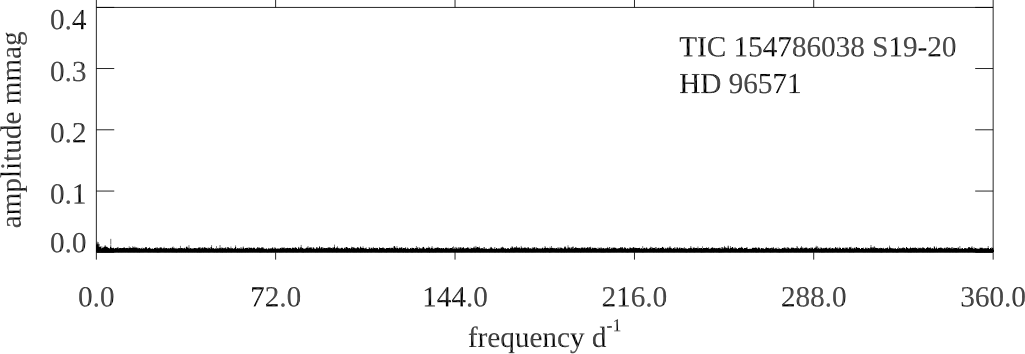}
  \includegraphics[width=0.45\linewidth,angle=0]{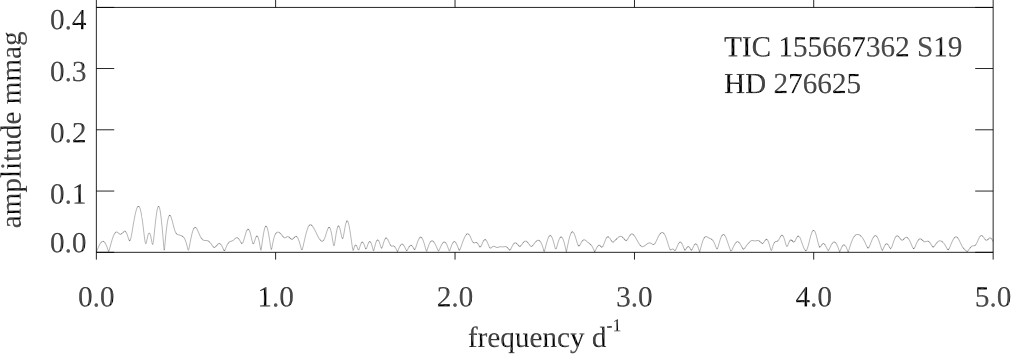}
  \includegraphics[width=0.45\linewidth,angle=0]{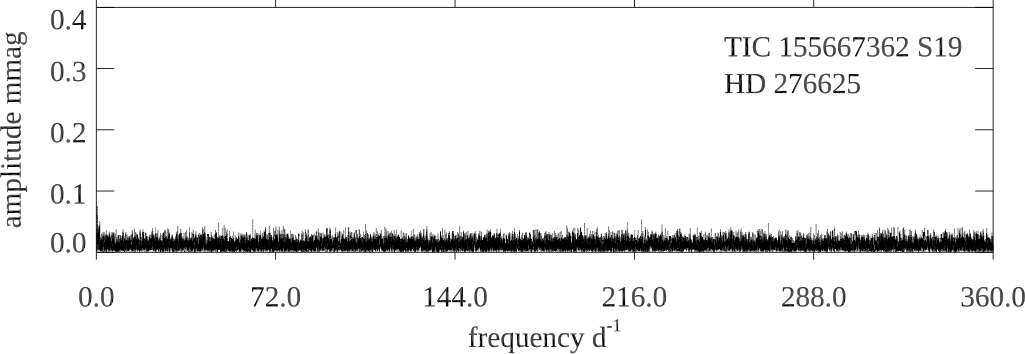}
  \caption{Amplitude spectra for the long-period Ap stars, continued. TIC\,127959761 is a $\delta$~Sct star (Section\,\ref{deltasct} and Fig.\,\ref{fig:deltasct}).}
  \label{fig:ssrAp3}
  \end{figure*}

\begin{figure*}[p]
  \centering
  \includegraphics[width=0.45\linewidth,angle=0]{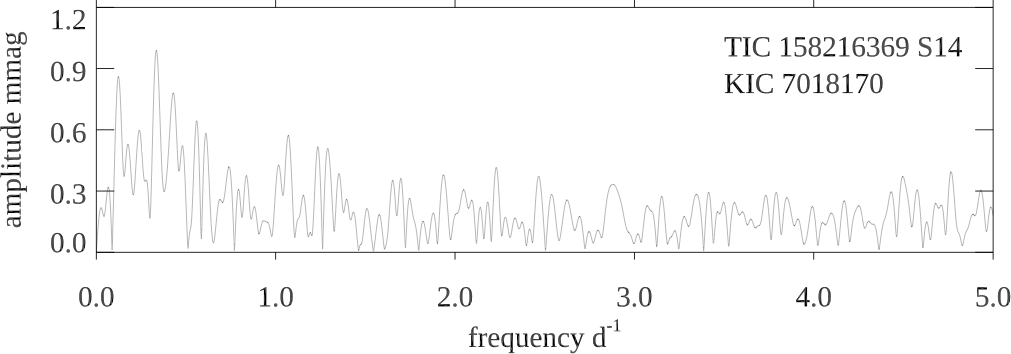}
  \includegraphics[width=0.45\linewidth,angle=0]{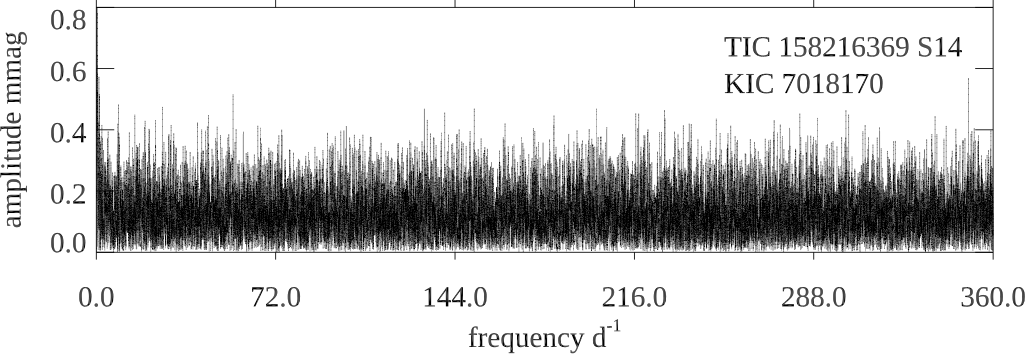}
  \includegraphics[width=0.45\linewidth,angle=0]{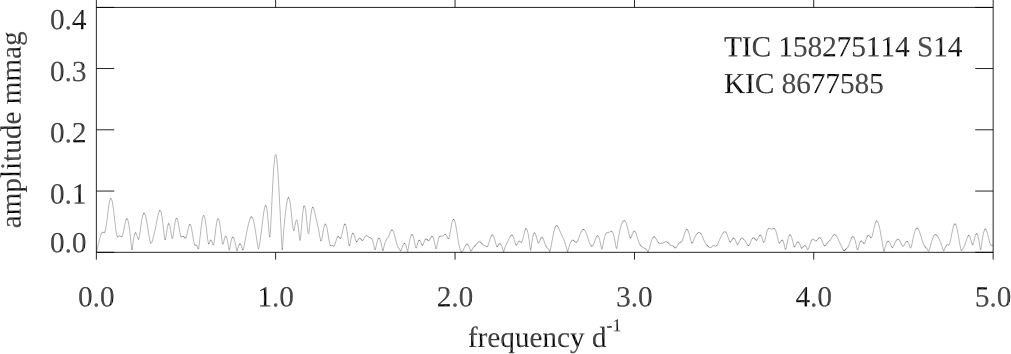}
  \includegraphics[width=0.45\linewidth,angle=0]{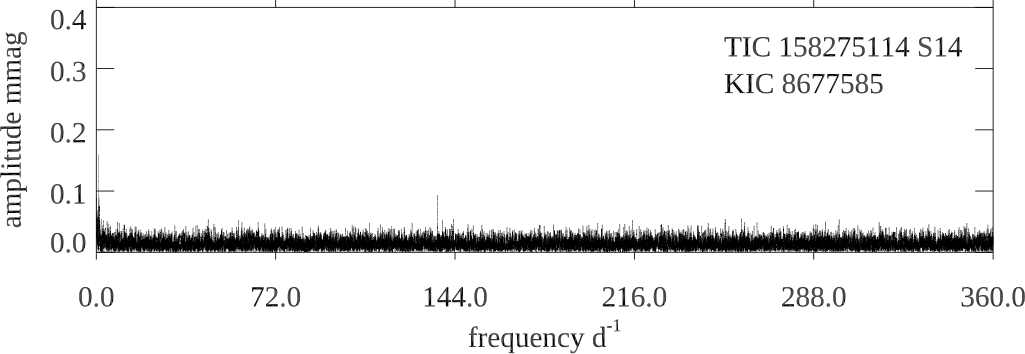}
  \includegraphics[width=0.45\linewidth,angle=0]{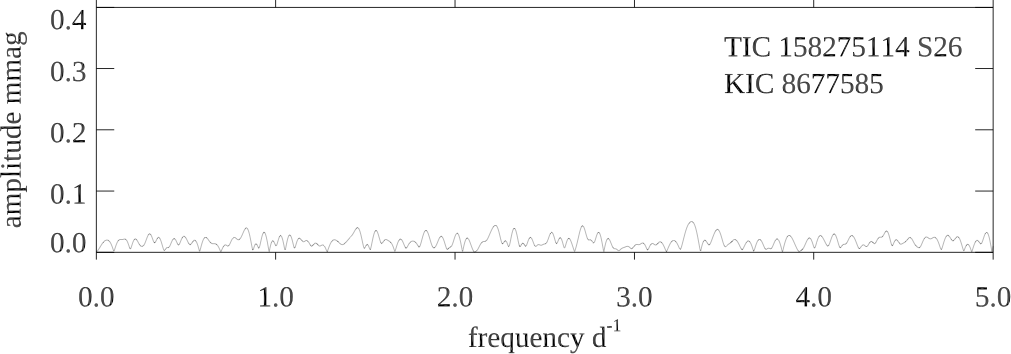}
  \includegraphics[width=0.45\linewidth,angle=0]{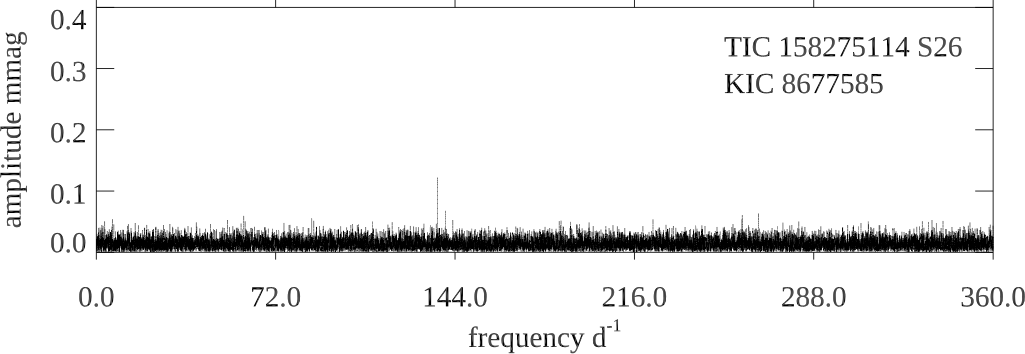}
  \includegraphics[width=0.45\linewidth,angle=0]{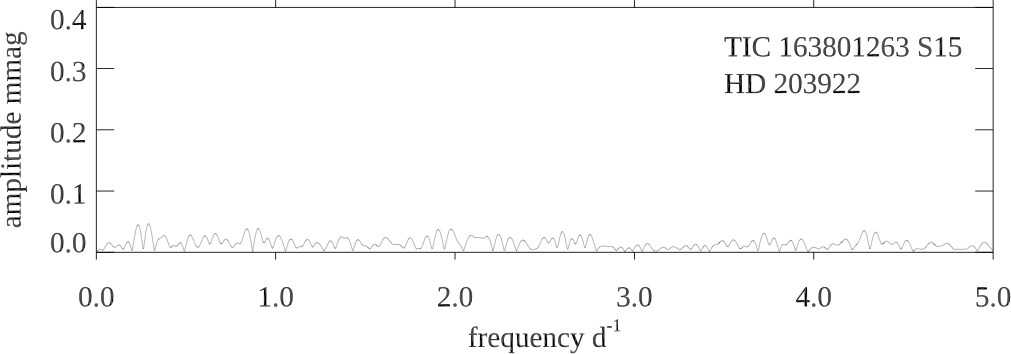}
  \includegraphics[width=0.45\linewidth,angle=0]{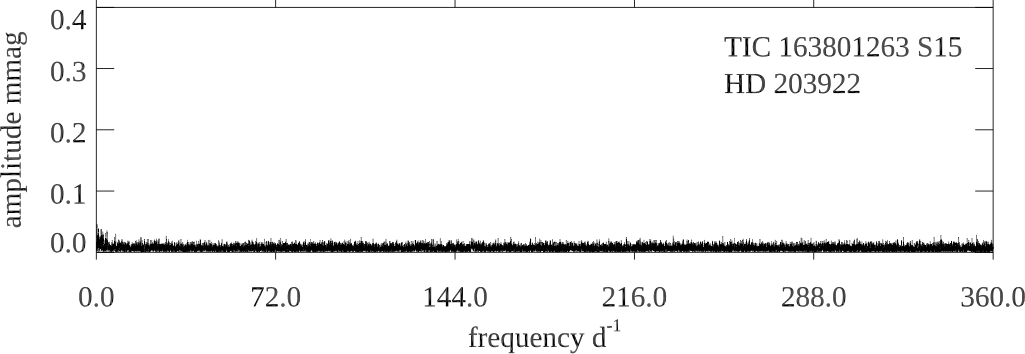}
  \includegraphics[width=0.45\linewidth,angle=0]{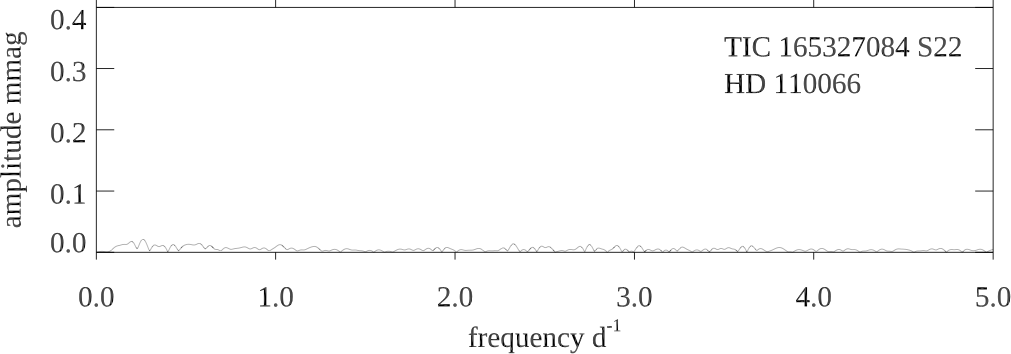}
  \includegraphics[width=0.45\linewidth,angle=0]{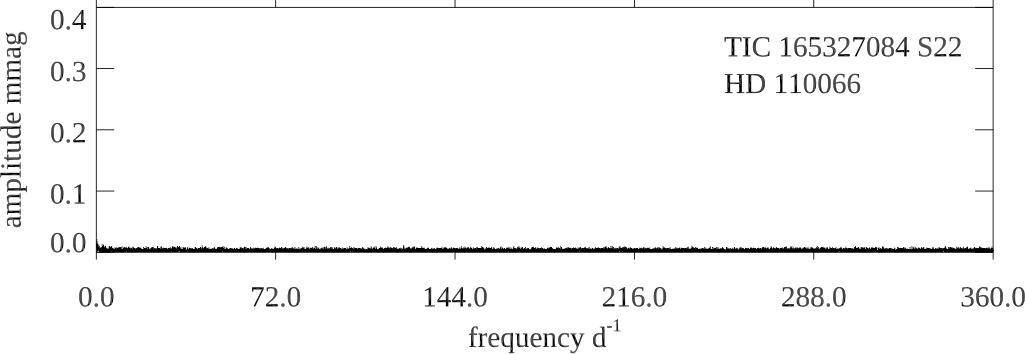}
  \includegraphics[width=0.45\linewidth,angle=0]{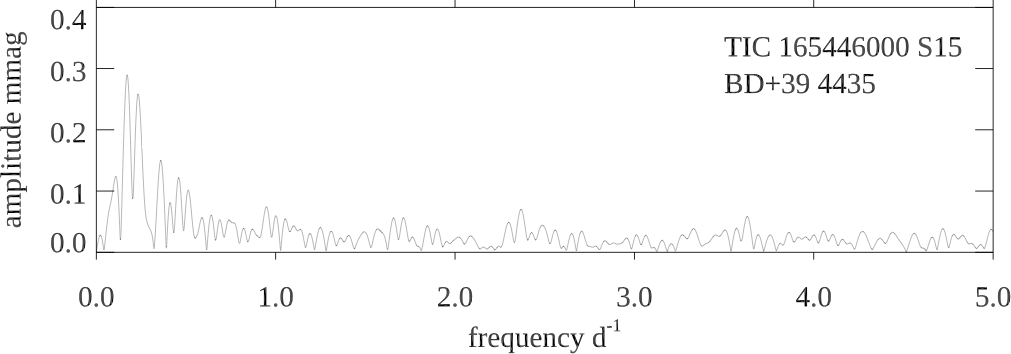}
  \includegraphics[width=0.45\linewidth,angle=0]{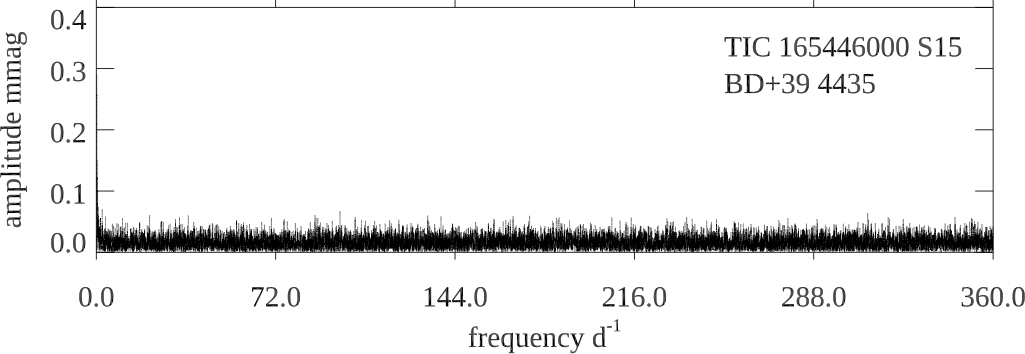}
  \includegraphics[width=0.45\linewidth,angle=0]{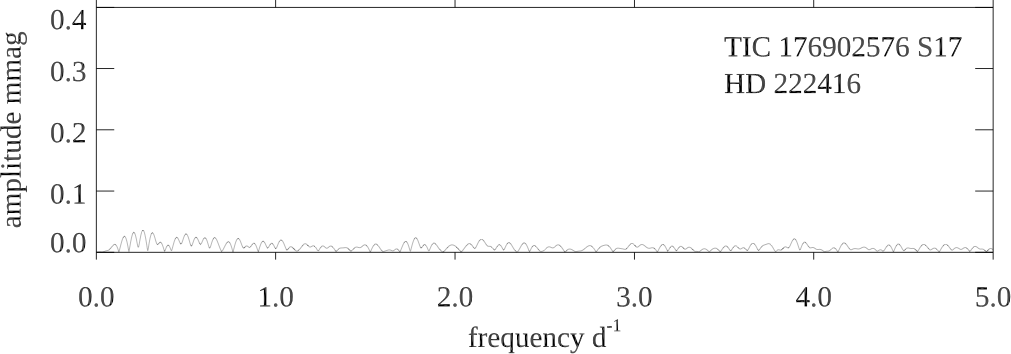}
  \includegraphics[width=0.45\linewidth,angle=0]{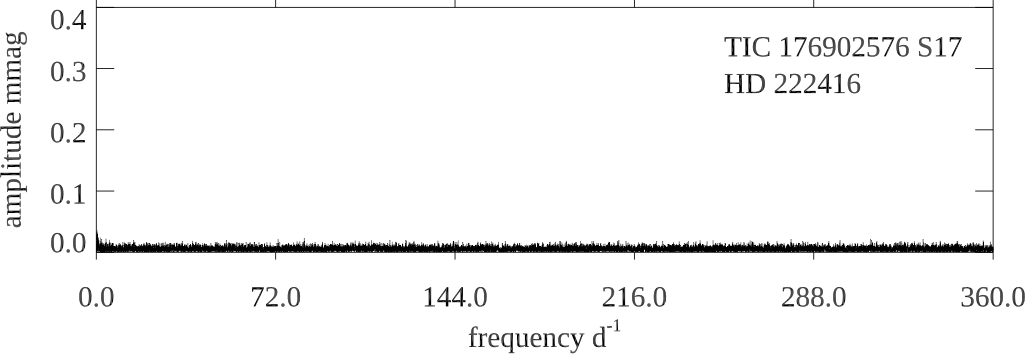}
  \caption{Amplitude spectra for the long-period Ap stars, continued. For TIC\,158275114, S14 and S26 are plotted separately, showing that the low frequency peak near 1\,d$^{-1}$ is not present in both sectors, hence is not a rotation signal. This star is a well-studied roAp star, KIC\,8677585 \citep{2013MNRAS.432.2808B} (section\,\ref{roAp}).}
  \label{fig::ssrAp4}
  \end{figure*}

\begin{figure*}[p]
  \centering
  \includegraphics[width=0.45\linewidth,angle=0]{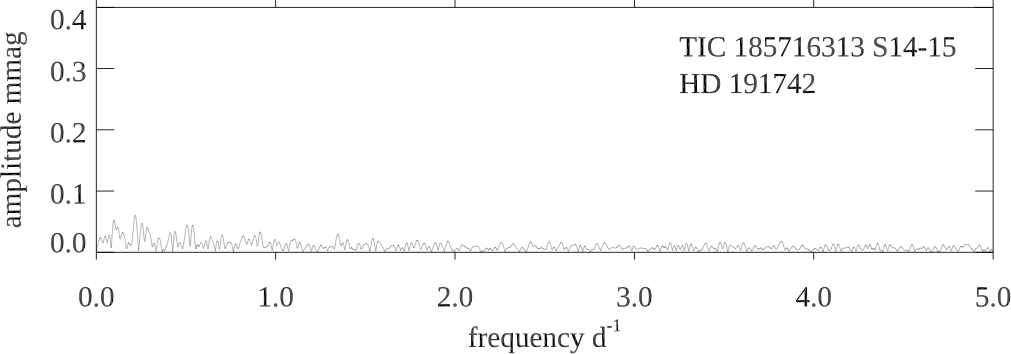}
  \includegraphics[width=0.45\linewidth,angle=0]{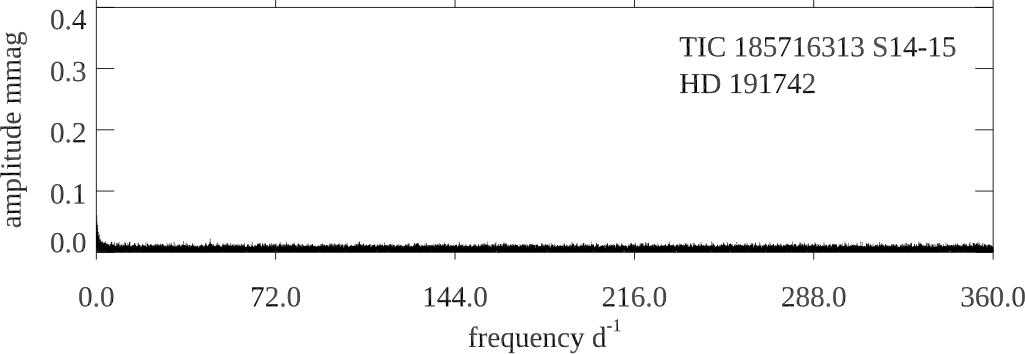}
  \includegraphics[width=0.45\linewidth,angle=0]{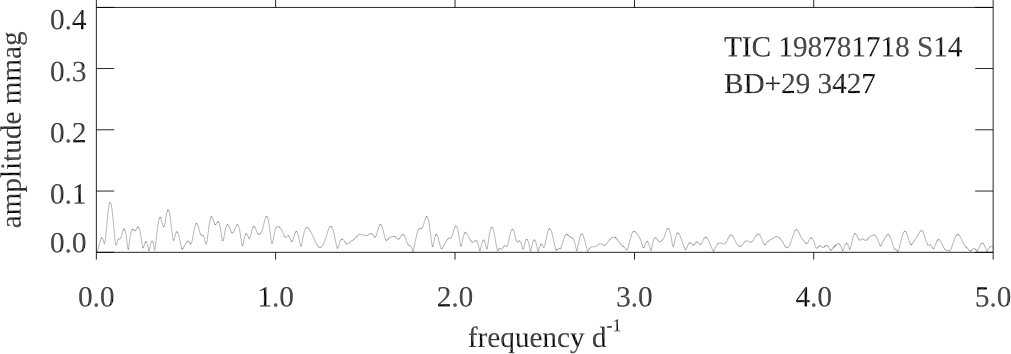}
  \includegraphics[width=0.45\linewidth,angle=0]{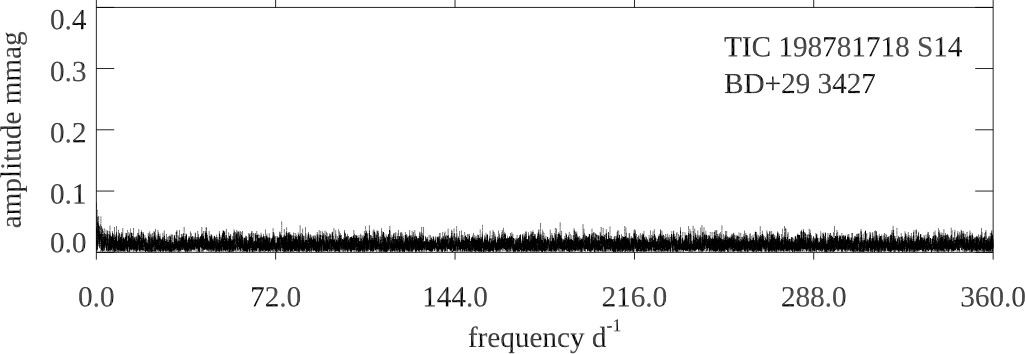}
  \includegraphics[width=0.45\linewidth,angle=0]{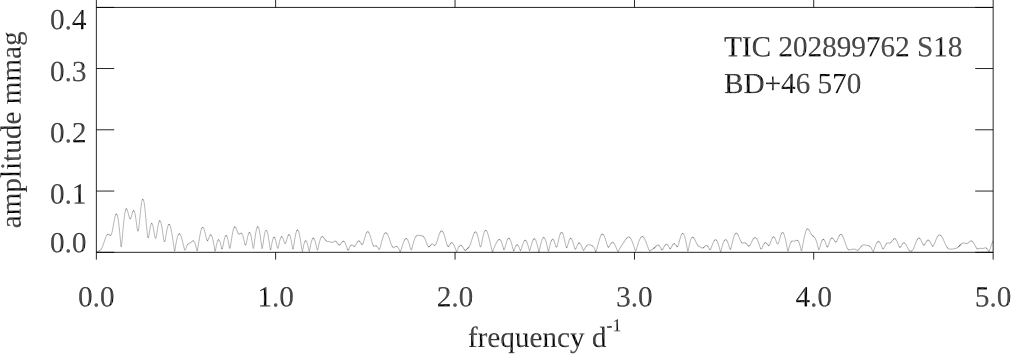}
  \includegraphics[width=0.45\linewidth,angle=0]{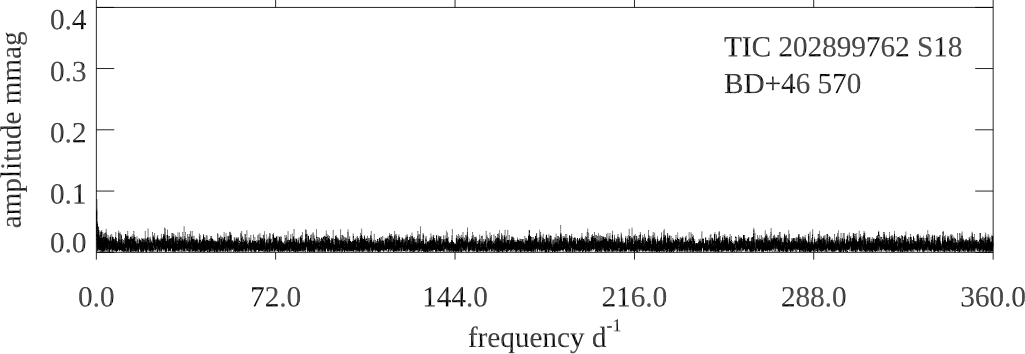}
  \includegraphics[width=0.45\linewidth,angle=0]{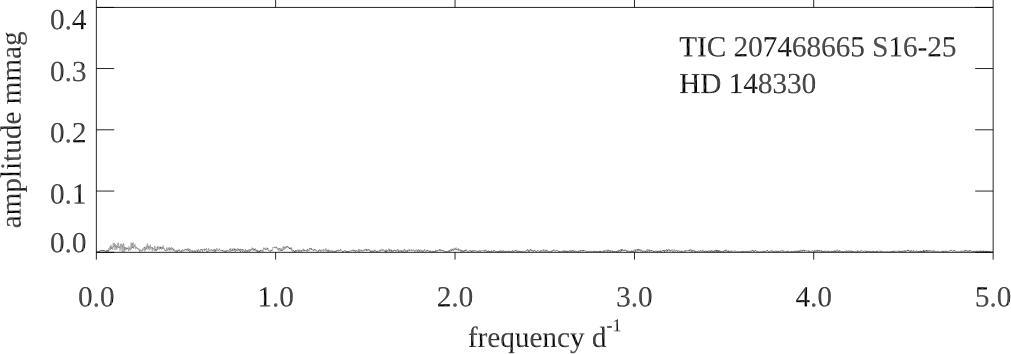}
  \includegraphics[width=0.45\linewidth,angle=0]{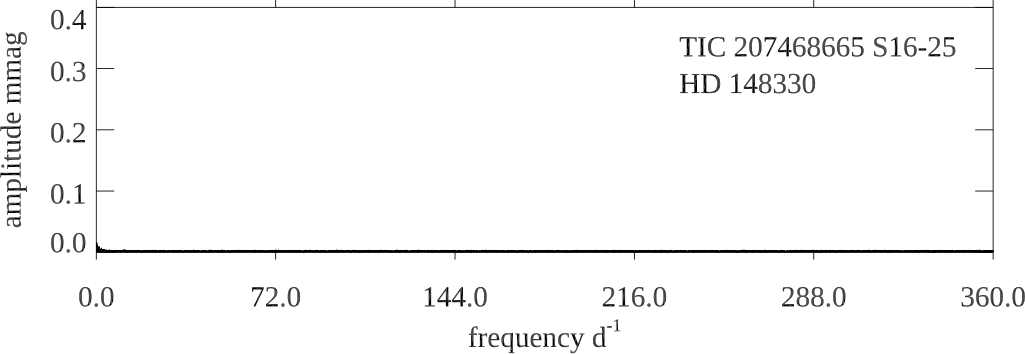}
  \includegraphics[width=0.45\linewidth,angle=0]{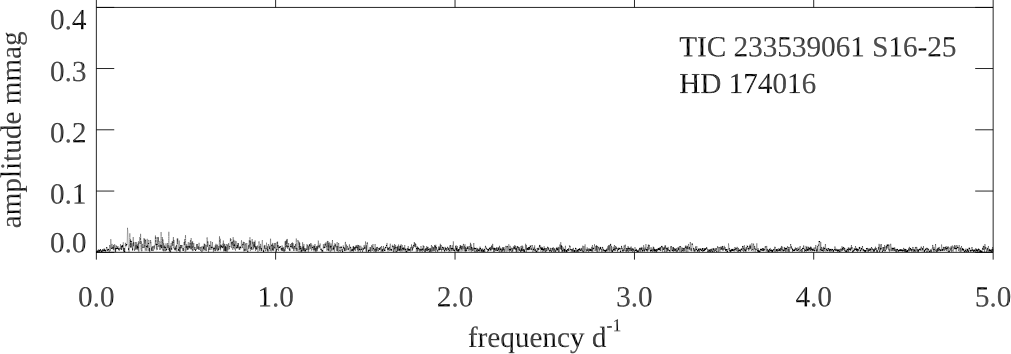}
  \includegraphics[width=0.45\linewidth,angle=0]{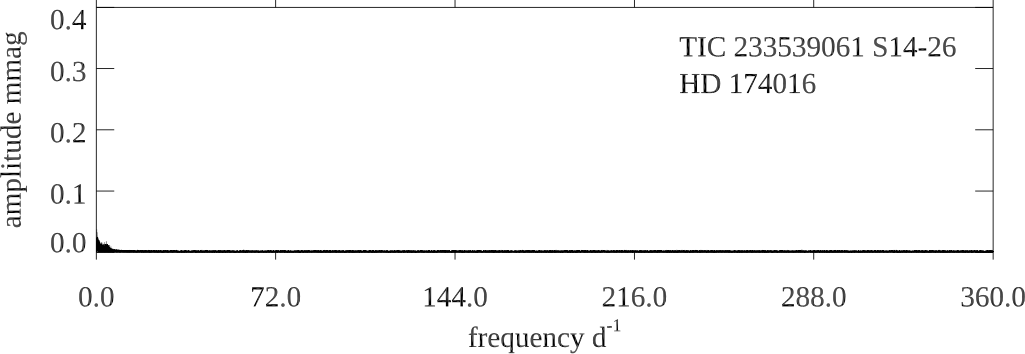}
  \includegraphics[width=0.45\linewidth,angle=0]{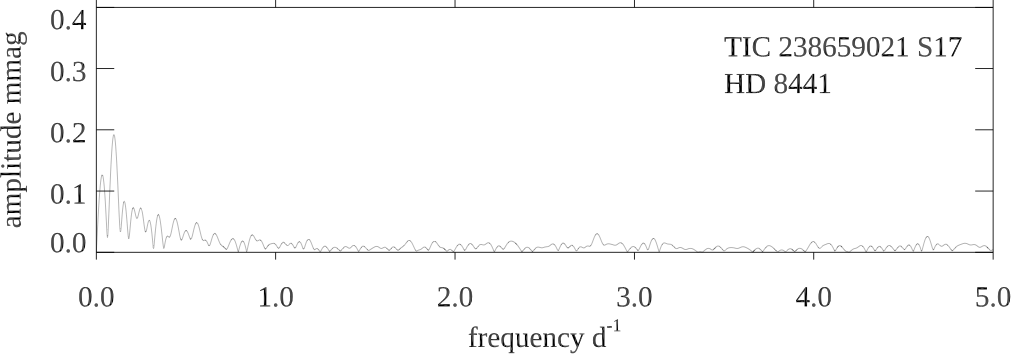}
  \includegraphics[width=0.45\linewidth,angle=0]{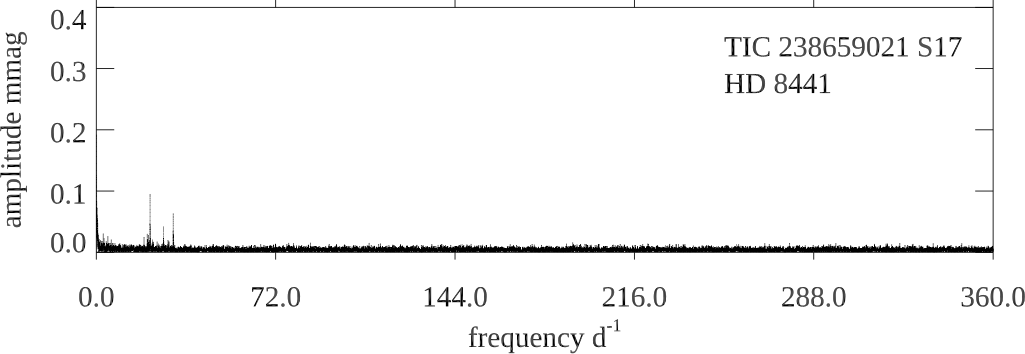}
  \includegraphics[width=0.45\linewidth,angle=0]{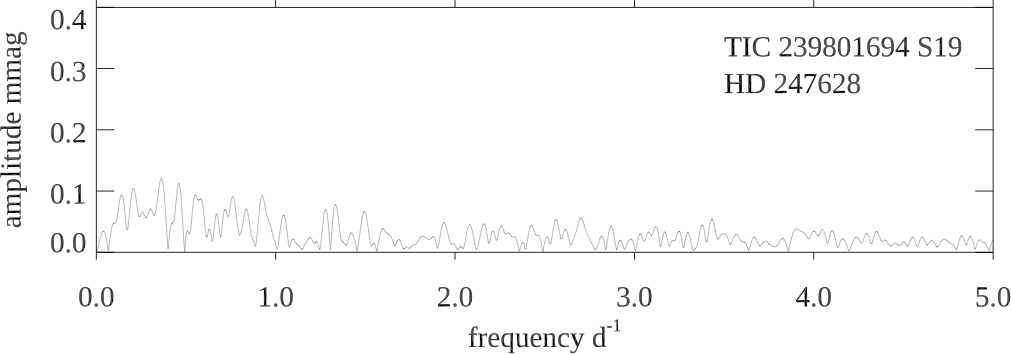}
  \includegraphics[width=0.45\linewidth,angle=0]{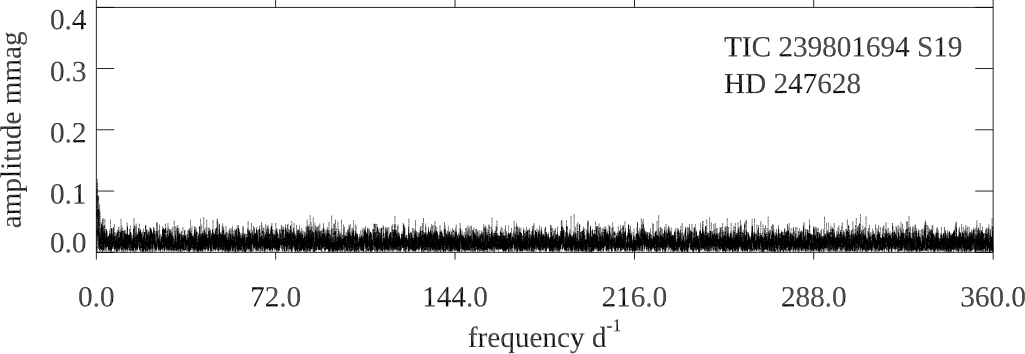}
  \caption{Amplitude spectra for the long-period Ap stars, continued. TIC\,238659021 is a $\delta$~Sct star (Section\,\ref{deltasct} and Fig.\,\ref{fig:deltasct}).}
  \label{fig:ssrAp5}
  \end{figure*}

\begin{figure*}[p]
  \centering
  \includegraphics[width=0.45\linewidth,angle=0]{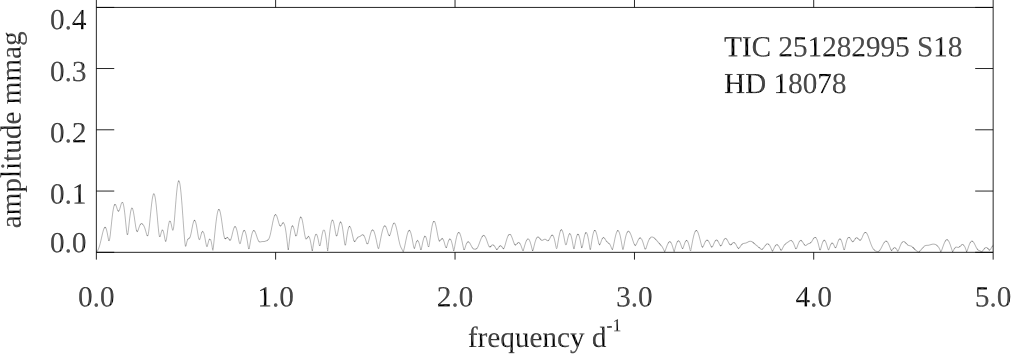}
  \includegraphics[width=0.45\linewidth,angle=0]{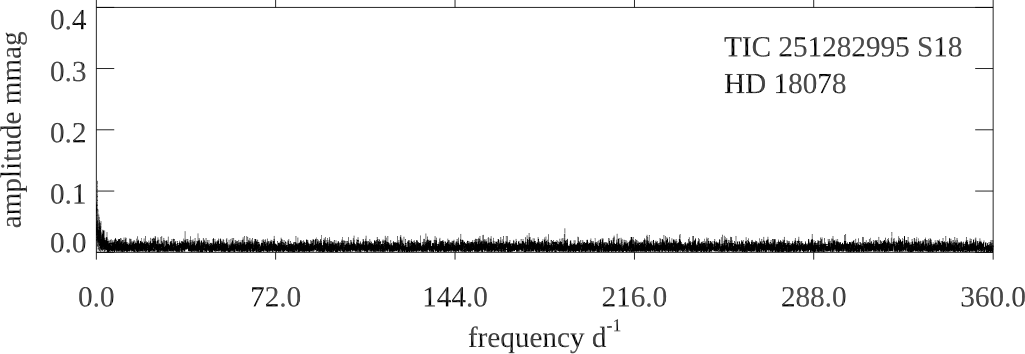}
  \includegraphics[width=0.45\linewidth,angle=0]{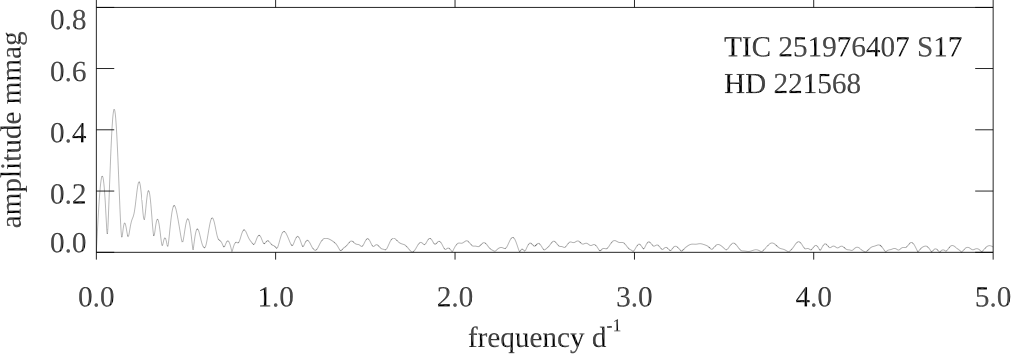}
  \includegraphics[width=0.45\linewidth,angle=0]{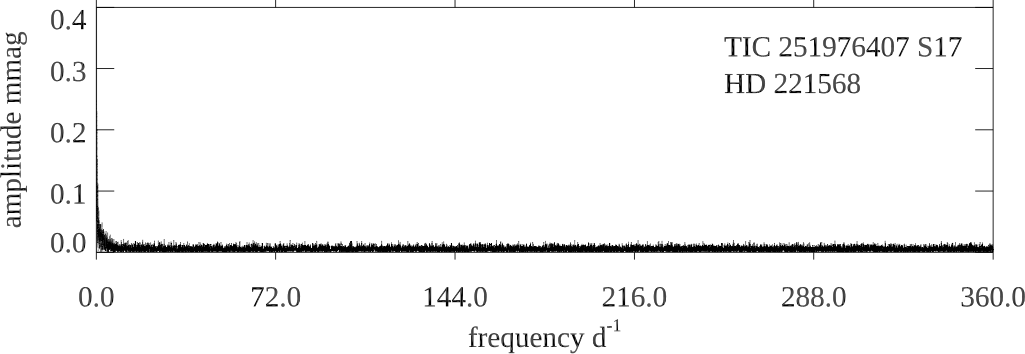}
  \includegraphics[width=0.45\linewidth,angle=0]{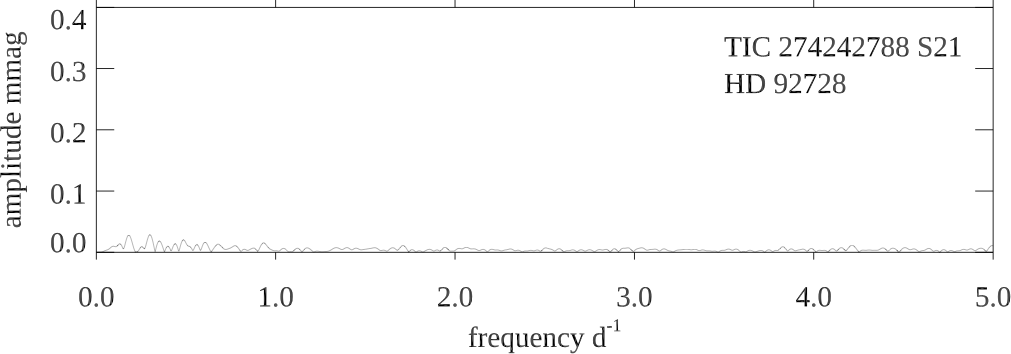}
  \includegraphics[width=0.45\linewidth,angle=0]{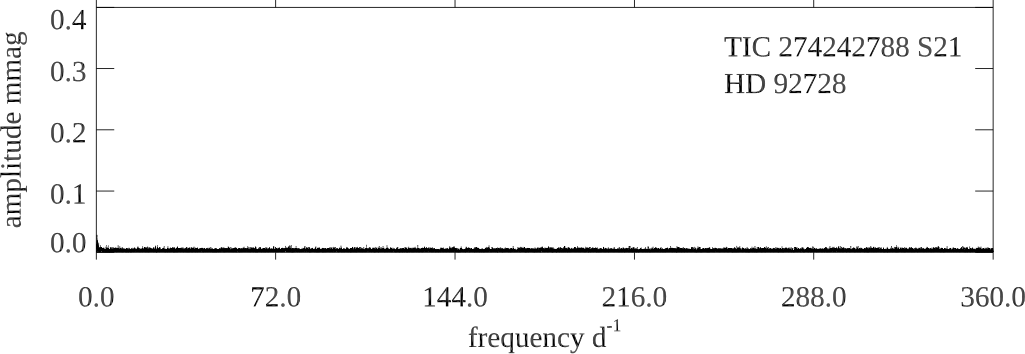}
  \includegraphics[width=0.45\linewidth,angle=0]{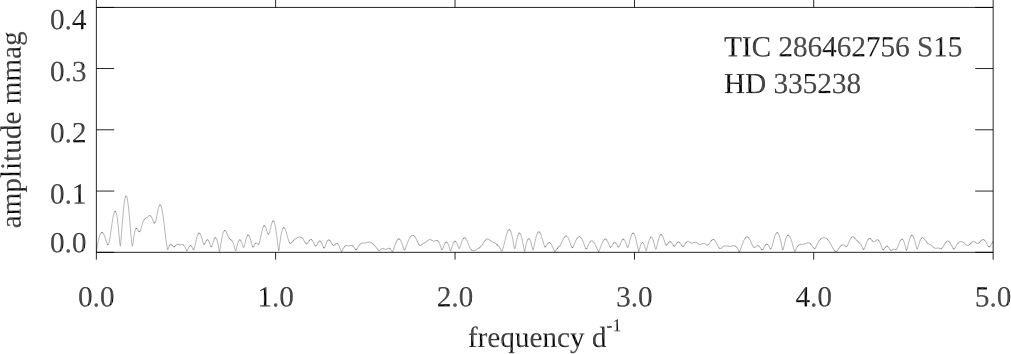}
  \includegraphics[width=0.45\linewidth,angle=0]{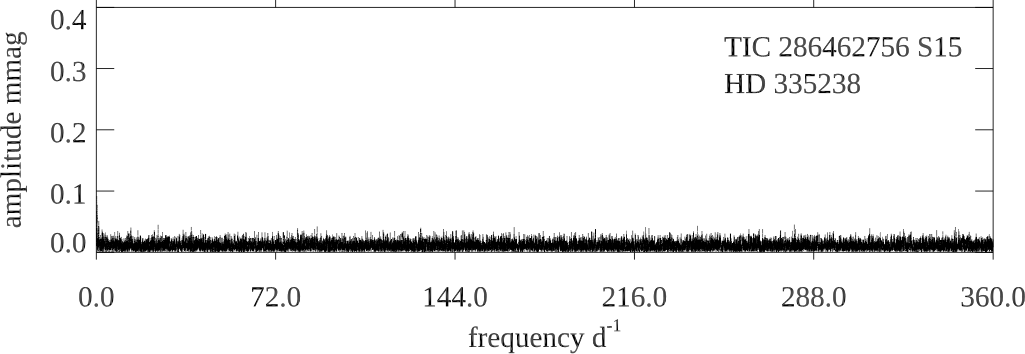}
  \includegraphics[width=0.45\linewidth,angle=0]{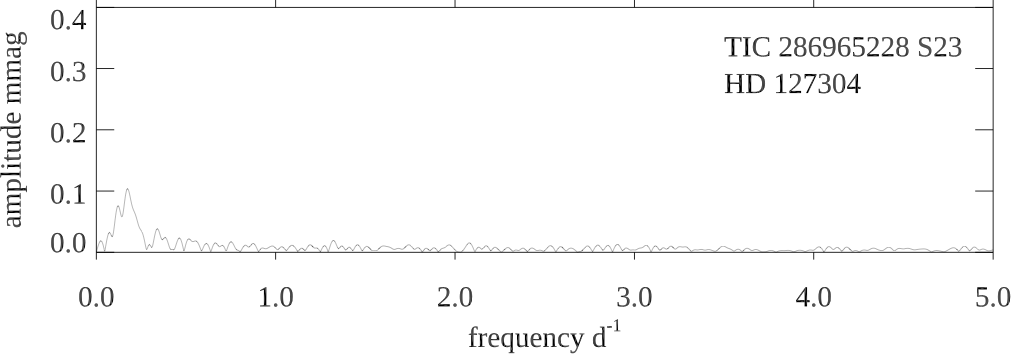}
  \includegraphics[width=0.45\linewidth,angle=0]{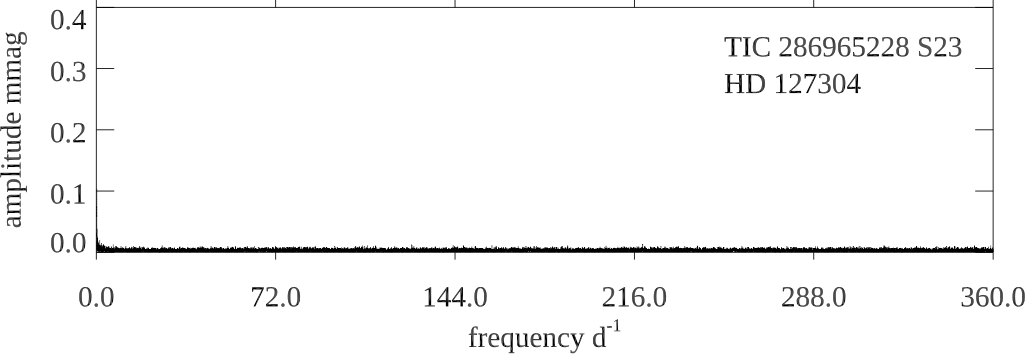}
  \includegraphics[width=0.45\linewidth,angle=0]{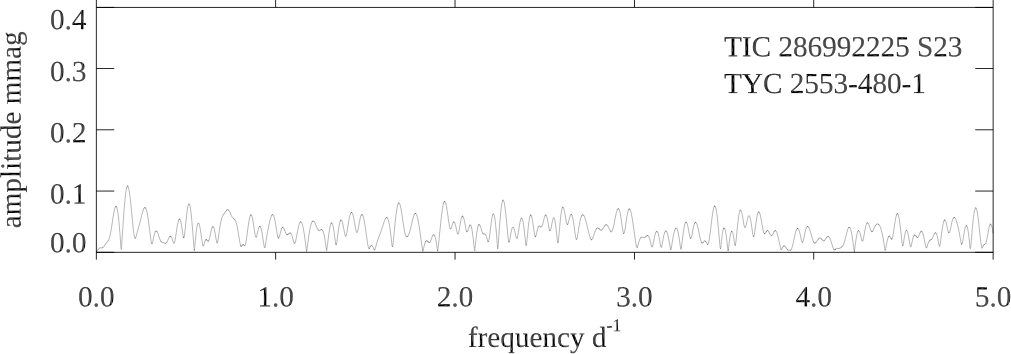}
  \includegraphics[width=0.45\linewidth,angle=0]{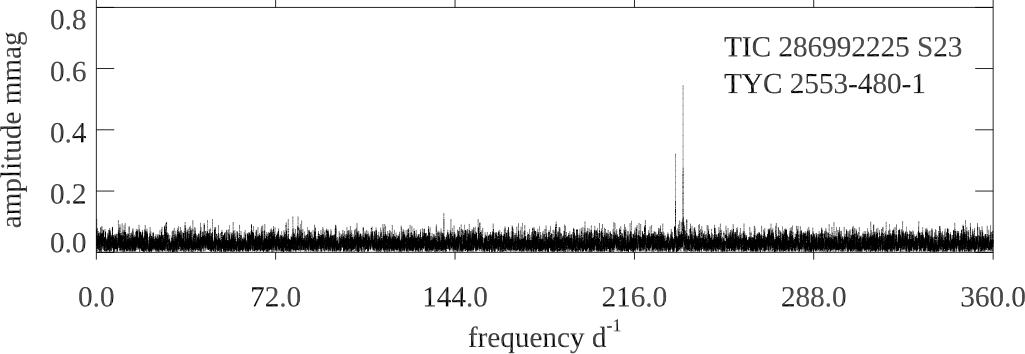}
  \includegraphics[width=0.45\linewidth,angle=0]{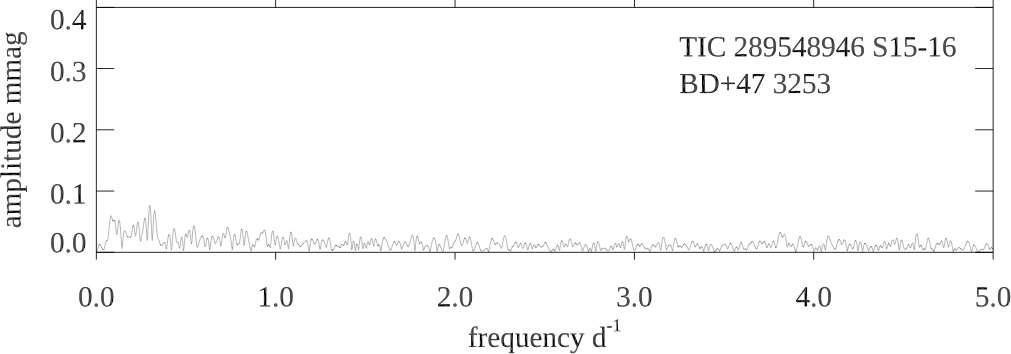}
  \includegraphics[width=0.45\linewidth,angle=0]{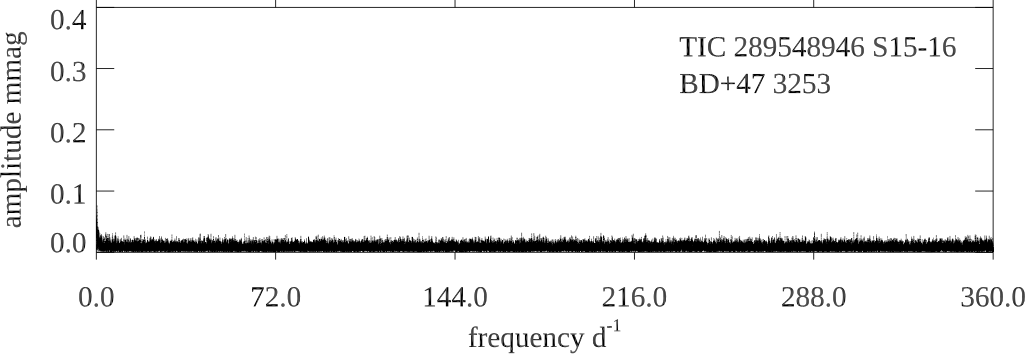}
  \caption{Amplitude spectra for the long-period Ap stars, continued. TIC\,286992225  is an roAp star (section\,\ref{roAp}).}
  \label{fig:ssrAp6}
  \end{figure*}

\begin{figure*}[p]
  \centering
  \includegraphics[width=0.45\linewidth,angle=0]{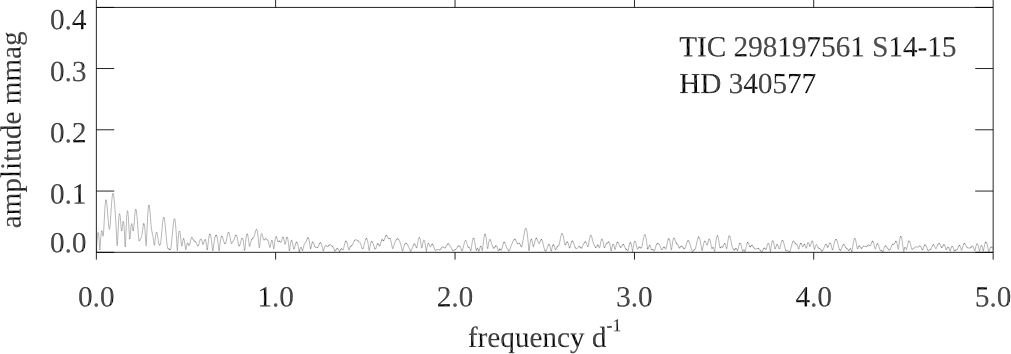}
  \includegraphics[width=0.45\linewidth,angle=0]{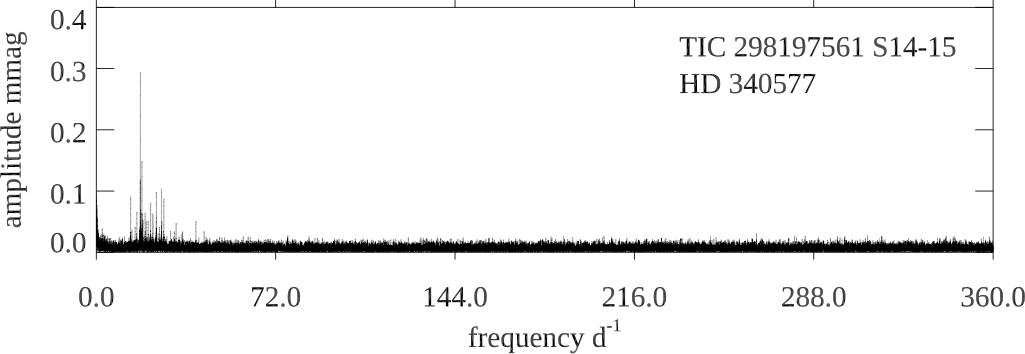}
  \includegraphics[width=0.45\linewidth,angle=0]{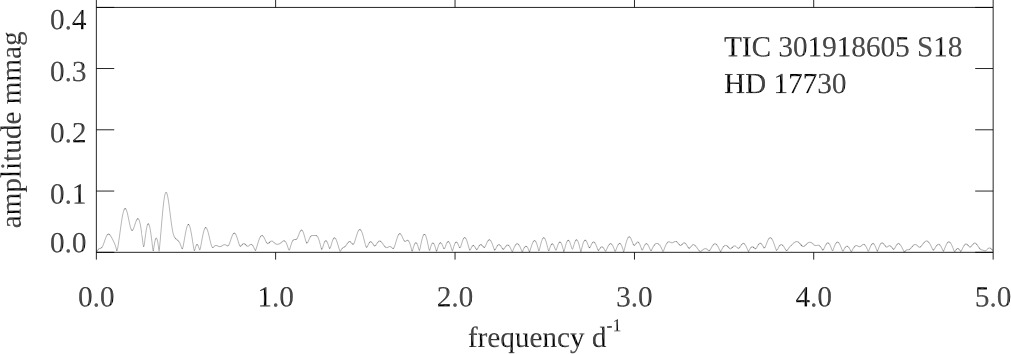}
  \includegraphics[width=0.45\linewidth,angle=0]{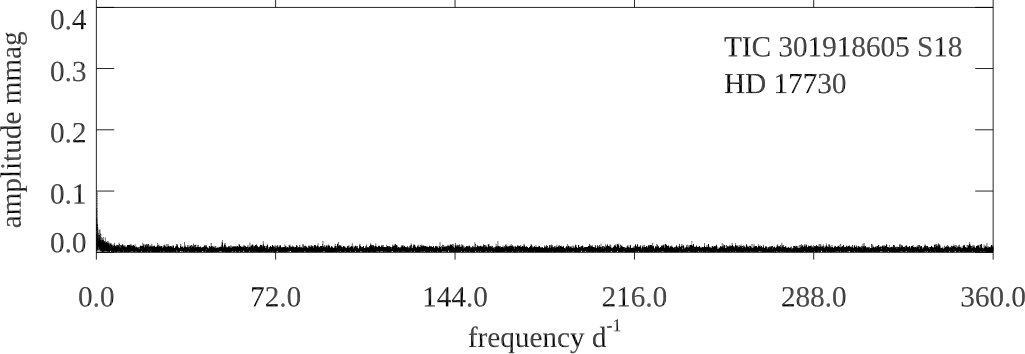}
  \includegraphics[width=0.45\linewidth,angle=0]{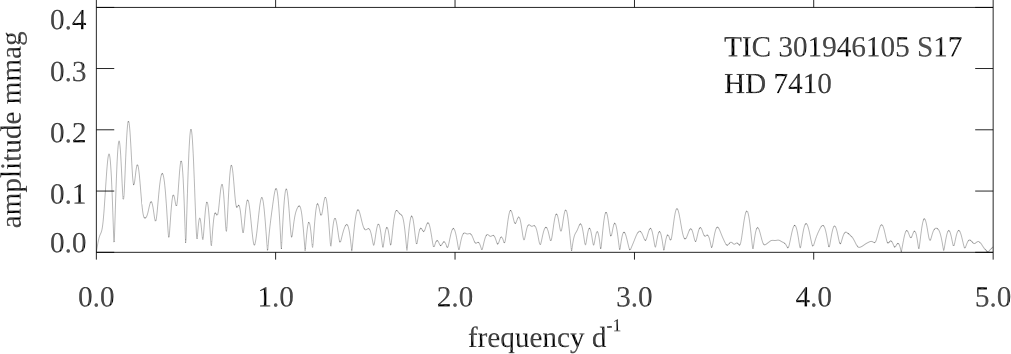}
  \includegraphics[width=0.45\linewidth,angle=0]{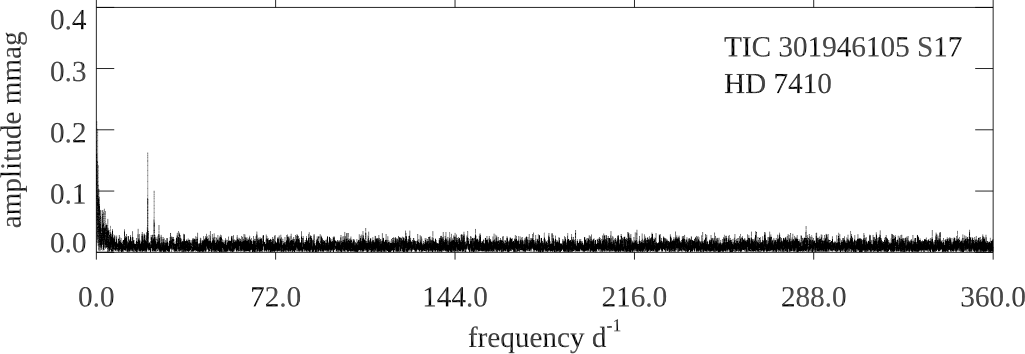}
  \includegraphics[width=0.45\linewidth,angle=0]{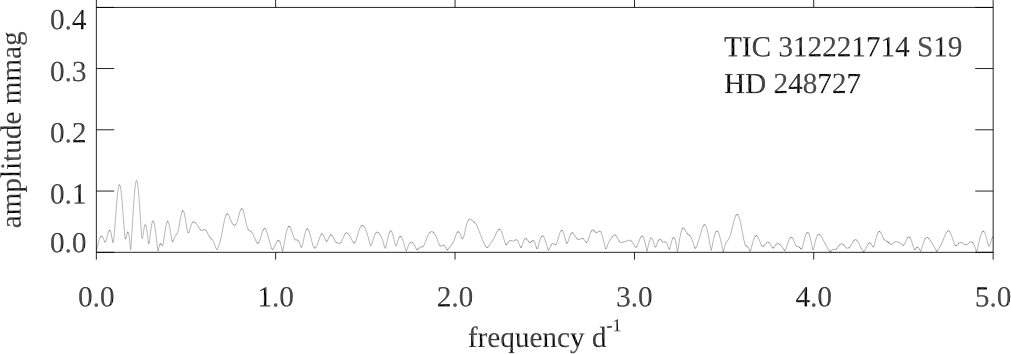}
  \includegraphics[width=0.45\linewidth,angle=0]{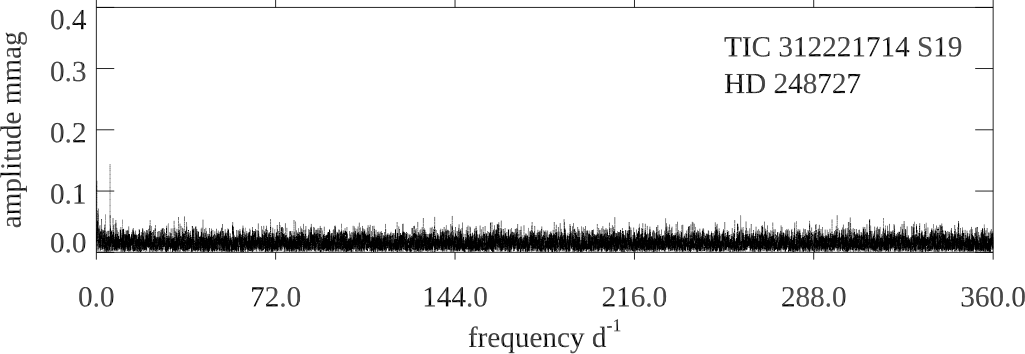}
  \includegraphics[width=0.45\linewidth,angle=0]{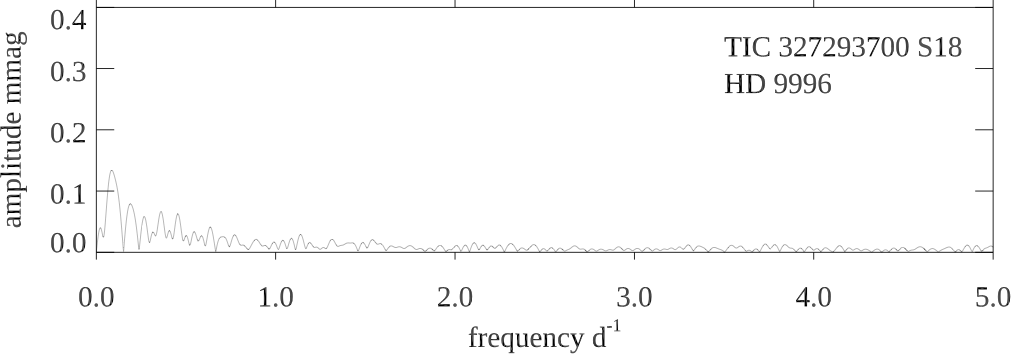}
  \includegraphics[width=0.45\linewidth,angle=0]{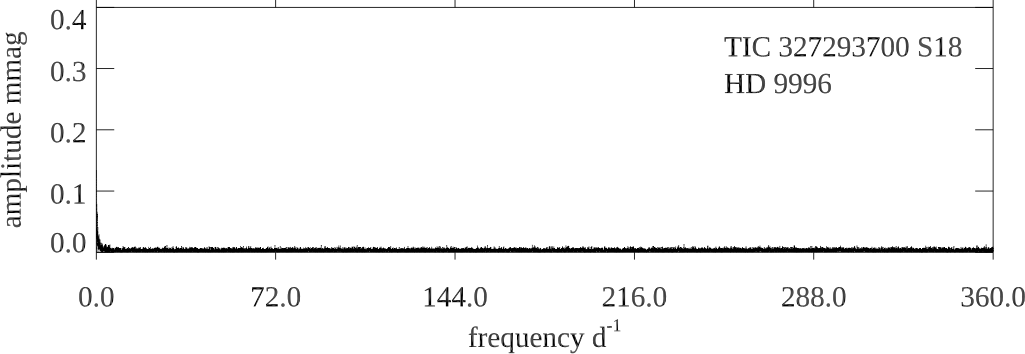}
  \includegraphics[width=0.45\linewidth,angle=0]{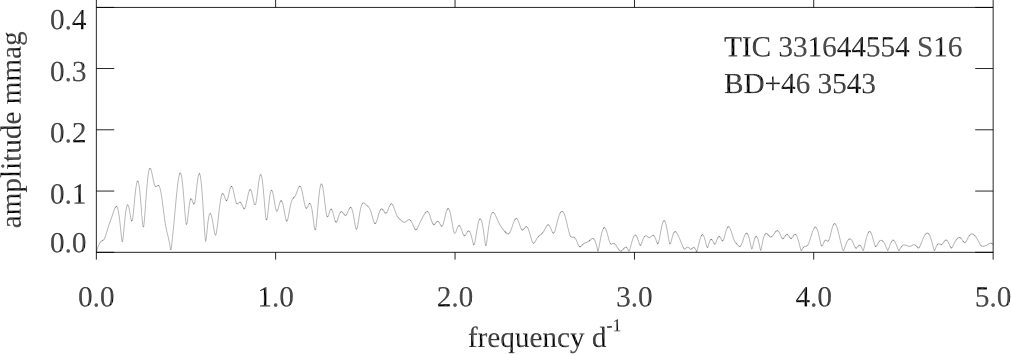}
  \includegraphics[width=0.45\linewidth,angle=0]{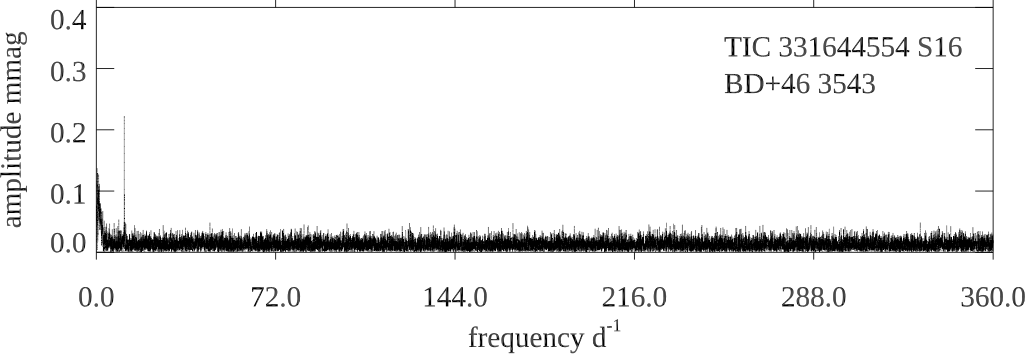}
  \includegraphics[width=0.45\linewidth,angle=0]{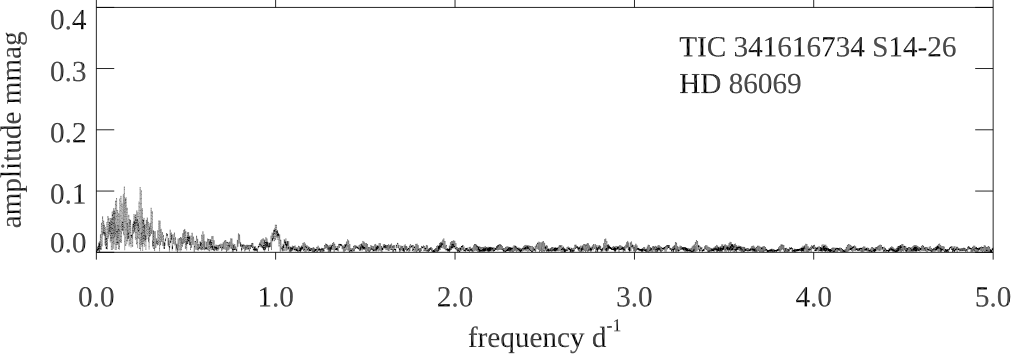}
  \includegraphics[width=0.45\linewidth,angle=0]{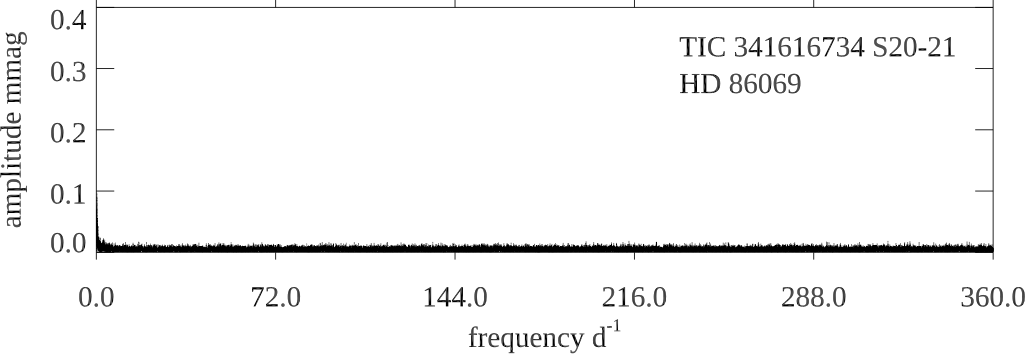}
  \caption{Amplitude spectra for the long-period Ap stars, continued. TIC\,298197561, TIC\,301918605, and TIC\,331644554 are $\delta$~Sct stars (Section\,\ref{deltasct} and Fig.\,\ref{fig:deltasct}). TIC\,312221714 has a peak at 5.5\,d$^{-1}$ that is between the expected frequency ranges for g~modes and p~modes in these stars; perhaps it is a mixed mode. }
  \label{fig:ssrAp7}
  \end{figure*}

\begin{figure*}[p]
  \centering
  \includegraphics[width=0.45\linewidth,angle=0]{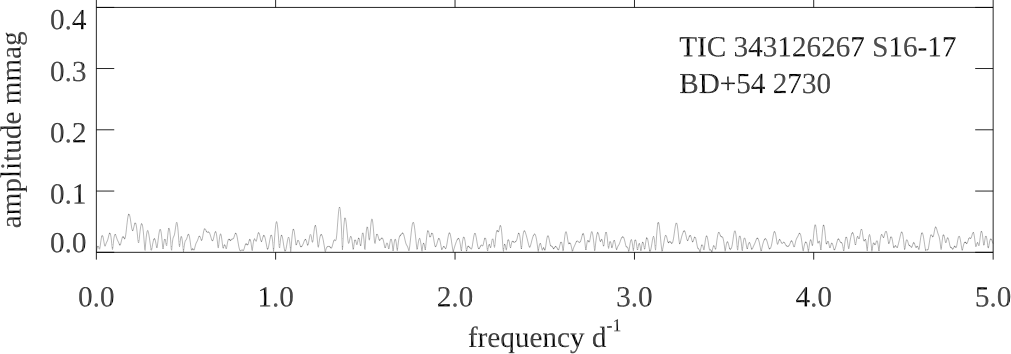}
  \includegraphics[width=0.45\linewidth,angle=0]{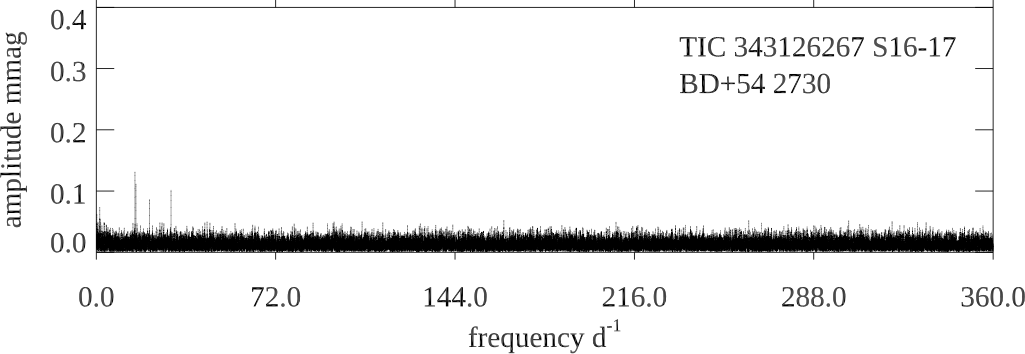}
  \includegraphics[width=0.45\linewidth,angle=0]{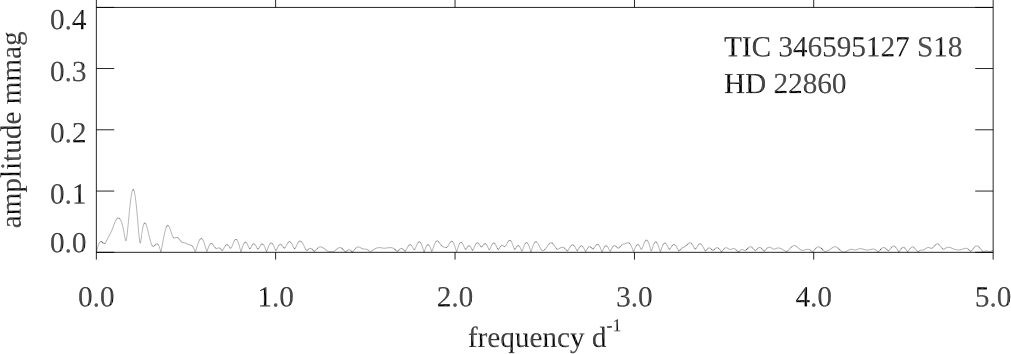}
  \includegraphics[width=0.45\linewidth,angle=0]{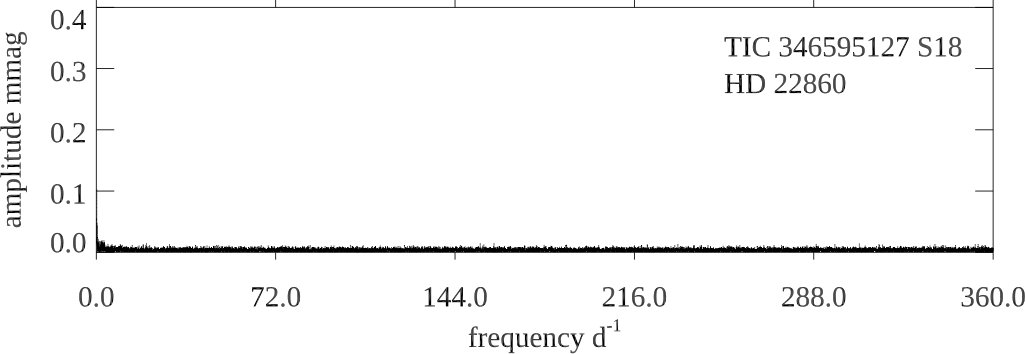}
  \includegraphics[width=0.45\linewidth,angle=0]{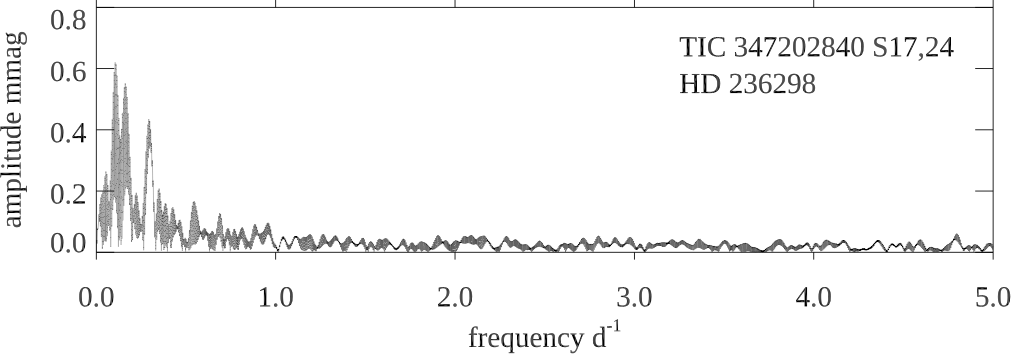}
  \includegraphics[width=0.45\linewidth,angle=0]{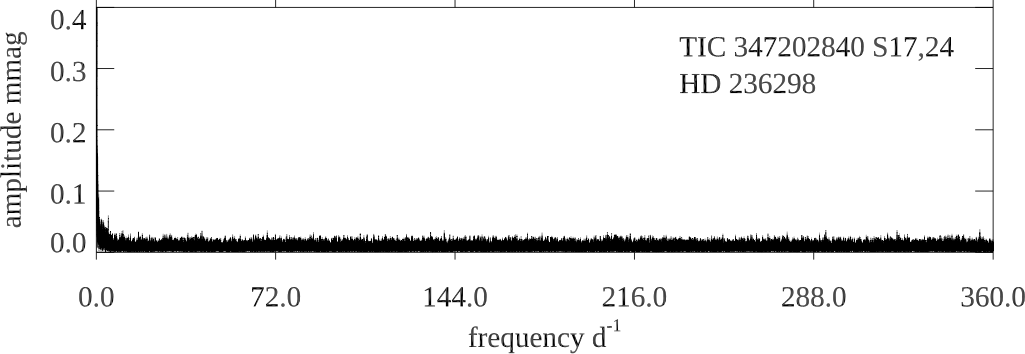}
  \includegraphics[width=0.45\linewidth,angle=0]{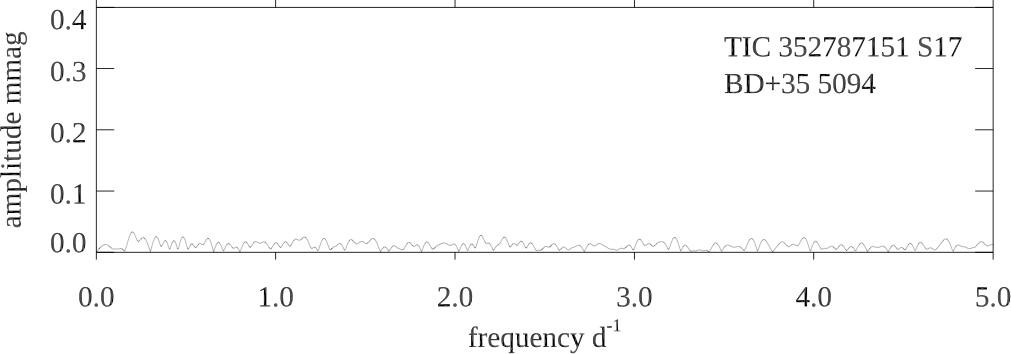}
  \includegraphics[width=0.45\linewidth,angle=0]{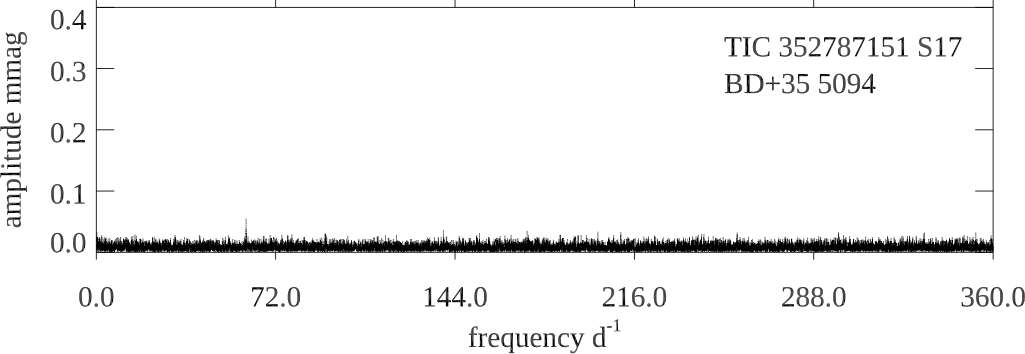}
  \includegraphics[width=0.45\linewidth,angle=0]{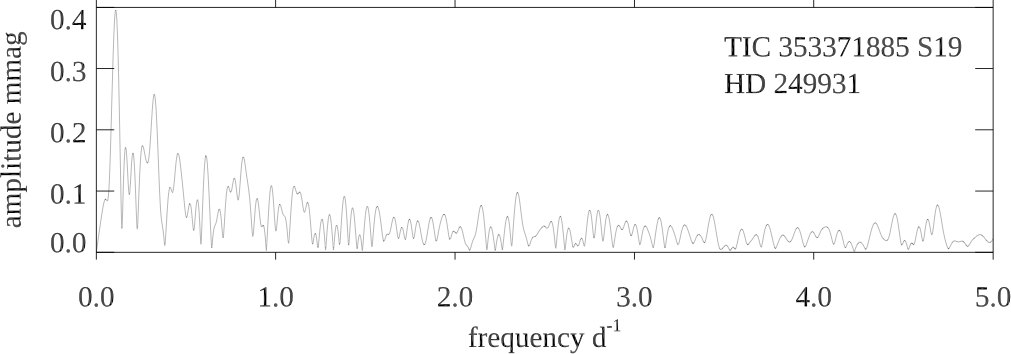}
  \includegraphics[width=0.45\linewidth,angle=0]{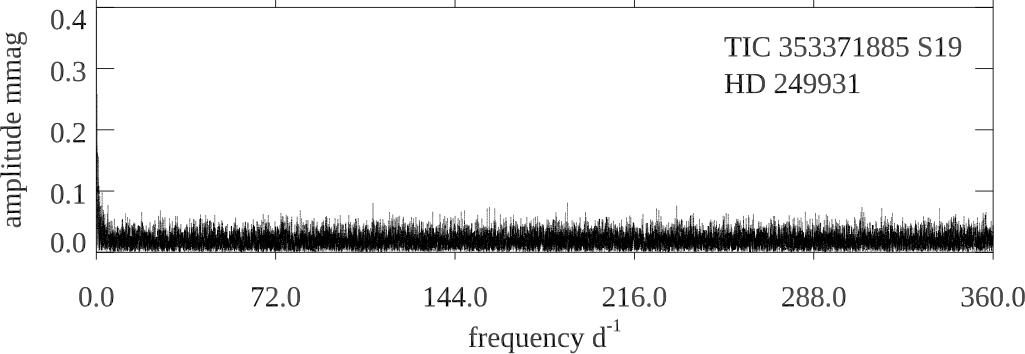}
  \includegraphics[width=0.45\linewidth,angle=0]{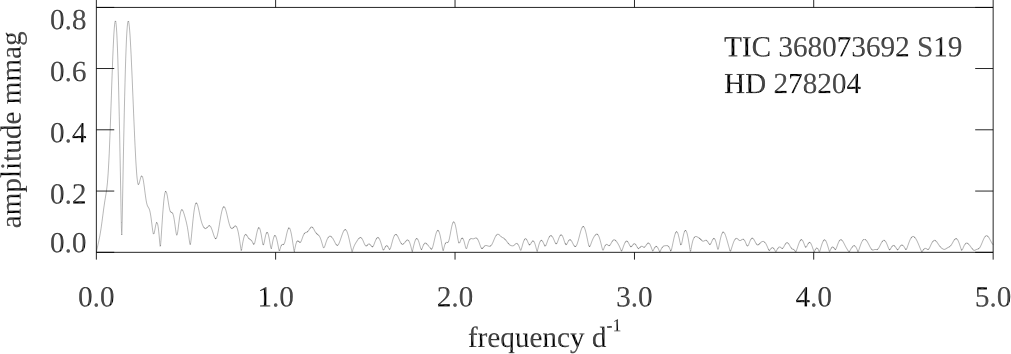}
  \includegraphics[width=0.45\linewidth,angle=0]{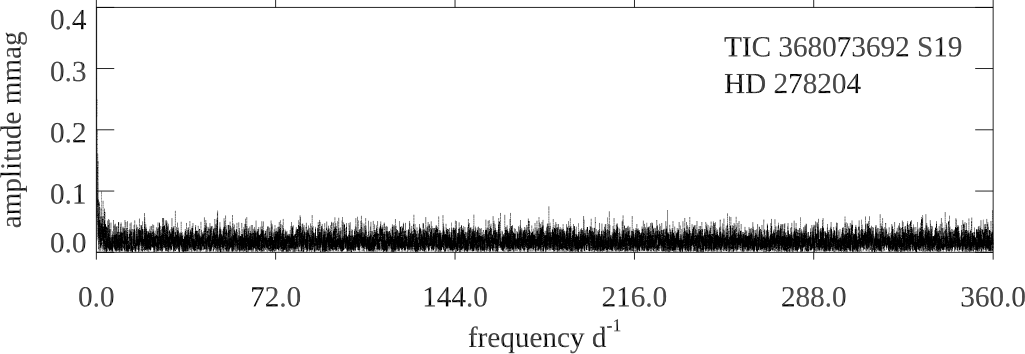}
  \includegraphics[width=0.45\linewidth,angle=0]{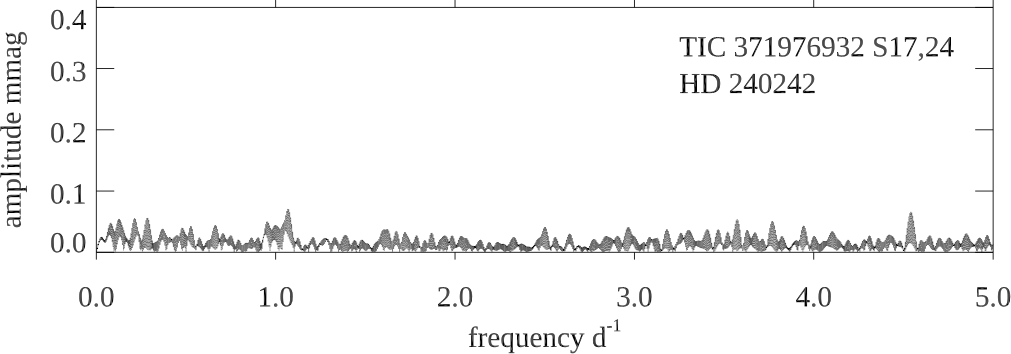}
  \includegraphics[width=0.45\linewidth,angle=0]{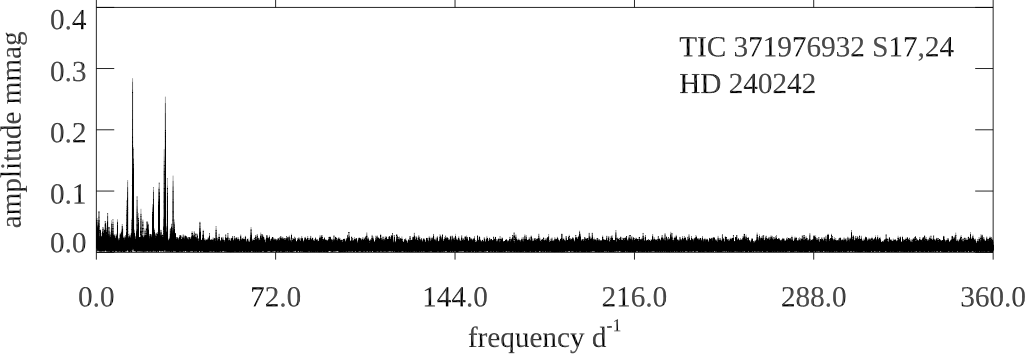}
  \caption{Amplitude spectra for the long-period Ap stars, continued. TIC\,343126267 and TIC\,371976932 are $\delta$~Sct stars (Section\,\ref{deltasct} and Fig.\,\ref{fig:deltasct}). TIC\,352787151 is a possible roAp star (section\,\ref{roAp}). The unexplained low frequency variability in TIC\,347202840 is different in the two sectors, hence is not periodic and is not rotational. }
  \label{fig:ssrAp8}
  \end{figure*}

\begin{figure*}[p]
  \centering
  \includegraphics[width=0.45\linewidth,angle=0]{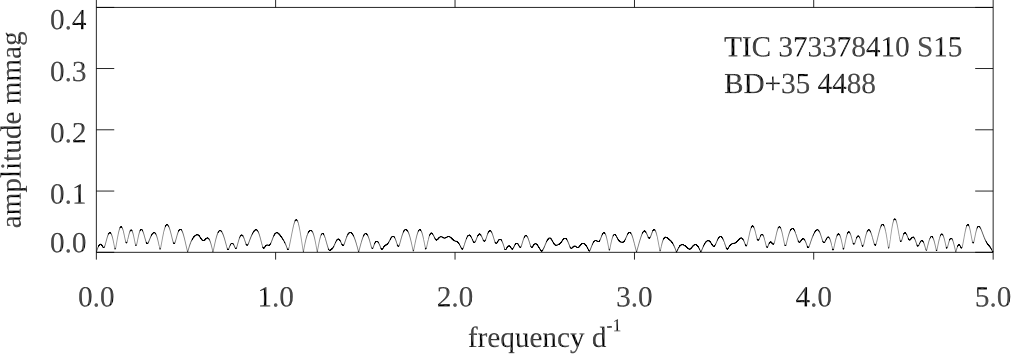}
  \includegraphics[width=0.45\linewidth,angle=0]{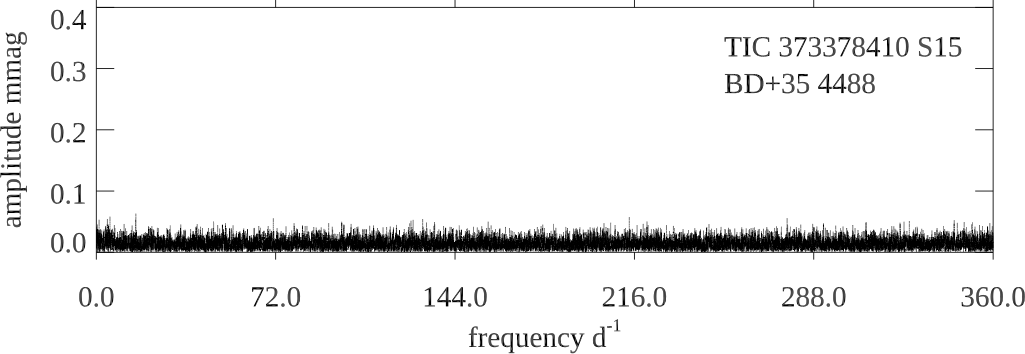}
  \includegraphics[width=0.45\linewidth,angle=0]{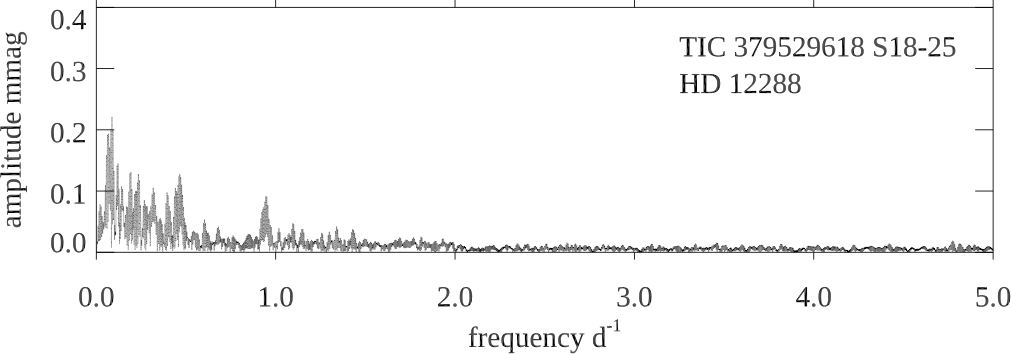}
  \includegraphics[width=0.45\linewidth,angle=0]{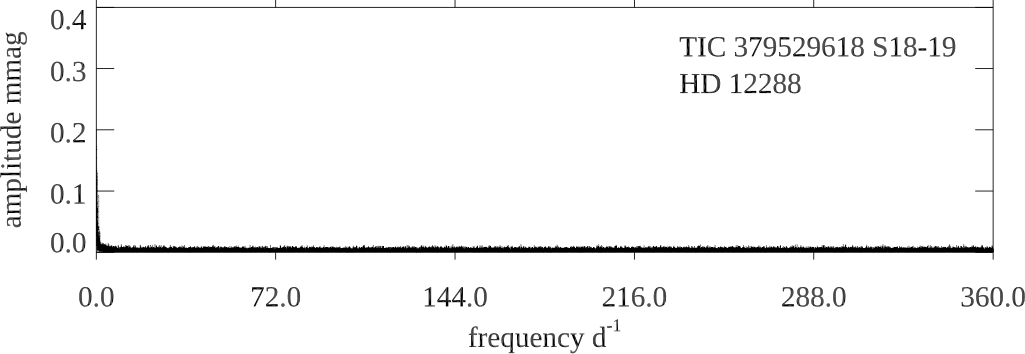}
  \includegraphics[width=0.45\linewidth,angle=0]{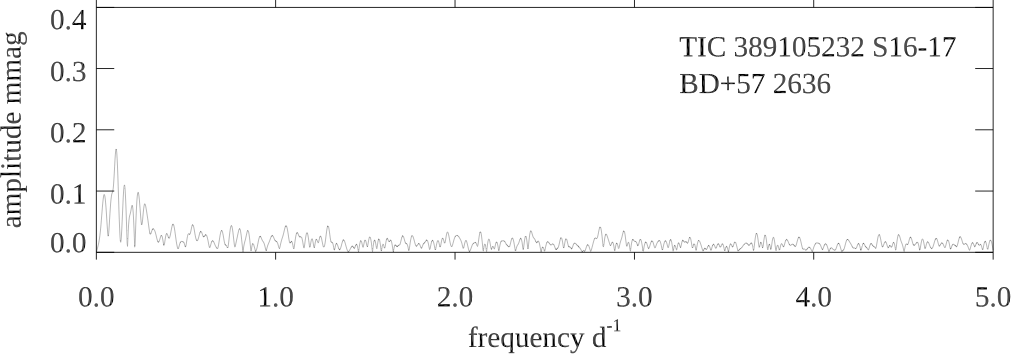}
  \includegraphics[width=0.45\linewidth,angle=0]{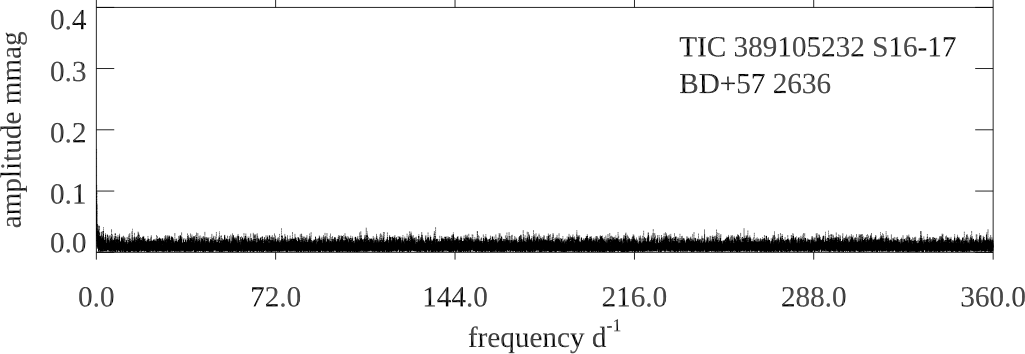}
  \includegraphics[width=0.45\linewidth,angle=0]{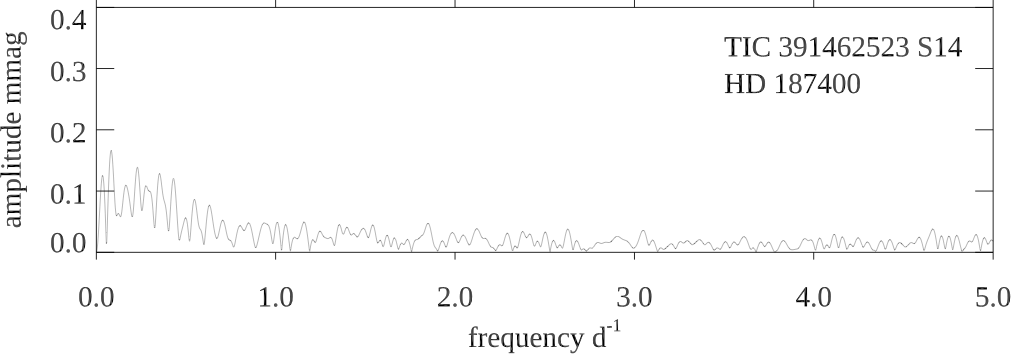}
  \includegraphics[width=0.45\linewidth,angle=0]{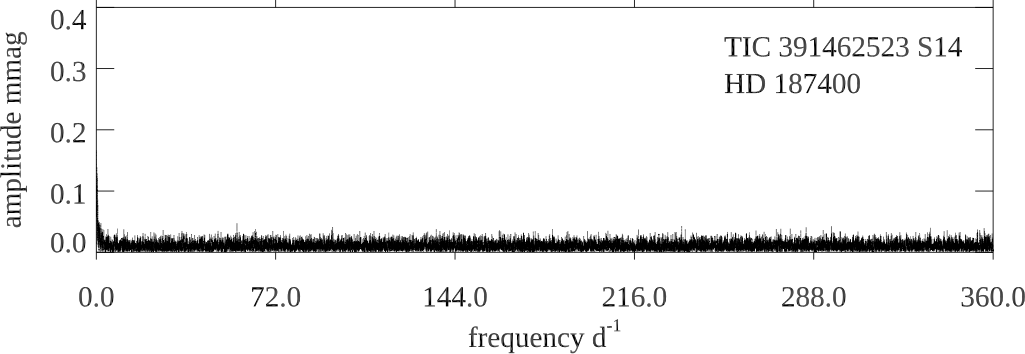}
  \includegraphics[width=0.45\linewidth,angle=0]{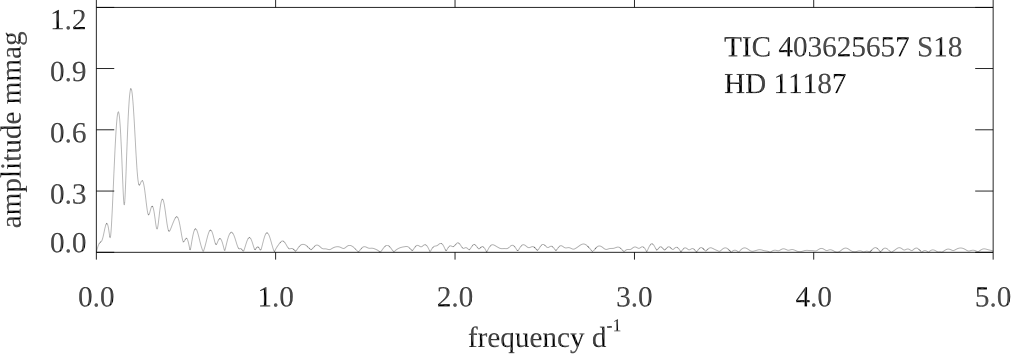}
  \includegraphics[width=0.45\linewidth,angle=0]{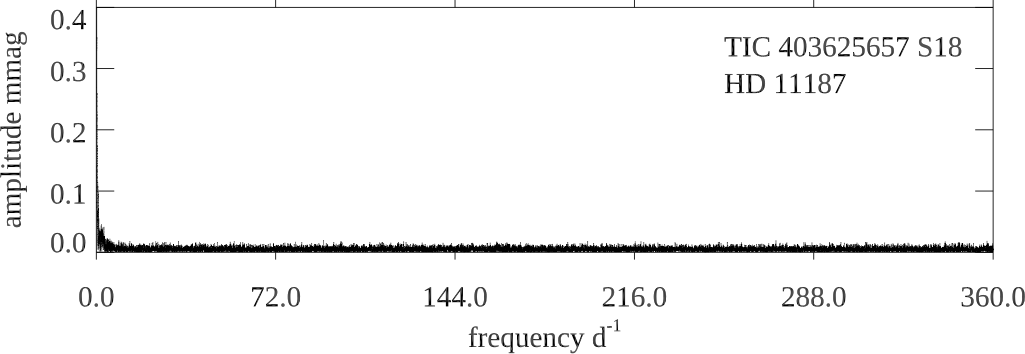}
  \includegraphics[width=0.45\linewidth,angle=0]{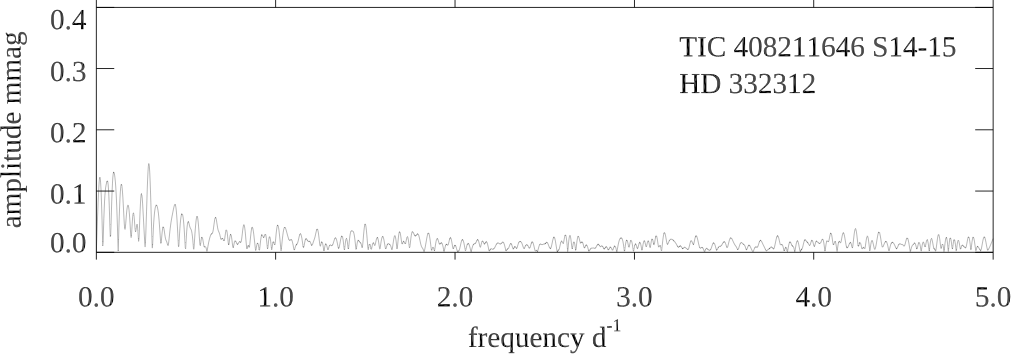}
  \includegraphics[width=0.45\linewidth,angle=0]{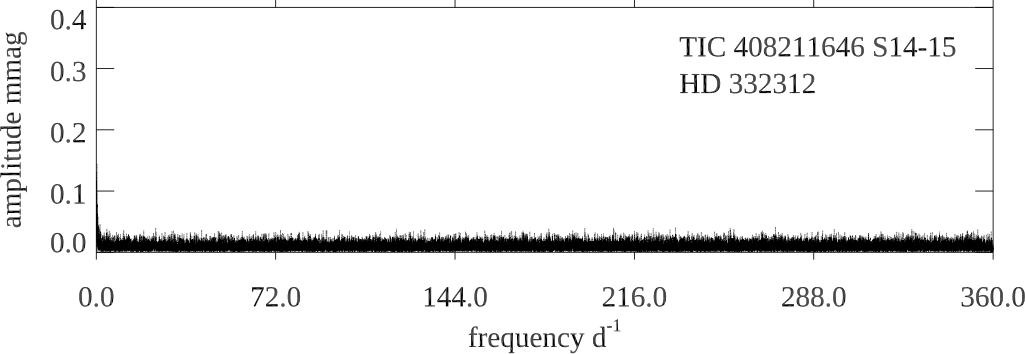}
  \includegraphics[width=0.45\linewidth,angle=0]{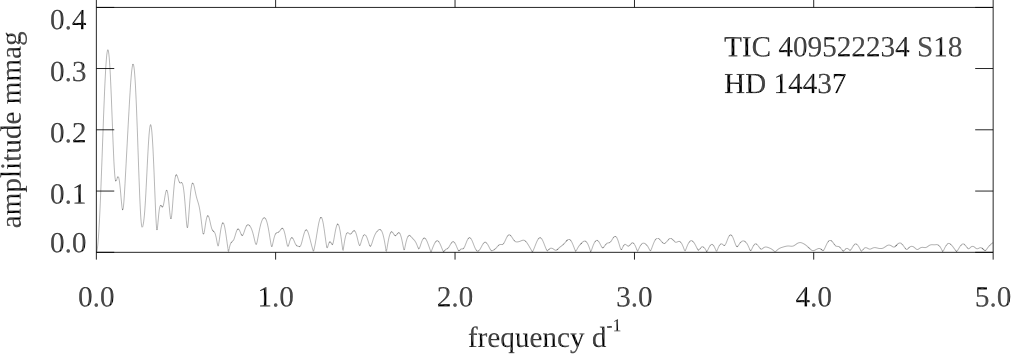}
  \includegraphics[width=0.45\linewidth,angle=0]{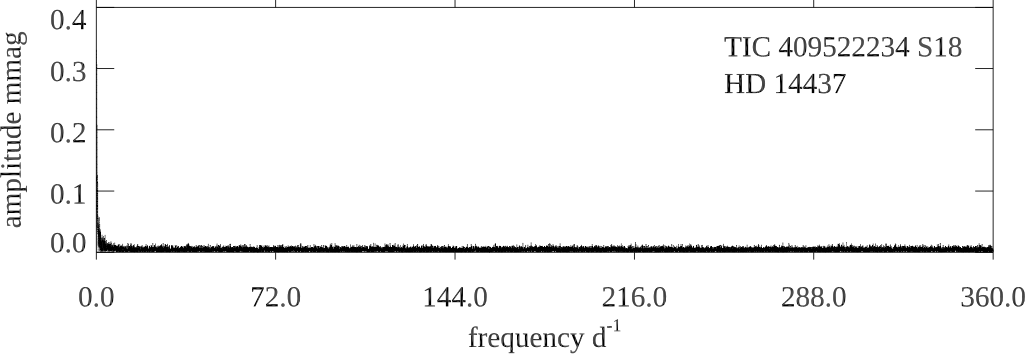}
  \caption{Amplitude spectra for the long-period Ap stars, continued. TIC\,379529618 shows some low-frequency variability that is not rotational. The peak at 0.96\,d$^{-1}$ is coherent across two sectors, but does not appear in the third sector. It is of unknown origin. }
  \label{fig:ssrAp9}
  \end{figure*}

\begin{figure*}[p]
  \centering
  \includegraphics[width=0.45\linewidth,angle=0]{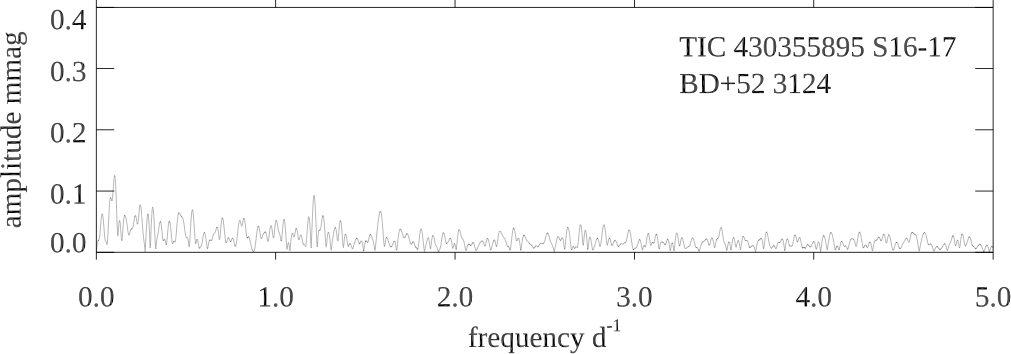}
  \includegraphics[width=0.45\linewidth,angle=0]{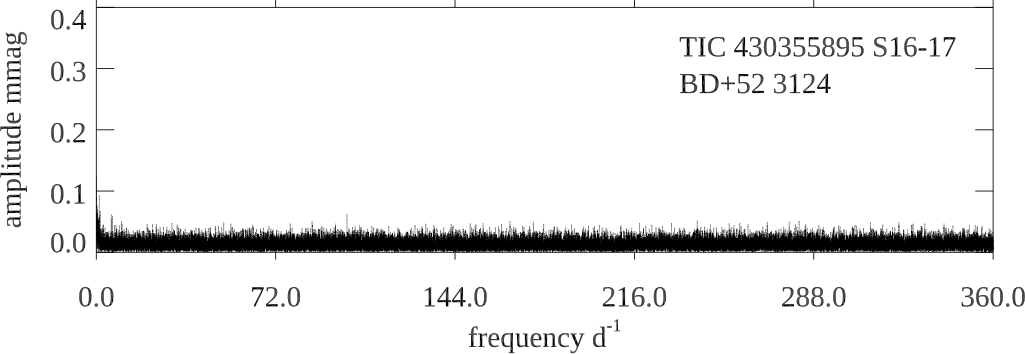}
  \includegraphics[width=0.45\linewidth,angle=0]{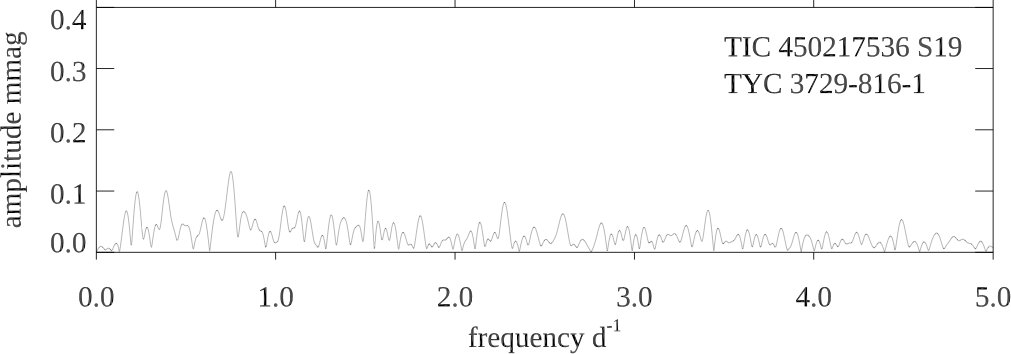}
  \includegraphics[width=0.45\linewidth,angle=0]{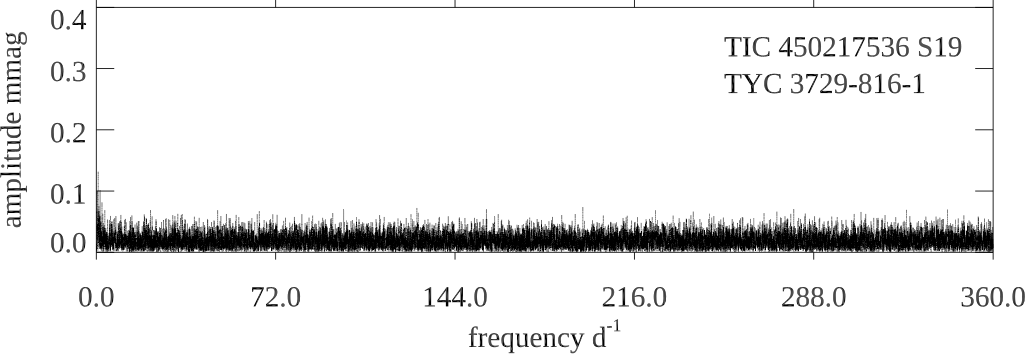}
  \includegraphics[width=0.45\linewidth,angle=0]{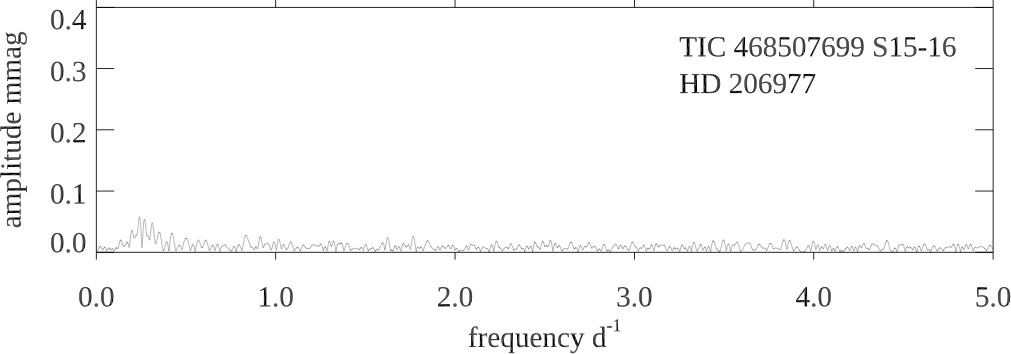}
  \includegraphics[width=0.45\linewidth,angle=0]{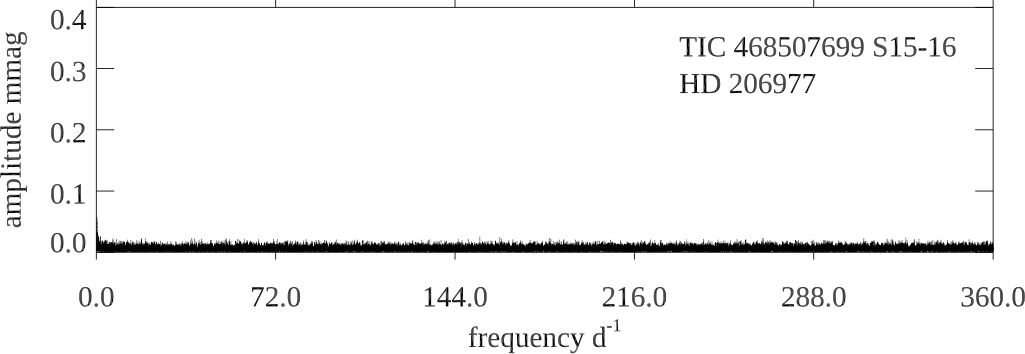}
  \includegraphics[width=0.45\linewidth,angle=0]{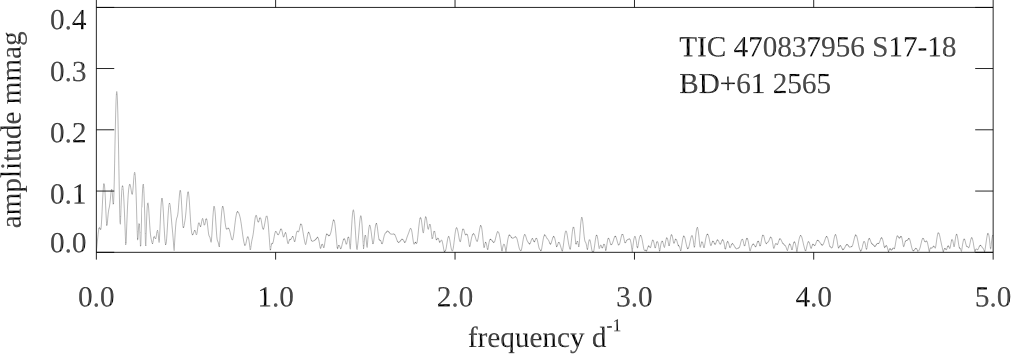}
  \includegraphics[width=0.45\linewidth,angle=0]{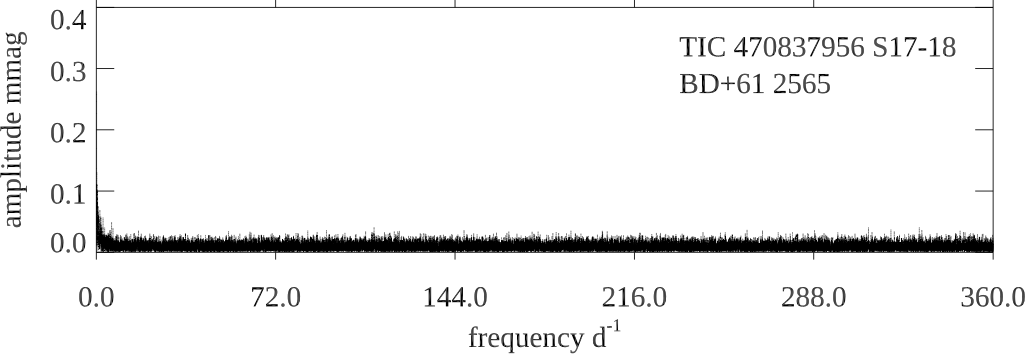}
  \includegraphics[width=0.45\linewidth,angle=0]{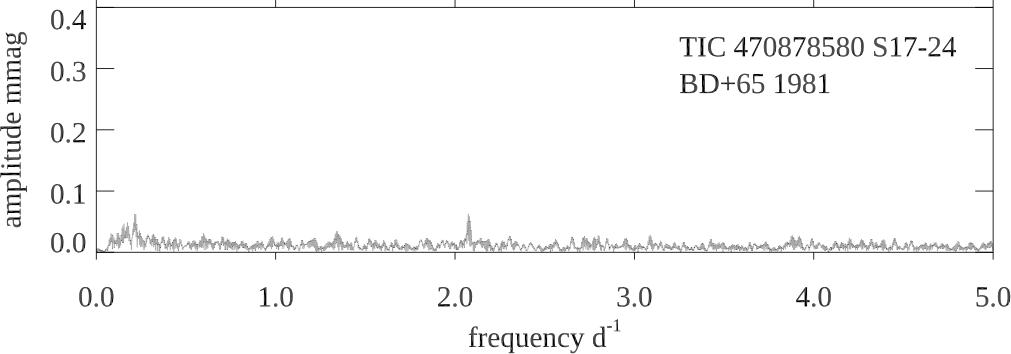}
  \includegraphics[width=0.45\linewidth,angle=0]{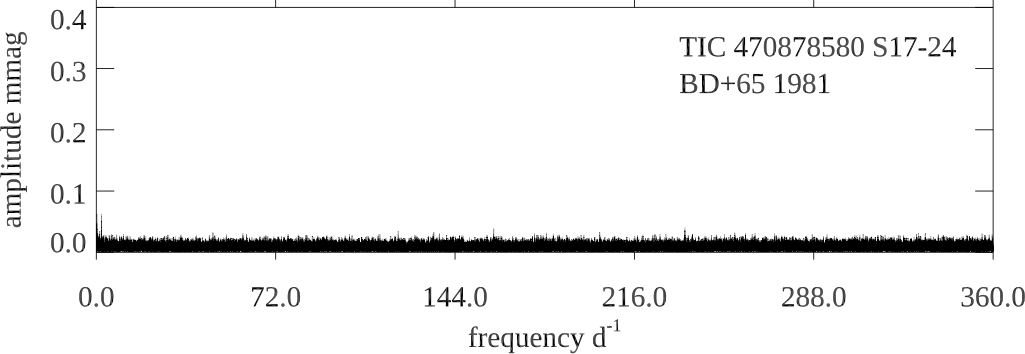}
  \caption{Amplitude spectra for the long-period Ap stars, continued.}
  \label{fig:ssrAp10}
  \end{figure*}

Figures~\ref{fig:ssrAp1} to \ref{fig:ssrAp10} show the amplitude
spectra of those Ap stars that we have identified as having long
rotation periods and that are not shown in Fig.~\ref{fig:ssrAp2}.

\afterpage{\clearpage} 

\afterpage{\clearpage} 

\afterpage{\clearpage} 

\afterpage{\clearpage} 

\afterpage{\clearpage} 

\afterpage{\clearpage} 

\afterpage{\clearpage} 

\afterpage{\clearpage} 

\end{document}